\documentclass{aastex63}

\usepackage[ruled,vlined]{algorithm2e}
\usepackage[utf8]{inputenc}
\usepackage{comment}
\usepackage{dsfont}
\usepackage{natbib}
\usepackage{tikz}

\usepackage{tabulary}

\usetikzlibrary{angles,quotes}

\usepackage{hyperref}
\hypersetup{
    colorlinks=true,
    linkcolor=blue,
    filecolor=magenta,      
    urlcolor=cyan,
}

\usepackage{amsmath}
\usepackage{amssymb}
\usepackage{amsfonts}
\usepackage{algorithmic}

\usepackage{graphicx}
\usepackage{textcomp}
\usepackage{booktabs}
\usepackage{bbm}
\usepackage{xcolor}
\usepackage{svg}

\newcommand*{\vertbar}{\rule[-1ex]{0.5pt}{2.5ex}}

\newcommand{\ip}[2]{\left\langle#1,#2\right\rangle}

\DeclareMathOperator*{\argmax}{arg\,max}
\DeclareMathOperator*{\argmin}{arg\,min}

\DeclareMathOperator{\rank}{rank}
\DeclareMathOperator{\diag}{diag}
\DeclareMathOperator{\mvec}{vec}

\received{}
\revised{}
\accepted{\today}

\submitjournal{AJ}

\begin{document}

\title{Robust Detrending of Spatially Correlated Systematics in Kepler Light Curves Using Low-Rank Methods }


\correspondingauthor{Jamila Taaki}
\email{xiaziyna@gmail.com}

\author[0000-0001-5475-1975]{Jamila S. Taaki}
\affiliation{Department of Electrical and Computer Engineering,
University of Illinois at Urbana-Champaign\\
306 N. Wright St. MC 702, Urbana, IL 61801-2918}

\author[0000-0001-6233-8347]{Athol J. Kemball}
\affiliation{Department of Astronomy, University of Illinois at Urbana-Champaign\\
1002 W. Green Street, Urbana, IL 61801-3074}
\affiliation{School of Physics, University of the Witwatersrand, PO Box Wits, Johannesburg, South Africa}

\author{Farzad Kamalabadi}
\affiliation{Department of Electrical and Computer Engineering,
University of Illinois at Urbana-Champaign\\
306 N. Wright St. MC 702, Urbana, IL 61801-2918}

\begin{abstract}
Light curves produced by wide-field exoplanet transit surveys such as CoRoT, Kepler, and TESS are affected by sensor-wide systematic noise which is correlated both spatiotemporally and with other instrumental parameters such as photometric magnitude.  Robust and effective systematics mitigation is necessary to achieve the level of photometric accuracy required to detect exoplanet transits and to faithfully recover other forms of intrinsic astrophysical variability. We demonstrate the feasibility of a new exploratory algorithm to remove spatially-correlated systematic noise and detrend light curves obtained from wide-field transit surveys. This spatial systematics algorithm is data-driven and fits a low-rank linear model for the systematics conditioned on a total-variation spatial constraint. The total-variation constraint models spatial systematic structure across the sensor on a foundational level. The fit is performed using gradient descent applied to, a variable reduced least-squares penalty and a modified form of total-variation prior; both the systematics basis vectors and their weighting coefficients are iteratively varied. The algorithm was numerically evaluated against a reference principal component analysis, using both signal injection on a selected Kepler dataset, as well as full simulations within the same Kepler coordinate framework. We find our algorithm to reduce overfitting of astrophysical variability over longer signal timescales (days) while performing comparably relative to the reference method for exoplanet transit timescales. The algorithm performance and application is assessed and future development outlined.

\end{abstract}


\section{Introduction} \label{sec:intro}

Transit surveys have contributed significantly to exoplanet science over the past two decades, uncovering exoplanet population statistics at current survey completeness limits \citep{Batalha12647, bryson2020probabilistic}. Space-based surveys, including CoRoT \citep{auvergne2009corot}, Kepler \citep{borucki}, and TESS \citep{ricker2014transiting} have led to the discovery of over  5400\footnote{https://exoplanetarchive.ipac.caltech.edu/} exoplanets to date. These wide-field surveys include a large number of candidate systems and 
require a photometric precision after standard calibration and processing (residual photometric precision) adequate to detect exoplanet transits \citep{Deeg2018}. Achieving low residual photometric noise requires the optimization of both the instrument design and the associated algorithms for instrumental noise removal. In this paper, we describe a method for addressing spatially-correlated systematic noise across a wide-field imaging sensor used for exoplanet transit detection.\\

If an exoplanet transit induces a fractional flux density variation $\triangle F$ in observations, with a differential photometric precision $\sigma_p$ (after processing) then the signal-to-noise ratio is $S/N \sim \frac{\triangle F}{\sigma_p}$ \citep{Deeg2018}. For example, an Earth-Sun transit requires $\sigma_p \sim 80$ parts-per-million (ppm) over the transit duration $T \sim$ hours \citep{Caldwell_2010}. The Kepler telescope achieved $\sigma_p \sim 30$ ppm over $T= 6.5$h for stars with Kepler magnitude $K_p \sim 12$ \citep{koch, Gilliland_2011, cdpp_christ}. TESS achieved $\sigma_p \sim 230$ ppm over $T \sim 1$ hr for a star with TESS magnitude $Tp \sim 10$ which is sufficient to detect super-Earths around bright stars \citep{tess_sec_1}. For reference, the Hubble Space Telescopes Space Telescope Imaging Spectrograph (HST/STIS) can attain $\sigma_p \sim 120$ ppm over $T= 45$ min \citep{hst} and non-wide-field ground-based telescopes can reach a precision sufficient to detect large Jovian exoplanets \citep{ground, Stefansson_2017}. \\

The instrumental response will vary over a range of timescales and across various spatial scales producing systematic noise effects, as expected from general principles \citep{mclean2008electronic}. The primary data products for the Kepler and TESS telescopes include light curves derived using optimized simple aperture photometry (SAP) or image data providing pixel-level light curves \citep{kep_data_handbook, tess_data_handbook}. Algorithmic innovations to suppress the residual differential photometric precision $\sigma_p$, arising from unmodeled systematic noise effects, are therefore critical to detect weak exoplanet signals. These systematic effects have diverse physical origins \citep{kep_data_handbook, tess_data_handbook}. A single pixel exhibits a non-uniform sensitivity across its surface \citep{intrapixel, multiwavelength}. In addition, there are variations in pixel-to-pixel sensitivity and instrumental pixel response \citep{kep_handbook} including CCD pattern, read, photon, and quantization noise \citep{mclean2008electronic, Gilliland_2011, Caldwell_2010}; instrumental CCD terms are expected to vary by CCD output channel or module. A net pixel response function (PRF) \citep{Bryson_2010} captures the spatially-variant response across the detector including the spatial variation of the optical PSF and instrumental detector sensitivity, amongst related factors \citep{kep_data_handbook}. The spatial response of the detector may, however, include unmodeled error due to pointing jitter, focus changes, uncorrected differential velocity aberration terms, and their interaction \citep{kep_handbook}. Thermal effects as well as discrete spacecraft guidance and downlink control events, including momentum dumps, may produce temporal and spatial variation in the systematics, some abrupt \citep{kep_handbook, tess_handbook}. Further, bright astrophysical sources may produce pixel saturation or broadening of the PRF \citep{kep_handbook, tess_handbook}.\\

The first correction for noise effects is typically applied in early robust calibration pipelines \citep{kep_data_handbook, tess_data_handbook}. However residual spatially- and time-variant systematics unavoidably remain in the calibrated data products given the complex instrumental response at the current level of photometric accuracy. We denote these residual systematics contributions to each light curve $i$ at discrete time sample $n$ as $\mathbf{l}_{i}[n]$ (Section~\ref{sec: decompose}) where the data may be either pixel or SAP light curves. The systematics $\mathbf{l}_i [n]$ has the same discrete time sampling as the measured light curve. The residual systematics vary spatially over the sensor, reflected in equivalent notation $\mathbf{l}_{x,y}[n]$, where light curve $i$ maps to sensor position $(x,y)$. \citet{Petigura_2012} show an example of spatially-varying residual systematic noise in Kepler data by characterizing correlations across the sensor. \citet{moreno} similarly characterize spatial systematics correlation in Kepler/K2 data. An a priori analytic instrumental model for $\mathbf{l}_{x,y}[n]$ is intractable in practice and data-driven approaches, informed by physically-realistic instrumental assumptions, provide the most effective approach to mitigate the residual systematic noise in the target light curves. There is a rich history in the literature on this subject which we review briefly only to place our method in context.\\

A class of methods (hereinafter {\it external parameter decorrelation} methods \citep{Bakos2007}) model the functional form of the residual systematics $\bold{l}_{x,y}[n]$ as correlates of other explanatory variables such as pointing error estimates or ancillary engineering data (PDC-LS, \citealt{Twicken2010}). \citet{charbonneau_detection_2005} demonstrated Spitzer\footnote{https://www.spitzer.caltech.edu/} aperture photometry correction using decorrelation with pointing errors estimated using target centroiding. There has been substantial development in this latter area, especially inspired by the pointing challenges of the Kepler K2 mission and pixel-level detrending, including the work described by \citet{precise_prf, huang, aigrain, Lund_2015} and \citet{Crossfield_2015}. External parameter decorrelation methods implicitly include spatiotemporal variability in the residual systematics via the encoding of their functional form in the proxy variables.\\

A second class of methods (hereinafter {\it cotrending} methods) makes the foundational assumption that a set of measured light curves, or their computed vector basis, comprise an efficient basis of regressors over which to expand the unknown functional form of $\bold{l}_{x,y}[n]$ as a low-rank linear model. 
This implicitly assumes that the residual systematics are temporally and spatially 
correlated across the sensor; this is similarly a physically reasonable assumption. The Trend-Filtering-Algorithm (TFA, \citealt{tfa}) directly uses a uniform selection of external light curves as regressors in a least-squares fit to detrend an individual target light curve. The cotrending approach is analogous to the joint coupled least-squares solution for the product of stellar extinction coefficients and airmasses as posed in a global solution for wide-field multi-object photometry (Sysrem, \citealt{zucker}); this solution is equivalent to a Principal Component Analysis (PCA) of target observations. A refinement including additive external parameters is provided by SARS \citep{sars}.\\

Cotrending methods have seen significant algorithmic refinement. The PDC-MAP algorithm \citep{stumpe2012kepler, smith2012kepler} constructs a set of cotrending basis vectors (CBV) using singular value decomposition (SVD) on a template set of light curves selected for their high degree of correlation (therefore capturing foundational instrumental trends) and quiescence. Each CBV is fit against the target light curve using Bayesian maximum a posteriori (MAP) methods to obtain relative coefficient weightings for each basis term, the sum of which is then subtracted from the target time series to remove systematic effects. An empirical Bayesian prior is used to constrain overfitting, constructed on the variation of coefficient value over stellar magnitude and spatial position on the sensor. The CBV are orthogonal by mathematical construction and, therefore, do not map directly to constituent instrumental effects; this dilutes the variable dependence in the empirical coefficient prior.  Instrumental effects are often separated by characteristic timescale and this mapping can be improved (MS-MAP, \citealt{msmap}) by deriving separate CBV for different timescale wavelet sub-bands; this provides better temporal separation in the spatiotemporal dependence of $\bold{l}_{x,y}[n]$. A net composite systematic correction is applied as the multiscale combination across each sub-band. The Astrophysically Robust Correction algorithm (ARC; \citealt{rob1}, \citealt{aig}) is related conceptually to PDC-MAP but uses automatic relevance coefficient priors to maximize model evidence for instrumental trends. \\ 

Additional cotrending methods include the Causal Pixel Model (CPM, \citealt{cpm}) that uses a template (or training) set of regressor pixel-data light curves that are separated in time from the target light curve sample being corrected for systematics. The training set is spatially separated from the target but the target light curve itself is included to include autoregressive information. Both measures suppress residual variability except on short transit timescales. A successor method is described by \citet{Hattori_2022}. The method described by \citet{foreman2015systematic} uses a large training set to derive an expanded CBV but uses joint estimation of systematics and the transit signal in order to better constrain the problem. A related Bayesian formulation is described by \citet{Taaki_2020}. The Pixel Level Decorrelation algorithm (PLD, \citealt{deming}) was developed to implicitly correct photometric decorrelation with pointing error in Spitzer data. PLD uses pixel light curve regressors within a summed aperture to model SAP light curves. The regressors encode pointing errors without the need for explicit centroiding as in external parameter decorrelation approaches. The PLD method was extended by \citet{Luger_2016} for application to Kepler K2 data by including higher-order terms in the SAP flux dependence on pointing error, by modeling astrophysical variability using a Gaussian process, and by constructing a PCA basis from the regressors to better constrain the model. The method was further improved \citep{everest2} by including L2 coefficient regularization, which was also used in CPM, and by incorporating PLD vectors from nearby stars (nPLD) to improve sensitivity to pointing error for faint stars.\\

The cotrending methods described above address challenges in estimating intrinsic astrophysical variability. Foundationally it is difficult to perfectly separate systematics and astrophysical variability a priori in a light curve regressor or derived basis set. Therefore the linear systematics model may include and absorb true astrophysical variability (overfitting) or inject spurious variability into detrended light curves. There may also be incidental correlation between a relatively clean systematics basis and astrophysical variability \citep{Smith_2018}. Overfitting can be reduced by judicious use of prior constraints that represent a physically realistic instrumental response; these constraints may also significantly improve the condition of the problem.\\

In the cotrending methods described above, spatial structure is incorporated in the systematic noise model using several different approaches. In the simplest form, the spatial variation constraint is imposed implicitly by cotrending targets across a discrete sensor region as a whole (e.g. a CCD output channel) \citep{ smith2012kepler, rob1,  msmap, everest2, Lund_2021}. 
Spatial variation has also been incorporated by restricting regressors by proximity to the target light curve \citep{everest2, Lund_2021}, or by directly introducing a parametric dependence of fitted systematics on sensor position in an empirical prior \citep{smith2012kepler,msmap}.\\

In this paper, we present an exploratory algorithm to solve for a cotrending low-rank linear systematics model while incorporating a spatial constraint across the sensor at a foundational level. We refer to this as the {\it spatial systematics method} in what follows. The spatial constraint is of generalized total variation (TV) form \citep{Rudin92nonlineartotal}. This TV constraint on basis vector coefficients promotes the correlation of adjacent neighbors on the sensor and weighted spatially across the sensor, while permitting discontinuities as found in wide-field imaging sensors at module or channel edges. As such, the spatial constraint has a realistic physical motivation and therefore a strong potential to reduce overfitting. The fit over all SAP light curves is performed by minimizing the sum of the least-squares residual between the light curves and the linear systematics model, and, the total variation spatial constraint. We use variable elimination \citep{Golub2007TheDO} to reformulate the optimization problem as a function of the weighted coefficients only, introducing stability to the minimization \citep{Golub_2003, Shearer_2013}. An approximate closed-form gradient of the objective function is derived which we minimize via gradient descent. In this iterative solution both the weighted coefficients, and the basis vectors, which functionally are defined by the coefficients, are varied. The algorithm as implemented is available as a public Python package on the Github repository\footnote{\url{github.com/xiaziyna/spatial-detrend}} or via PyPi under package name {\it spatial-detrend}\ \footnote{\url{https://pypi.org/project/spatial-detrend}}. \\

We numerically evalaute the performance of our method in the context of the Kepler sensor and against PCA as a reference method. We use both injected signals in real long-cadence (LC) Kepler data and fully-simulated data in this evaluation. As noted above \citep[PDC-MAP]{smith2012kepler} Kepler systematics depend on stellar magnitude. This is also found for CoRoT light curves \citep{2009mazeh, sars}. This effect likely arises due to pixel saturation effects. Kepler operated in the magnitude band Kp $\in$ 9 to 15 \citep{koch}. In this work for simplicity we do not include a prior on stellar magnitude dependence but instead select light curves from a narrow fixed magnitude band Kp $\in$ 12 to 13 within which magnitude-dependent systematics such as saturation are not expected to substantially vary \citep{smith2012kepler}; this therefore allows the spatial variability systematics to be isolated. Our numerical evaluation demonstrates that the spatial systematics algorithm has reduced overfitting of astrophysical variability, particularly on day-long timescales, compared to the reference PCA method. \\

The paper is organized as follows. Section~\ref{sec:Model} introduces the light curve signal model, our method for spatial systematics inference, an analysis of spatial correlation across the Kepler sensor, and describes our experimental tests to numerically evaluate the algorithm. In Section~\ref{sec: results} we report the results of our numerical evaluation. These results are discussed in Section~\ref{sec: discussion} and conclusions are presented in Section~\ref{sec: conclusions}. A table of commonly-used symbols is provided in Appendix \ref{ap: symbols} for reference.

\section{METHODS} \label{sec:Model}
As described above, the data products for wide-field exoplanet transit surveys may comprise: i) pixel-level image data either within target apertures or for full frames; and, ii) SAP target light curves \citep{kep_data_handbook, tess_data_handbook}. Here we use Kepler SAP target light curves with a sampling cadence of 30 minutes produced by the Kepler science data processing pipeline but not including post-processing by the Kepler Pre-Search Data Conditioning (PDC) module \citep{Twicken2010, smith2012kepler, stumpe2012kepler}; the latter light curves with full Kepler post-processing are distinguished as PDC-SAP light curves. We present the details of the spatial systematics method that is the subject of this paper.

\subsection{Matrix Notation}
We define common notation here that is used throughout the paper. The $i$-th row of matrix $\bold{M}$ is denoted as $[\bold{M}]_i$ and the $j$-th column as $[\bold{M}]_{\cdot, j}$. The matrix element corresponding to the $i$-th row and the $j$-th column is denoted as $[\bold{M}]_{i, j}$ or $M_{i,j}$. The transpose of a matrix $\bold{M}$ is denoted as $\bold{M}^T$. An identity matrix of size $N \times N$ is denoted as $\mathds{1}_N$.\\
The $L_p$ norm of a vector $\bold{x}$ of length $I$ is defined as: $\lVert \bold{x} \rVert_p=\left(\sum_{i=1}^I |x_i|^p \right)^{1/p}$. If $\bold{M} \in \mathbb{R}^{I \times J}$, we use  $\lVert \bold{M} \rVert_{p,q}$ to denote the $L_p$ norm applied to each column $[\bold{M}]_{\cdot, j}$, followed by the $L_q$ norm applied to this vector, such that $\lVert \bold{M} \rVert_{p,q} =\left( \sum_{j=1}^J \left( \sum_{i=1}^I |M_{i,j}|^p \right)^{q/p} \right)^{1/q}$. The Frobenius norm is defined as $\lVert \bold{M} \rVert_F = \lVert \bold{M} \rVert_{2,2}^1$, such that $\lVert \bold{M} \rVert_F^2 = \sum_{j=1}^J \sum_{i =1}^I |M_{i,j}|^2 $ \citep{golub2013matrix}.

\subsection{Light Curve Decompositon} \label{sec: decompose}
A collection of target light curves $\{ \bold{y}_i : i \in I\}$ are obtained on a sensor, each light curve $\bold{y}_i$ (over target index set $I$) is a length-$N$ time-series.  Each light curve is represented as the sum of a systematics term $\bold{l}_i$ and a statistical noise term $\bold{n_i}$:
\begin{align} \label{eq: decompose}
    \bold{y}_i =  \bold{l}_i + \bold{n}_i
\end{align}
In matrix form the data model takes the form $\bold{Y} =  \bold{L} + \bold{N}$, where $\bold{y}_i, {\bold{l}_i},$ and ${\bold{n}_i}$ form the columns of matrices $\bold{Y} \in  \mathbb{R}^{N \times I}$, $\bold{L} \in \mathbb{R}^{N \times I}$, and $\bold{N} \in \mathbb{R}^{N \times I}$ representing the light curves, systematic noise, and statistical noise respectively. The goal of this work is to form an accurate estimate of the systematic noise $\mathbf{l}_i$.
\\
Sources of systematic errors are described in Section~\ref{sec:intro}; see also \citet{corot_in_flight}, \citet{kep_handbook}, and \citet{tess_handbook}. Statistical noise originates due to a mixture of instrumental error and astrophysical variability. 
We approximate statistical noise $\bold{n}_i$ as white Gaussian noise $\bold{n}_i \thicksim \mathcal{N}(0, \sigma_i^2)$ since additive statistical noise sources may reasonably approach $\mathcal{N}(0, \sigma_i^2)$ under the central limit theorem \citep{grinstead2012introduction}. This is a common assumption in this domain and implicit in least-squares minimization approaches \citep{smith2012kepler, stumpe2012kepler, msmap, aig}.
The noise level $\sigma_i$ is unknown a priori, but may reasonably be estimated from filtered and coarsely-detrended light curves. In what follows, light curves are normalized by an estimate of $\sigma_i$ so that $\bold{n}_i \sim \mathcal{N}(0, 1)$ for mathematical convenience. \\

Light curves may include intermittent sources of systematic noise that produce outliers and are difficult to model. Their influence may be minimized by prior constraints (a further motivation for our approach here), but data-driven outlier filtering is generally necessary in this domain (Section~\ref{sec: exp}).

\subsection{Systematics Model} \label{sec: sys_model}
\paragraph{Low-rank model:}

We adopt a cotrending basis model:
\begin{align}
    \bold{l}_i [n] = \sum_{k=1}^K  c_i^k \bold{v}_k [n]
\end{align} where $(K \ll N)$ describes the rank of the systematic noise model, $\{\bold{v}_k: k \in K \}$ are a set of basis vectors shared by light curves, and each $c_i^k$ is a coefficient weighting of $\bold{v}_k$ for light curve $i$. The coefficient vector for light curve $i$ is $\bold{c}_i = [c_i^1, c_i^2, ... c_i^k]^T$. In matrix form:
\begin{align}
    \bold{L} = \bold{V} \bold{C}
\end{align} 
where $\bold{L} \in \mathbb{R}^{N \times I}$ is defined above, the columns of $\bold{V} \in  \mathbb{R}^{N \times K}$ are the basis vectors $\{\bold{v}_k : k \in K\}$, and the columns of $\bold{C} \in \mathbb{R}^{K \times I}$ are the coefficients $\{\bold{c}_i : i \in I\}$. Graphically,
\begin{equation} 
             \begin{bmatrix}
     \vertbar & \vertbar &        & \vertbar \\
    \mathbf{l}_1    & \mathbf{l}_2    & \ldots & \mathbf{l}_{I}    \\
    \vertbar & \vertbar &        & \vertbar 
  \end{bmatrix} 
         = 
    \begin{bmatrix}
     \vertbar & \vertbar &        & \vertbar \\
    \mathbf{v}_1    & \mathbf{v}_2    & \ldots & \mathbf{v}_K    \\
    \vertbar & \vertbar &        & \vertbar 
  \end{bmatrix}     \begin{bmatrix}
     \vertbar & \vertbar &        & \vertbar \\
    \mathbf{c}_1    & \mathbf{c}_2    & \ldots & \mathbf{c}_I    \\
    \vertbar & \vertbar &        & \vertbar 
  \end{bmatrix} 
\end{equation}
We define a column-normalized coefficient matrix $\bold{\bar{C}}$ of the form $\bar{\bold{c}}_i = \frac{\bold{c}_i}{\| \bold{c}_i \|_2}$:
\begin{align}
\bold{\bar{C}} = 
\begin{bmatrix}
     \vertbar & \vertbar &        & \vertbar \\
    \frac{\mathbf{c}_1}{\| \mathbf{c}_1 \|_2}    & \frac{\mathbf{c}_2}{\| \mathbf{c}_2 \|_2}    & \ldots & \frac{\mathbf{c}_I}{\| \mathbf{c}_I \|_2}    \\
    \vertbar & \vertbar &        & \vertbar 
  \end{bmatrix} 
  \label{eq: normc}
\end{align}
\\
The rank of $\bold{L}$ may be less than the number of independent systematic noise sources. Since $\rank(\bold{V}\bold{C}) \leq \min \{\rank (\bold{V}), \rank(\bold{C}) \}$, it is possible to have a more expansive basis set $\bold{V}$ representing many individual noise effects, but for the relative weightings $\bold{C}$ between light curves to have a degenerate and therefore low-rank structure and consequently for $\bold{L}$ to be low rank. 

\paragraph{SVD/PCA systematics estimation:} 
Singular Value Decomposition (SVD) and Principal Component Analysis (PCA) are integral to many cotrending approaches
\citep{zucker, thatte, Petigura_2012, stumpe2012kepler, smith2012kepler} and are summarized in Appendix~\ref{ap: PCA}. Under a white noise model for $\bold{N}$ the maximum likelihood (ML) estimate of a matrix $\bold{L}$ (not necessarily low-rank) given light curves $\bold{Y} = \bold{N} + \bold{L}$ is equivalent to minimizing the least-squares residual $f(\bold{V},\bold{C})=|| \bold{Y}-\bold{L}||_F^2$:

\begin{align}\label{eq: pca_lsq}
    \argmax_{\bold{L}} p(\bold{Y}|\bold{L}) \equiv \argmin_{\bold{L}} || \bold{Y}-\bold{L}||_F^2
\end{align}
where $p(\bold{Y}|\bold{L})$ is the likelihood of $\bold{Y}$ given $\bold{L}$ \citep{srebro2004learning}.

For the data model in Equation \ref{eq: decompose}, without any further constraints, a rank-$K$ optimal least-squares solution of $\mathbf{L}$ can be found by rank-thresholding the SVD or PCA decomposition of $\bold{Y}$ per the Eckart–Young–Mirsky theorem (Appendix~\ref{ap: PCA}) and in this case takes the form: $\bold{L} =  \bold{V}_K \bold{C}_K$.
In what follows we use the SVD/PCA method primarily as a comparative method (Section~\ref{sec: exp}). 

\paragraph{Spatial systematics model:}
As described in Section~\ref{sec:intro}, the residual systematics are spatially correlated across the sensor. In addition, the white noise model for $\bold{N}$ is an idealization and the systematics cannot then be perfectly separated by rank thresholding alone. In our spatial systematics algorithm, we retain a low-rank systematics data model $\mathbf{L}$ while recognizing the incomplete model for $\mathbf{N}$. We introduce a spatial side constraint of total variation measure (Appendix~\ref{ap: tv}) on the normalized coefficient matrix $\bold{\bar{C}}$ to facilitate more robust separation of the unknown astrophysical signals. The total variation constraint admits a small number of bounded discontinuities within generally smooth functions \citep{Rudin92nonlineartotal, vogel_book} and promotes correlation between neighboring light curve systematics. This choice of spatial constraint is experimentally motivated in Section \ref{sec: exp}. In this formulation, the estimate for $\bold{L} = \bold{V}\bold{C}$ is obtained by minimizing an objective function comprising the least-square residual $f(\bold{V},\bold{C})=|| \bold{Y}-\bold{V}\bold{C}||_F^2$ and the total variation penalty constraint $g(\bold{C})=\|\bold{D}_{\bold{W}}  \bold{\bar{C}} \|_{2,p}^p$ with $p \in [1,2]$ (Appendix~\ref{ap: tv}):

\begin{align} \label{eq: obj1}
    \argmin_{\bold{V}, \bold{C}\; : \; rank(\bold{V}\bold{C}) \leq K} \left\{ f(\bold{V},\bold{C}) + g(\bold{C}) \right\} = 
    \argmin_{\bold{V}, \bold{C}\; : \; rank(\bold{V}\bold{C}) \leq K} \left\{ ||\bold{Y} - \bold{V}\bold{C} ||_F^2  + ||\bold{D}_{\bold{W}}  \bold{\bar{C}} ||_{2, p}^p \right\}
\end{align}

Here $\bold{D}_{\bold{W}}=\bold{W}\bold{D}$ where $\bold{D}$ is a difference operator, as defined in Appendix~\ref{ap: tv}, and $\bold{W}$ is a diagonal matrix of weights $w^{x, y}$ between $[0,1]$ to model non-uniform spatial correlation across the sensor. We assume each light curve $i$ has a pixel position $i \to (x,y) \in (X, Y)$, where $X, Y \in \mathbb{Z}^+$. Further, we assume uniform spatial cells of size $(\Delta x, \Delta y)$ across the sensor and that there is a one-to-one mapping between light curve $i$ and spatial cell.

\begin{figure}[htbp] 
  \centering
  \includegraphics[width=.5\linewidth]{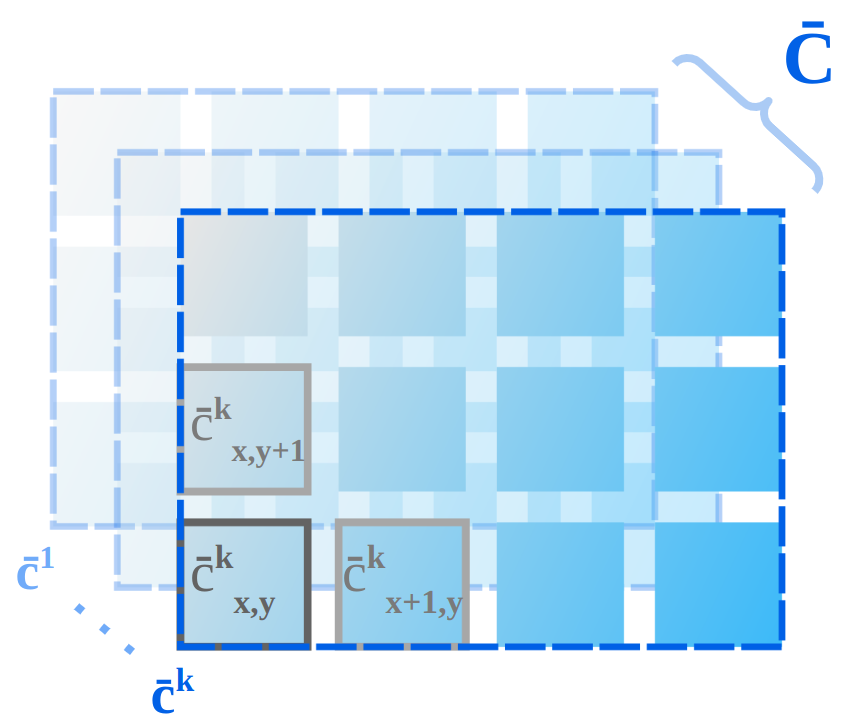}
    \caption{A visual representation of the coefficient matrix $\mathbf{\bar{C}}$. Each image layer $\mathbf{\bar{c}}^k$ represents the coefficient weights for a basis term $k$, ($[\mathbf{\bar{C}}]_k$) organized by 2D sensor cell position. The mapping of light curve $i$ to cell position $i\ \to (x,y)$ is implicit in this notation, as described in the main text. \label{fig: coeff_sensor}} 

\end{figure}

The weighted difference operator $\bold{D}_{\bold{W}}$ applied to $\bold{\bar{C}}$ computes for each $k \in K$ and cell $(x,y)$, the weighted difference of neighbouring coefficients
$[\bar{c}_{x, y}^k - \bar{c}_{x, y+1}^k, \bar{c}_{x, y}^k - \bar{c}_{x+1, y}^k] \in \mathbb{R}^{2}$ as the columns of $\bold{D}_{\bold{W}}\bold{\bar{C}} \in \mathbb{R}^{2 \times K\cdot X \cdot Y}$ and where $\bar{c}^k_{x,y}=\bar{c}^k_{i\ \to (x,y)}$. This is depicted graphically in Figure \ref{fig: coeff_sensor}. In expanded form $ \|\bold{D}_{\bold{W}}\bold{\bar{C}}\|_{2,p}^p = \sum_{k \in K} \sum_{(x,y)} w^{x, y} ( |\bar{c}_{x, y}^k - \bar{c}_{x, y+1}^k|^2 + |\bar{c}_{x, y}^k - \bar{c}_{x+1, y}^k|^2 )^{\frac{p}{2}}$.  

When $p=2$ minimizing the total variation spatial constraint is equivalent to maximizing the correlation between neighboring coefficient vectors; this is shown in Appendix \ref{ap: corr_same}. 

Placing the spatial prior on the normalized coefficients $\bold{\bar{C}}$ instead of the full systematics model $\bold{L}  = \mathbf{V} \mathbf{C}$ has mathematical advantages as described in Section \ref{sec:floats} and is shown in Appendix \ref{ap: corr_distortion} to promote correlation between neighboring systematics estimates by proxy.

The prior is calculated using normalized coefficients $\bold{\bar{C}}$ to avoid the objective being minimized for a trivial minima. If instead the prior was of the form $\|\bold{D}_{\bold{W}}  \bold{C} \|_{2,p}^p$ any solution $\bold{C}$ can be replaced with a solution $\bold{C}' = \alpha\bold{C}$ for $\alpha < 1$ that produces a lower value of the spatial prior $\|\bold{D}_{\bold{W}}  \bold{C}' \|_{2,p}^p = \alpha\|\bold{D}_{\bold{W}}  \bold{C} \|_{2,p}^p$. Taking $\bold{V}' = \frac{\bold{V}}{\alpha}$, there is no change to the least-squares penalty $\|\bold{Y} - \bold{V}\bold{C} \| = \|\bold{Y} - \bold{V}'\bold{C}' \|$, therefore a trivial minima can be found as $\alpha \to 0$. 

Basis vectors obtained with SVD and PCA are constrained to be orthogonal, however, orthogonality is not a necessary constraint as a property of a systematics model, nor to achieve a minimal optimization cost. Our model does not restrict $\bold{V}$ or $\bold{C}$ to be orthogonal.
 
\subsection{Systematics Inference} \label{sec:floats}

The systematics $\bold{L}=\bold{VC}$ are inferred by minimizing the objective function defined in Equation~\ref{eq: obj1}. This function is defined over variables  $\bold{V} \in  \mathbb{R}^{N \times K}, \bold{C} \in \mathbb{R}^{K \times I}$, and comprises a least squares residual $f(\bold{V}, \bold{C}) = \|\bold{Y} - \bold{V}\bold{C} \|_F^2$, which depends on both variables $\bold{V}, \bold{C}$, and the total variation penalty constraint $g(\bold{C}) = \|\bold{D}_{\bold{W}}  \bold{\bar{C}} \|_{2,p}^p$, which only depends on $\bold{C}$. The least-squares residual $f(\bold{V}, \bold{C})$ is not separable over $\bold{V}, \bold{C}$.

Coupled least-squares problems of this type can be solved by iterative alternating solutions for $\bold{V}$ and $\bold{C}$ \citep{Wold1973, zucker} or by using variable elimination to remove dependence on one of the variables \citep{Golub_2003}. There is no evidence that either is preferable and we adopt the latter variable projection method here. 

For fixed $\bold{V}$ or fixed $\bold{C}$ the least-squares residual $f(\bold{V}, \bold{C})$ has an analytic solution for the complementary free variable. Therefore we can eliminate variable dependence on $\bold{V}$ in the least-squares residual so that the objective can be minimized over $\bold{C}$ alone. For a value of $\bold{C}$ the minimizing value $\bold{V}$ of the least-squares residual is denoted as a function $h(\bold{C})$:
\begin{align}
h(\bold{C}) = \argmin_{\bold{V}} f(\bold{V}, \bold{C}) 
\end{align}

This function $h(\bold{C})$ can be inserted as $\bold{V}$ into the least-squares residual so that the original objective depends only on $\bold{C}$:
\begin{align}
    \min_{\bold{V}, \bold{C}} \{f(\bold{V}, \bold{C}) + g(\bold{C})\}\; \to \; \min_{\bold{C}} \{f(h(\bold{C}), \bold{C}) + g(\bold{C})\}
\end{align}

Using the theory described in \citet{Golub_2003}, as elaborated in Appendix~\ref{ap: vp}, the variable-reduced objective takes the form:
\begin{align} \label{eq: obj2}
    \argmin_{ \bold{C}\; : \; \rank(\bold{C}) \leq K} \{||(\mathds{1}_{I} - \bold{C}^T\bold{C}^\dagger)\bold{Y}^T||_F^2  + ||\bold{D}_{\bold{W}}  \bold{\bar{C}} ||_{2,p}^p \}
\end{align}

The minimum of the derived objective does not have an analytic solution but can be obtained numerically using gradient descent optimization methods. However, the $L_p$ norm in the total variation constraint $g(\bold{C})$ is not directly differentiable since it is discontinuous for zero-valued entries of $\bold{D}_\bold{W} \bold{\bar{C}}$ when $p=1$. Proximal methods are standard for obtaining minima of objective functions with non-differentiable penalties such as $L_1$ norms, as described in \citet{parikh2013proximal}. In particular, the Split-Bregman method \citep{split-breg} is well suited to total-variation regularized problems. A preferred approach is to replace the non-differentiable $L_p$ norm with a continuously-differentiable Huber loss function as an approximation \citep{vogel_book}. Our general gradient-descent algorithm to minimize the variable-reduced objective is described in Algorithm \ref{alg: gd}. We refer to the variable reduced objective function in Equation~\ref{eq: obj2} as $w (\cdot)$, the gradient of $w(\cdot)$ with respect to $\bold{C}$ is denoted $\nabla w(\cdot)$. The gradient of the variable-reduced least-squares term $f'(\bold{C}) = f(h(\bold{C}), \bold{C}) = ||(\mathds{1}_{I} - \bold{C}^T\bold{C}^\dagger)\bold{Y}^T||_F^2$ in the objective is denoted $\nabla f'(\cdot)$ and the gradient of the total variation constraint term $g(\bold{C})=||\bold{D}_{\bold{W}}  \bold{\bar{C}} ||_{2,p}^p$ as $\nabla g(\cdot)$. These gradients are used used to iteratively update $\bold{C}^t$ at each timestep $t$. We note that this variable-projection algorithm also updates $\bold{V}$ implicitly, denoted $\bold{V}_{|\bold{C}}$. The details of the Huber loss function and the derived gradients are described in Appendix \ref{ap: grad}.\\

\begin{algorithm}[H] \label{alg: gd}
\SetAlgoLined
 Initialize $\bold{C}$ for rank $K$ and weight matrix $\bold{W}$. The step-size $\alpha^t$ may be fixed or vary at each step $t$. \\
 \While{ $|w(\bold{C}^{t+1}) - w(\bold{C}^{t})| \geq \epsilon$}{
    1) $\nabla w(\bold{C}^t) = \nabla f'(\bold{C}^t) + \nabla g(\bold{C}^t)$ \\

    2) $\bold{C}^{t+1} = \bold{C}^t  - \alpha^t \nabla w(\bold{C}^t)$ \\ }
    $\mathbf{L} = \mathbf{V}_{|\mathbf{C}} \mathbf{C} $ with $\bold{V}_{|\mathbf{C}} = \mathbf{Y} \mathbf{C}^T(\mathbf{C}\mathbf{C}^T)^{-1} $. 
 \caption{Variable Projection Gradient Descent for the Spatial Systematics Algorithm}
\end{algorithm}

\subsection{Algorithm Implementation and Evaluation} \label{sec: impeval} \label{sec: exp} \label{sec: simulation} \label{sec: implementation}

In this section, we describe further details concerning the implementation and evaluation of the spatial systematics algorithm presented in this paper. As noted above, this method has been developed to explore new algorithmic approaches to exoplanet transit detection. Our prime focus is to demonstrate the feasibility of the algorithm and to provide an initial assessment of its performance. Accordingly, this work is not designed to support or claim optimality of this method over existing specialized detrending approaches. 

The spatial systematics algorithm was evaluated numerically on a selected subset of Kepler test data using both injection tests with time-variable signals (Section~\ref{sec: exp1}) and fully-simulated light curves in the same Kepler coordinates (Section~\ref{sec: exp2}). Further, a high-level comparison between the spatial systematics algorithm and several standard detrending methods was performed over the same Kepler test data (Section~\ref{sec: compare}). The Kepler test data are described in Section~\ref{sec: testdata}.  \\

\begin{figure}[htbp] 
  \centering
  \includegraphics[width=.5\linewidth]{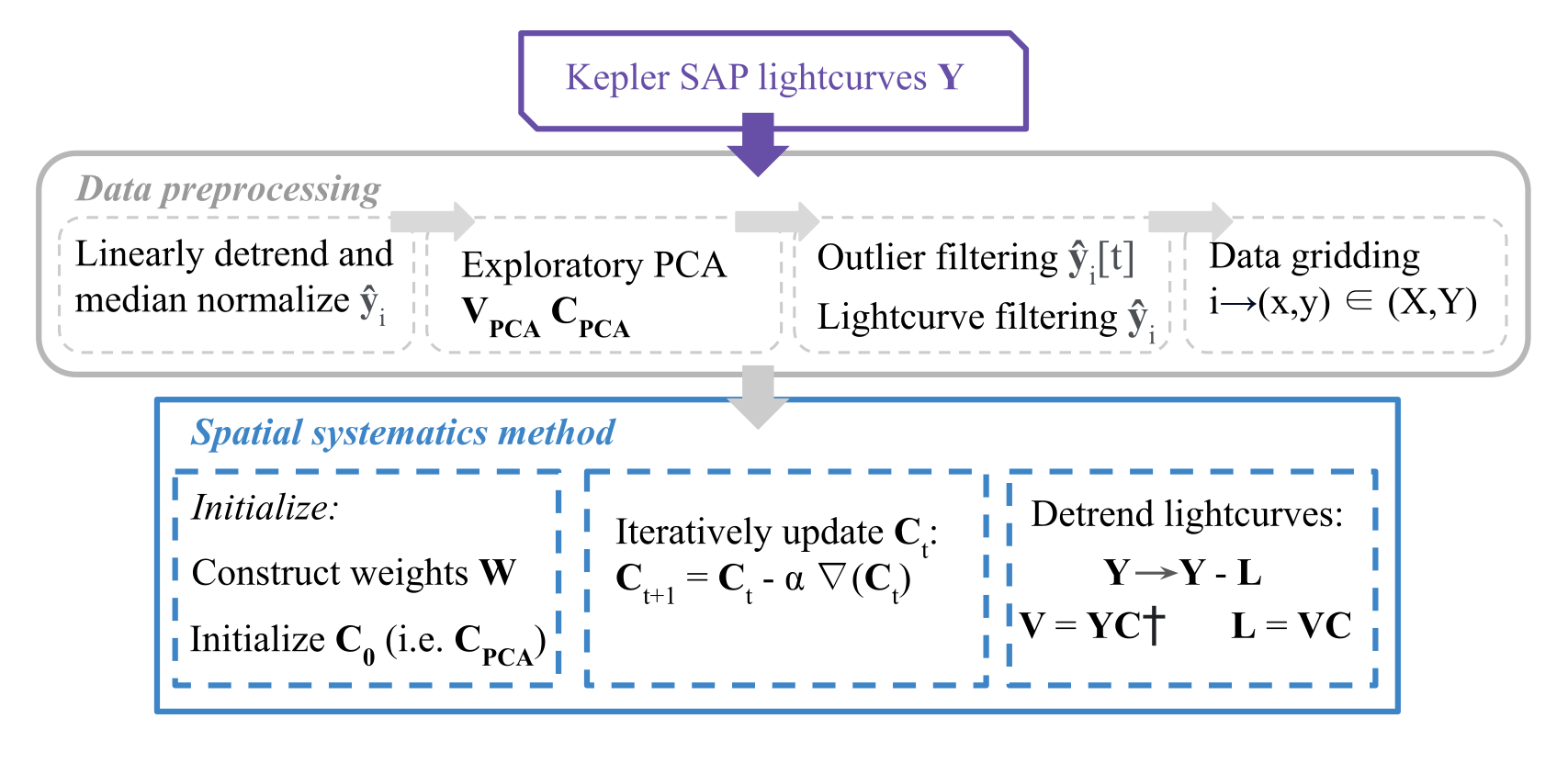}
    \caption{Data analysis framework for numerical evaluation of the spatial systematics algorithm. The data flow includes input data (purple box), pre-processing (grey box; Section~\ref{sec: testdata}), and application of the spatial systematics algorithm (blue box; Sections~\ref{sec:floats} and \ref{sec: impeval}).}
    
    \label{fig: flowchart}
\end{figure}

The data analysis framework used for the numerical evaluation tests is depicted in Figure~\ref{fig: flowchart}. The free parameters to be chosen, shown in our core Algorithm~\ref{alg: gd} include the initial value of $\bold{C}_0$, the model rank $K$, the total variation norm parameter $p$, the gradient step size $\alpha$, and the spatial weighting matrix $\bold{W}$. The analysis framework and initial values are discussed in further detail below.\\

The data analysis was performed in Python and using a single CPU core. The computational time complexity of the spatial algorithm acting on a collection of light curves $I = X \times Y$, is quadratic in $|I|$ and linear in the number of gradient step iterations, and the model rank $K$. The leading order time-complexity of a gradient step in Algorithm~\ref{alg: gd} defined in Appendix ~\ref{ap: grad} is $O(|I|^2 K)$. Since for difference matrices $\bold{D}_x \in \mathbb{R}^{Y(X-1) \times YX}$, $\bold{D}_y \in \mathbb{R}^{(Y-1)X \times YX}$, per gradient step, a matrix multiplication of $\bold{D}_x^T \bold{D}_x \in \mathbb{R}^{I \times I}$ and $\bold{D}_y^T \bold{D}_y \in \mathbb{R}^{I \times I}$ with $\mathbf{C}^T \in \mathbb{R}^{I \times K}$ is performed. The difference matrices $\mathbf{D}_x$ and $\mathbf{D}_y$ are sparse in form and for either matrix approximately $\frac{2}{|I|}$ fraction of elements are non-zero. In our implementation, matrices $\bold{D}_x$ and $\bold{D}_y$ are generated and stored in compressed-sparse-row (CSR) format with Scipy sparse.
    
\subsubsection{Kepler Test Data} \label{sec: testdata}

The test data were selected from the Kepler long-cadence SAP light curves; these have an integration time of 29.4 min \citep{jenkins2010overview}. The mapping of a stellar target to approximate sensor position repeats every four Kepler quarters \citep{aig} and accordingly to investigate the consistency of spatial trends over several quarters, we selected quarters (Q hereinafter) Q6, Q10, and Q14 separated at this cadence. Quarter Q6 was chosen to match \citet{Petigura_2012} and allow inter-comparison with their prior work on spatial correlation; the decision to include successor quarters (Q10, Q14) rather than preceding quarters (Q2) was also made due to calibration issues during early Kepler operations \citep{kep_q2}. Stellar targets were selected within a range of Kepler magnitude $Kp \in [12, 13]$ and with a combined differential photometric precision over 12 hours CDPP$_{12h} \leq 40$ \citep{cdpp_christ} \footnote{Light curve data were obtained from the MAST archive:  archive.stsci.edu/kepler/data$\_$search/search.php.}, resulting in 6179, 6286 and 6049 light curves for Q6, Q10 and Q14 respectively. These target selection criteria were informed by \citet{Petigura_2012}, however we do not believe however that our numerical algorithm evaluation is highly sensitive to these exact parameter choices. The narrow range in photometric magnitude was used to isolate the known dependence of systematics on magnitude \citep{msmap} and to allow a focus on spatial systematics in our algorithm. 

\begin{figure}[htbp] 
  \centering
  \includegraphics[width=1\linewidth]{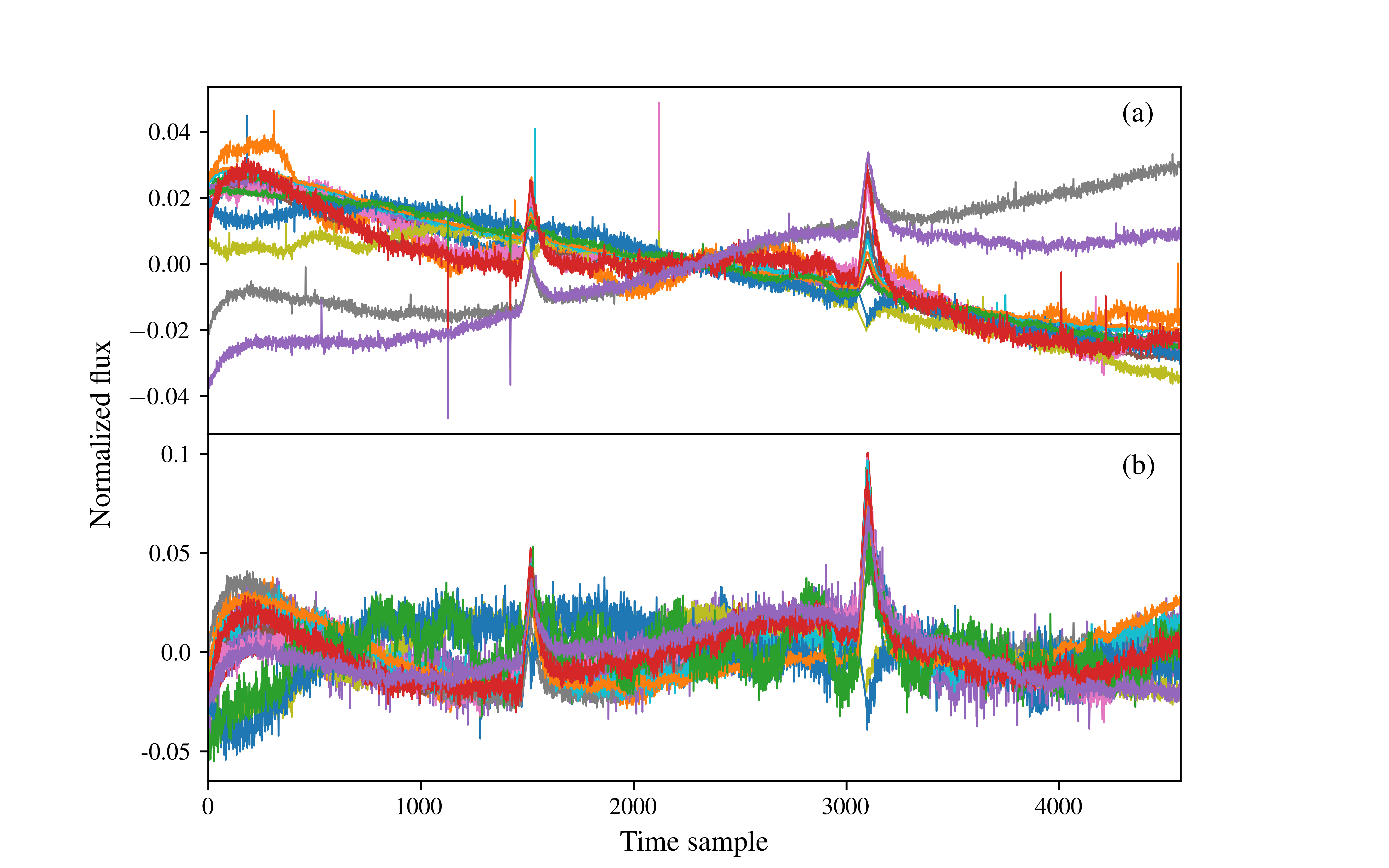}
    \caption{A representative sample of 15 light curves from Q10 shown before {\it (a)} and after  {\it (b)} linear detrending and outlier removal. The {\it x}-axis index is in the unit of long-cadence time samples $(\Delta t = 29.4{\rm min})$. The dominant systematics visible are due to the quarterly roll and earth point recoveries \citep{stumpe2012kepler}. }
     \label{fig: linear_dtrend}
\end{figure}

\paragraph{Pre-processing:} \label{sec: preprocess}
As noted in Section~\ref{sec: decompose} the assumption of Gaussian statistical noise $\bold{N}$ is an idealization and pre-processing is necessary to address several data conditioning issues, as described here. First, any missing data values, and four visually-identified outlier cadences, were substituted using linear interpolation between the preceding and following data samples spanning each gap. Median normalization was then applied to each light curve as $\bold{\hat{y}}=\bold{y} / {\rm med} (\bold{y})-1$, where ${\rm med} (\bold{y})$ is the median of $\bold{y}$, and the subscript on light curve index $i$ (Equation \ref{eq: decompose}) is dropped for clarity. Here, each pre-processing step is sequential with generic input $\bold{y}$ and output $\hat{\bold{y}}$. We then removed a first-order linear trend from each light curve as systematics are known to depend on timescale \citep{msmap}. This is evident in Figure \ref{fig: linear_dtrend}, where there is no clear correlation between short- and long-timescale effects. Removing the first-order trend allows a focus on spatial systematics in our algorithm. While it is possible to apply our algorithm to each timescale bandwidth separately in a decoupled manner that generalization is beyond the scope of the current paper. After linear detrending the first-order difference variance $\sigma_z^2={\rm var}(\{\bold{z}: y_n - y _{n-1}\ \forall\ n\})$, variance $\sigma^2={\rm var}(\bold{y})$ were computed for each light curve, where $n \in N$ here denotes time sample index in an individual light curve. The light curves were then filtered to retain only those that are in the lowest 90$\%$ in both $\sigma_z^2$ and $\sigma^2$, ranked separately. In a second filtering step, light curves are removed if they do not satisfy a minimum correlation requirement. Each light curve $\bold{y}_i$ must have a correlation coefficient $\rho$ greater than $0.6$ with at least 10 other light curves  $\bold{y}_j \; : \; j \neq i$, where the correlation coefficient is calculated as $\rho(\bold{y}_i, \bold{y}_j) = \frac{\bold{y}_i^T\bold{y}_j}{\| \bold{y}_i\| \| \bold{y}_j\|}$. To identify outlier samples the light curves were then coarsely detrended using exploratory PCA\footnote{https://scikit-learn.org/stable/} (Appendix \ref{ap: PCA}) and all samples exceeding a $3-\sigma$ point-to-point scatter were then flagged. This exploratory PCA detrending was not retained however; it was only used to flag and linearly interpolate outlier samples in the input data at this point in the pre-preprocessing. These initial pre-processing and filtering steps resulted in selecting 4749/6179 (76 \%), 4850/6286 (77 \%), and 4741/6049 (78 \%) of light curves for Q6, Q10, and Q14 respectively.

The final step in pre-processing concerned mapping the data to a regular spatial grid on the sensor. The Kepler sensor comprises 25 modules, of which 21 contain two CCDs across four output channels \citep{kep_handbook}. Each CCD is 2200 x 1024 pixels in size and each module contains a total of 2200 x 2048 CCD pixels \citep{kep_handbook}. 
The pixel position of each target light curve on a CCD was mapped to a rescaled square global pixel coordinate grid across the sensor of size $11000 \times 11000$ for ease of analysis. In this process each CCD row coordinate was rescaled by a factor 1100/1024 and gaps between CCDs were removed; this resulted in convenient module coordinates of size $2200 \times 2200$. The total variation constraint depends only on light curve adjacency, so this rescaling has no impact on our algorithm. 

\begin{figure}[htbp]
  \centering
  \includegraphics[width=.95\linewidth]{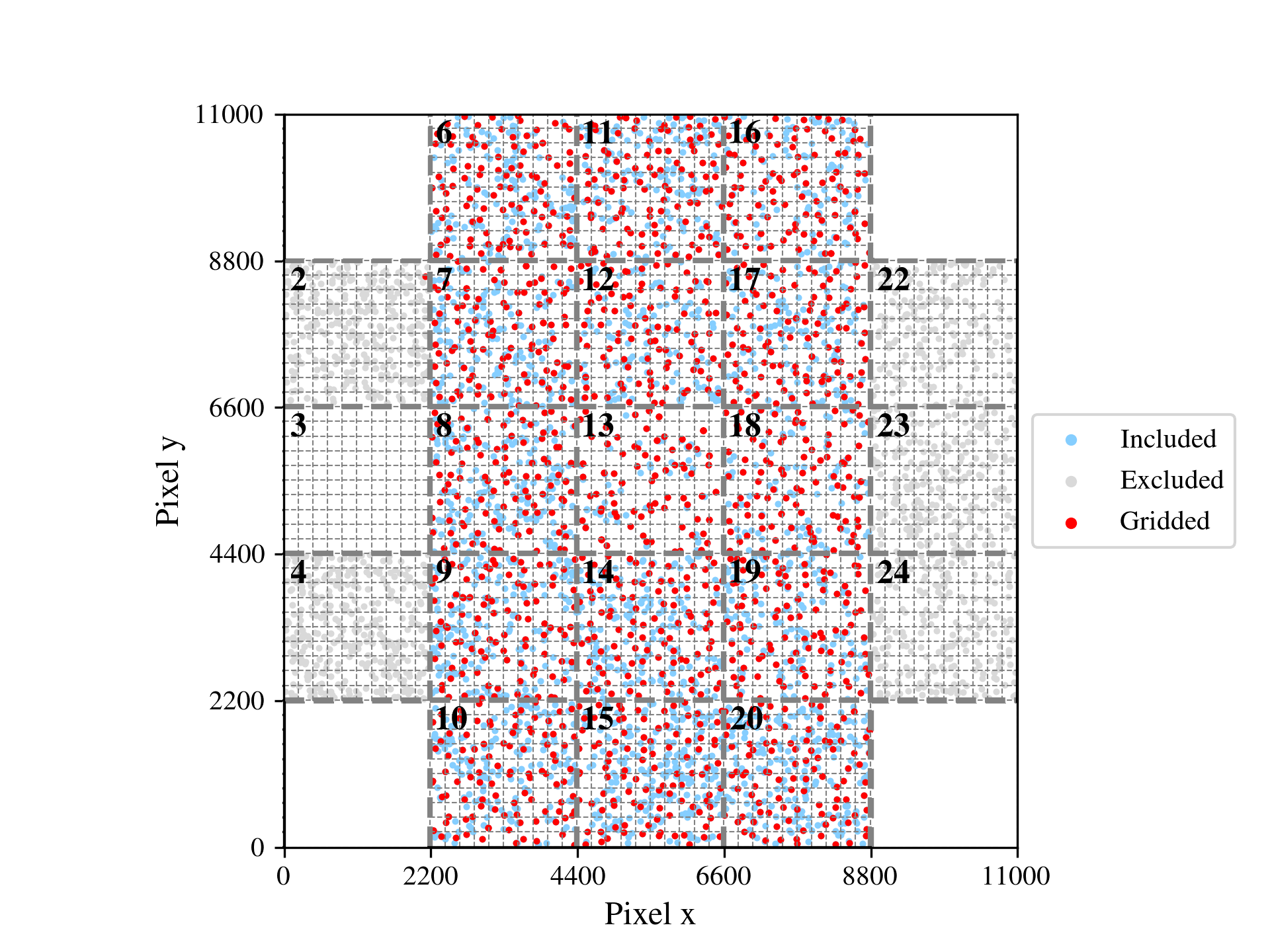}
    \caption{Positions of all (4850/6286) pre-processed Kepler test data light curves for Q10 plotted on the Kepler sensor in global pixel coordinates $(11000 \times 11000)$. Kepler modules are demarcated by bold dashed lines and labeled by module number. The $(30 \times 50)$ spatial grid is drawn in dashed lines with individual cells of size $(220 \times 220)$ pixels. All light curves with coordinates included in the rectangular gridded region are shown in blue. The subset of light curves that were gridded, and therefore processed by the spatial systematics algorithm, are shown in red at their original positions on the CCD. These light curves were either gridded to the $(30 \times 50)$ spatial cell in which they fall or to a nearby empty spatial cell, as described in the main text. After gridding, no spatial cell contained multiple light curves and there were no empty spatial cells. Light curves on modules excluded from the data grid are shown in grey. There are no targets in module 3 due to module failure.}\label{fig: sensor}
\end{figure}

The positions of the pre-processed test data light curves on the Kepler sensor are shown in blue in Figure \ref{fig: sensor}. This Figure uses global pixel coordinates and is further labeled by Kepler module number. The difference operators $\mathbf{D}_x$ and $\mathbf{D}_y$ are easier to implement on a rectangular data grid and light curves on modules (2, 3, 4, 22, 23, 24) were discarded, with light curves on a total of 15 modules retained. A rectangular grid is not an algorithmic requirement, however. As described in Section~\ref{sec: sys_model} our implementation of the total variation constraint requires a one-to-one mapping between light curve $i$ and individual cells in a regular spatial grid. We adopted a spatial cell size of $220 \times 220$ pixels for this gridding, comprising $(30 \times 50)$ cells across the remaining modules with $(10 \times 10)$ cells per module; this grid is shown in Figure~\ref{fig: sensor}. The light curve mapped to each cell was selected randomly from the pre-processed light curves falling within each cell. Where a cell contained no light curves, the light curve closest to the center of the cell was selected; this occurred for $12\%$ of $30 \times 50$ targets. The 1500 gridded light curves are shown in red in Figure~\ref{fig: sensor}.

\paragraph{Spatial structure on Kepler sensor:}
The pairwise neighbor correlation of the gridded light curve sample across the sensor was measured to inform our choice of algorithm parameter $\bold{W}$ (Equation~\ref{eq: obj1}). 
The correlation between vectors $\bold{p}$ and $\bold{q}$ is denoted $\rho(\mathbf{p}, \mathbf{q}) = \frac{\mathbf{p}^T \mathbf{q}}{\|\mathbf{p}\| \|\mathbf{q}\|}$. We denote the set of gridded cell positions within each module as $x \in X_M$ and $y \in Y_M$ respectively. The set of gridded light curves within a module is then the Cartesian set product $M = X_M \times Y_M$. The mean pairwise correlation $\bar{\rho}_M$ between all gridded light curves $(\bold{y}_m, \bold{y}_n)$ in a module was computed as: 
\begin{align}
    \bar{\rho}_{M} = \frac{1}{|M|^2 - |M|} \sum_{(m,n) \in M \times M, \; (m \neq n)} \rho(\mathbf{y}_m, \mathbf{y}_n)
\end{align}

\begin{figure}[htbp] 
  \centering
  \includegraphics[width=.8\linewidth]{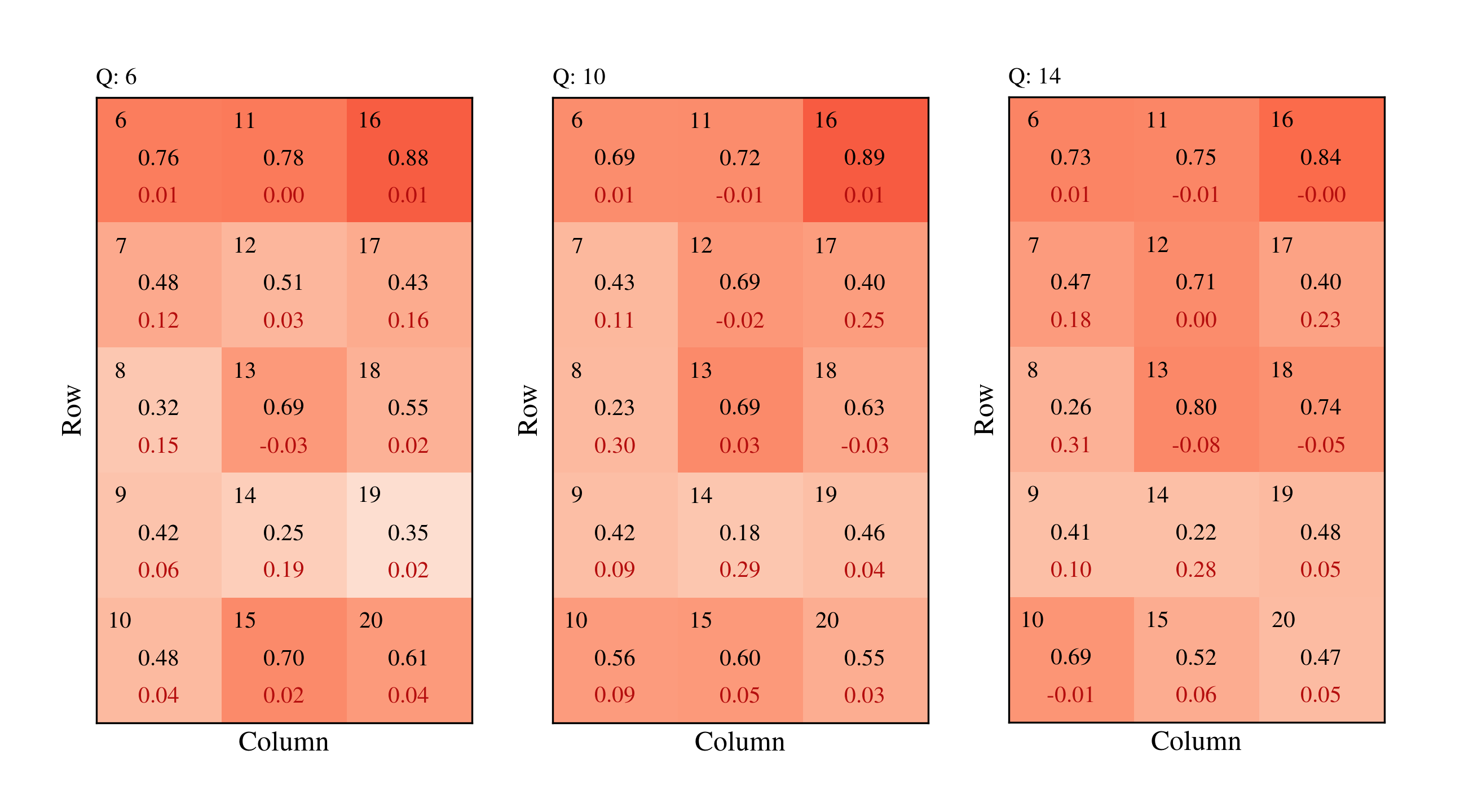}
    \caption{Spatial correlation structure within modules for Q6, Q10 and Q14 across the $(30 \times 50)$ gridded region of the Kepler sensor. The module number is shown in the top left corner of each module. The mean pairwise correlation $\bar{\rho}_M$ within each module is tabulated in black. The difference $\bar{\rho}_{M_{XY}}-\bar{\rho}_M$ is tabulated in red, where $\bar{\rho}_{M_{XY}}$ is the mean pairwise correlation of neighboring light curves within each module. The shaded color intensity is proportional to $\bar{\rho}_{M_{XY}}$.In most modules the correlation between neighbouring light curves is greater than the mean pairwise correlation between all pairs of light curves on the module.}
    \label{fig: mod_corr}
\end{figure}

The mean pairwise neighboring correlation within a module $ \bar{\rho}_{M_{XY}}$ was calculated as: 
\begin{equation}
    \bar{\rho}_{M_{XY}}  = \frac{1}{2 X_M Y_M - 2} \sum_{(x,y) \in (X_M -1) \times (Y_M -1)} \rho(\mathbf{y}_{x, y}, \mathbf{y}_{x, y+1}) + \rho(\mathbf{y}_{x, y}, \mathbf{y}_{x+1, y})
\end{equation}
The spatial correlation structure across the gridded $(30 \times 50)$ region of the Kepler sensor is shown in Figure~\ref{fig: mod_corr}. The mean pairwise correlation $\bar{\rho}_M$ for each module is tabulated in black in this Figure, and the difference $\bar{\rho}_{M_{XY}}-\bar{\rho}_M$ is shown in red. The mean pairwise neighbor correlation $\bar{\rho}_{M_{XY}}$ is generally higher than the mean pairwise correlation $\bar{\rho}_M$ across the module as a whole. The mean pairwise correlation for all gridded light curves over all modules was computed to be $\bar{\rho}=0.26$ and is notably lower than the per-module correlations. \citet{moreno} describe how time-dependent systematics, lagged with radius, may plausibly give rise to spatial correlations as observed in Figure \ref{fig: mod_corr}.

\begin{figure}[htbp]
  \centering
\includegraphics{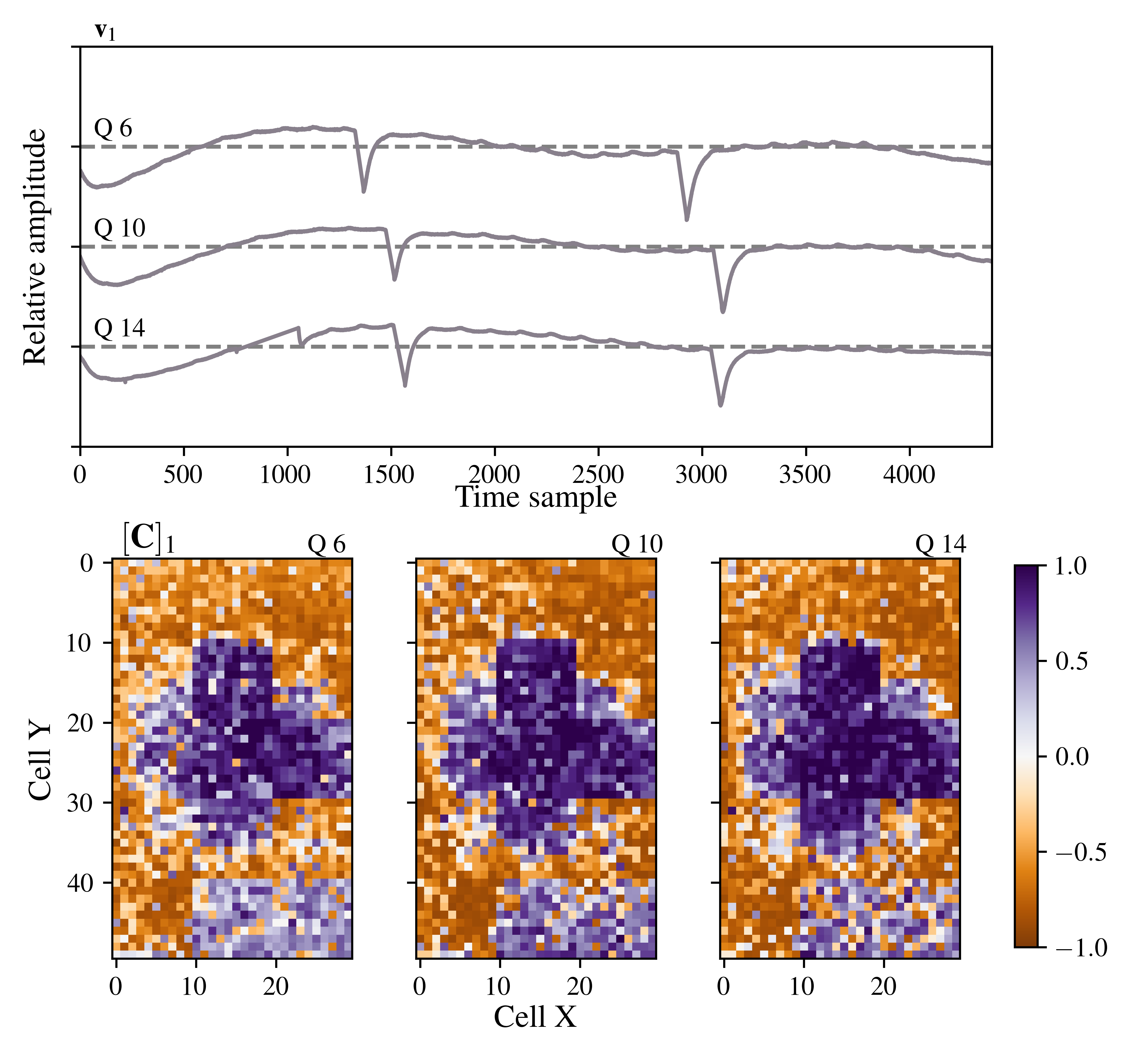}\caption{The leading PCA basis vector $\bold{v}_1$ for quarters Q6, Q10 and Q14) (top) and an associated color map of the leading PCA coefficient $c_{i \gets (x,y)}^1$ at each spatial cell position (bottom). The spatial structure of the leading coefficient term is persistent between quarters.}\label{fig: pca0} 
\end{figure}

We further explored the spatial structure on the Kepler sensor by performing an exploratory PCA decomposition of the gridded light curves for Q6, Q10, and Q14. The resulting leading PCA basis vector $\bold{v}_1$ for each quarter is shown in Figure \ref{fig: pca0} along with a color map of the leading coefficient value $c_i^1$ in each cell. We note the implicit mapping of light curve index $i$ to cell position $i \to (x,y) \in (X, Y)$ defined in Section~\ref{sec: sys_model}. The leading systematics basis vector is informative of general systematics as it is the strongest term. As shown in Figure~\ref{fig: pca0} the leading coefficient values are highly spatially correlated, with a blocked structure per module, with anti-correlations present between modules. This effect is explained by \citet{Petigura_2012} as due to PSF variation from the momentum cycle. The overall spatial structure is also persistent across these quarters. 

As discussed in Section~\ref{sec:intro} and in \citet{moreno}, \citet{Petigura_2012}, and \citet{Lund_2021}, the systematics exhibit spatial dependence that are key to our choice of a total variation constraint. 

\subsubsection{Algorithm Initial Parameter Values}\label{sec: init_val}
As described above, there are several free parameters to be chosen in our method, including the choice of initialization $\bold{C}^0$, the model rank $K$, the choice of prior norm $p$, the gradient step size $\alpha$ and the prior weighting matrix $\bold{W}$.

The weighting matrix $\bold{W}$ was formed as neighboring pairwise correlations of gridded light curves $\bold{y}_i$ scaled by a factor $0.2$ as $w^{x, y} = \frac{0.2}{2} [\rho(\bold{y}_{ x, y}, \bold{y}_{x+1, y}) + \rho(\bold{y}_{ x, y}, \bold{y}_{x, y+1})]$. A comprehensive parameter optimization was not performed, however heuristic evaluation for these data (Q6, Q10 and Q14) showed slightly improved convergence for this choice of $\bold{W}$ compared to uniform weighting.  

The exact model rank is unknown and for these Kepler test data, a model rank of $K = 20$ was chosen based on the singular value distribution of the light curve sample. This singular value distribution was consistent with that shown in \citet{smith2012kepler} where a Kepler CBV rank of $K=16$ was adopted.

The initial $\bold{C}^0$ was obtained from PCA applied to $\bold{Y}$. The choice of $p$ in the total variation constraint determines the degree of smoothness in the spatial correlation and the tolerance of discontinuities (Appendix~\ref{ap: tv}). In Appendix \ref{ap: prob} the choice of $p$ is interpreted in a Bayesian framework, where for $p=1$ the spatial constraint is equivalent to a Laplacian prior on the difference of coefficients, and for $p=2$ the spatial constraint is equivalent to a Gaussian on the difference of coefficients. Referring to Figure \ref{fig: pca0}, PCA fitted coefficients representative of the underlying systematic structure show a mostly spatially uniform structure with discontinuities at module edges. We adopted a value $p = 1.1$, chosen empirically as the observed coefficients are closer to Laplacian in form.

Gradient step size $\alpha^t$ at iteration $t$ was set using a backtracing line search \citep{linesearch} starting from $\alpha^t = 0.1$. This was found to allow the minimization to progress adequately during initial iterations but with fine-tuning in later iterations, thereby saving computation time. Our stopping condition was based on the difference in our cost function in Equation \ref{eq: obj2} between successive iterates falling to a negligible level $\| w(\bold{C}^{t+1}) - w(\bold{C}^{t}) \| \leq 10^{-5}$. 

\subsubsection{Simulated Astrophysical Signals} \label{sec: sim_sig}

We describe here the functional form of the simulated transient and variable signals used in the injection tests and full simulations described below. We denote an individual simulated astrophysical signal in light curve form as vector $\bold{a}$ defined on $ n \in \{1,..,N\}$ and relative to normalized light curve $\bold{y} \ : \ \|\mathbf{y}\|=1$. 
The functional form of the simulated signals was chosen to be either a sinusoid ($\bold{a}_s$), a periodic exoplanet transit ($\bold{a}_t$) or a transient flare ($\bold{a}_f$). Figure \ref{fig: simulated_signal} shows a simulated example signal of each type. These cover a representative range of timescales. The simulated signals were randomly generated relative to light curve $\bold{y}$ as follows:
\begin{itemize}
    \item Sine wave: $\bold{a}_s[n]= A \sin(2 \pi \beta n)$. The amplitude $A$ was drawn from the uniform distribution $U(0.005\|\mathbf{y}\|,\ 0.015\|\mathbf{y}\|)$. The angular frequency $\beta$ was drawn from $U\left(\frac{4}{N}, \frac{8}{N}\right)$. 
     
    \item Simulated exoplanet transit signal: $\bold{a}_{t}[n]$. A transit signal is simulated as a periodic repeating transit profile $b[n]$, with randomized parameters: orbital period $P$, epoch $n_0$, duration $d$, and transit depth $\delta$.
    \begin{equation}
    \bold{a}_{t} [n] = 
    \begin{cases}
         \delta \cdot b[(n-n_0) \;\mathrm{mod}\ P ] ,& \text{if } (n-n_0) \;\mathrm{mod}\ P \leq d \\ 
        0,              & \text{otherwise}
    \end{cases} \label{eq:transit}
    \end{equation}
    The transit profile $b[n]$ is an empirical tapered symmetric profile, defined on the first half of the transit $n \in [1, \frac{d}{2}]$ as: $b[n] =0.4 \exp(-0.3 n) - 1$ and on the second half $n \in [\frac{d}{2} + 1, d]$ as $b[n] =0.4 \exp(-0.3 (d-n) ) - 1$. The depth of this profile generally ranges from $-0.6$ at transit ingress and egress to $-1$ at the midpoint. Transits were simulated to allow at minimum three transits to occur in the light curve. The period $P$ was drawn from the uniform distribution $U(96, 360)$, which corresponds to a range of $4$ to $15$ days. The transit epoch $n_0$ is drawn from the uniform distribution $U(0, P)$.  The duration $d$ of the transit profile was drawn from the uniform distribution $U(8,24)$, which corresponds to a range of $4$ to $12$ hours. The depth $\delta$ is selected between $U(0.002\|\bold{y}\|, 0.016 \|\bold{y}\|)$. 
    \item Simulated flare: $\bold{a}_{f}[n]$. The time sample index $n_p$ of the flare peak was drawn from the uniform distribution $U(1, N)$. The flare profile is simulated as a rising exponential up to $n_p$ followed by a slower decaying exponential. The overall signal is multiplied by a random walk stochastic process $x[n]$, generated as $x[n+1] = x[n] + w[n] \; : \; w[n] = \mathcal{N} (0, 1),\ x[0] = 0$. The simulated flare is defined on the interval $n \in [1, n_p]$ as $\bold{a}_{f}[n] = x[n] \exp{\left(-\frac{20 \cdot (n_p - n)}{N}\right)}$ and on the interval $n \in [n_p, N]$ as $\bold{a}_{f}[n] = x[n] \exp{\left(-\frac{10 \cdot (n -n_p)}{N}\right)}$.

\begin{figure}[htbp!] 
  \centering
  \includegraphics[width=\linewidth]{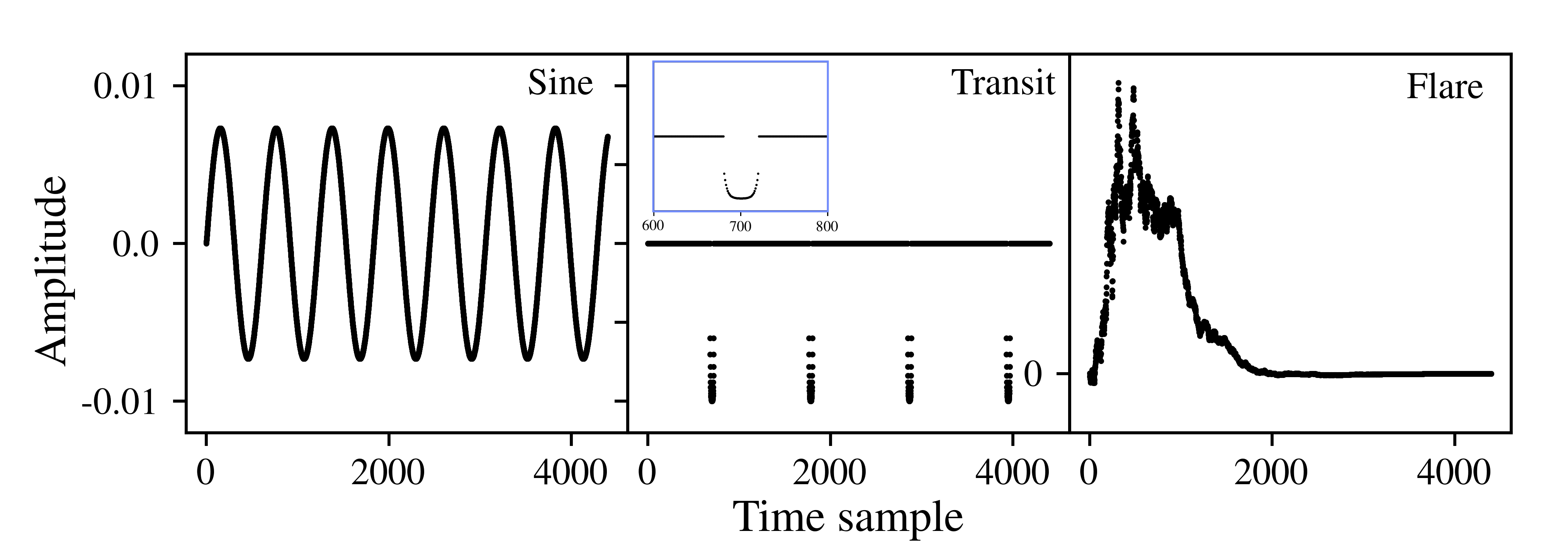}
    \caption{Example simulated sine $\mathbf{a}_s$, transit $\mathbf{a}_t$, and flare $\mathbf{a}_f$ signals as defined in Section~\ref{sec: sim_sig}. The inset panel (blue outline) shows the tapered transit profile at higher temporal resolution.}
    \label{fig: simulated_signal}
\end{figure}

\end{itemize}

The simulated signals vary in strength but are generally not weaker than the sample variance of the first-order difference across each light curve (Section \ref{sec: preprocess}), and not greater than the magnitude of systematics. Each injected signal has on average $30 \%$ of the energy of the systematics term $\frac{ \| \bold{a}_s\|}{\| \bold{y}\|} = 0.3$. 

\subsubsection{Experiment A: Injected Transient and Variable Signals} 
\label{sec: exp1}
In this experiment (A) we numerically evaluated the performance of the spatial systematics algorithm in the recovery of simulated signals (Section~\ref{sec: sim_sig}) injected into the gridded Kepler test data (Figure~\ref{fig: flowchart}), here selected for quarter Q10. We denote the pre-processed gridded Kepler test data here as matrix $\mathbf{Y}$ (Section~\ref{sec: sys_model}). We simulated nine astrophysical signals, with random parameters drawn as described in Section~\ref{sec: sim_sig}, divided equally into three each of the types: sinusoid, flare, and exoplanet transit. Each of the nine individual signals was randomly assigned to a separate randomly-selected light curve in the test data in a one-to-one mapping. Apriori the Kepler test data contain unknown astrophysical variability. Therefore choosing a small number of injected signals matches this realistic incidence and avoids significantly biasing the sample. We represent the injected signals as matrix $\mathbf{A}$, with the same shape as $\mathbf{Y}$, and where $\mathbf{A}$ is null except for nine randomly-selected columns containing the simulated injection signals. The post-injection data are denoted: $\mathbf{Y} \to \mathbf{Y}+\mathbf{A}$. 

A least-squares PCA solution was obtained using the data model $\mathbf{Y}=\mathbf{V}_{PCA}\mathbf{C}_{PCA}+\mathbf{N}$ (Section~\ref{sec: sys_model}), yielding detrended light curves using PCA: $\mathbf{Y'}_{PCA}=\mathbf{Y}-\mathbf{V}_{PCA}\mathbf{C}_{PCA}$. The post-injection Kepler test data $\mathbf{Y}$ were then detrended using the spatial systematics algorithm (Algorithm~\ref{alg: gd}) using $\mathbf{C}_{PCA}$ to initialize $\mathbf{C}$ and yielding detrended light curves $\mathbf{Y'}$. For both methods, we adopt a model rank $K=20$. This model rank is empirically sufficient to represent the systematics and is smaller than both the number of light curves $1500$ and the length $N \sim 4000$ of each light curve. 

This experiment was performed over an ensemble of ten instances of the random parameters defining the injected signals (Section~\ref{sec: sim_sig}) and random light curve assignment from $\bold{Y}$. The performance of both the spatial systematics and PCA algorithm was evaluated in terms of the estimated level of residual systematics after detrending and in terms of their recovery of injected signals, as described in further detail below.

\subsubsection{Experiment B: Fully-Simulated Light Curves} \label{sec: exp2}
\begin{figure}[htbp!]
  \centering
\includegraphics{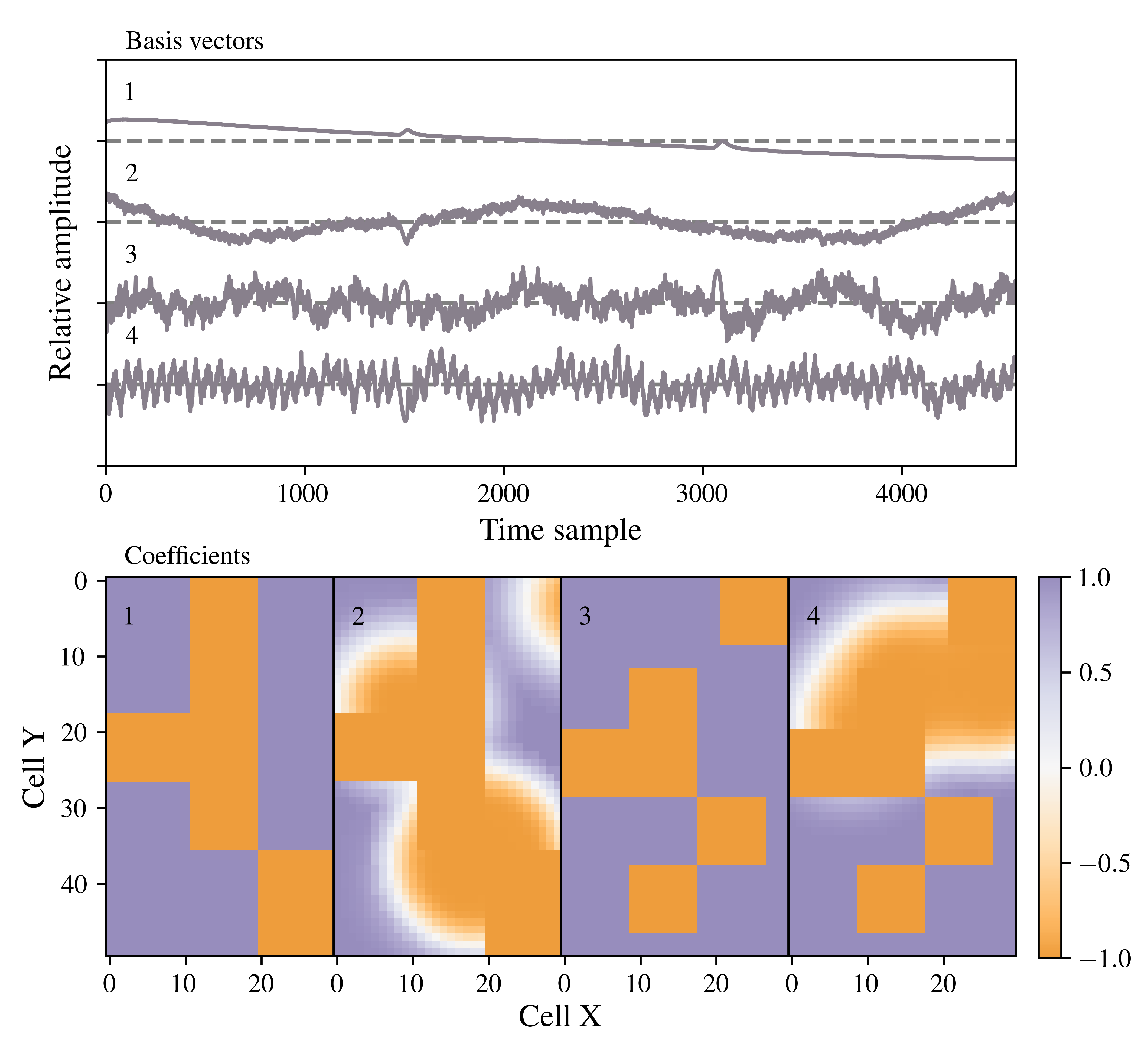}
\caption{Basis vectors $\mathbf{V}$ (upper) and coefficients $\mathbf{C}$ (lower) used for the full simulation of light curves in experiment B using a rank $K=4$ model.The basis vectors were selected at approximately equal spacing as index set (1, 5, 11, 16) from the 16 Kepler CBV published for Q10 and channel 30 \citep{stumpe2012kepler}. The coefficient spatial dependence was simulated as a mixture of random smooth features and sharp discontinuities. Both components are arbitrarily chosen and the discontinuities are not constrained to match the Kepler module gridding.} 

\label{fig: simulated_systematic}
\end{figure}

In this experiment (B) we fully simulated the light curves using only the coordinate framework of the Kepler test data. No simple generative model exists to fully simulate light curves due to the complex origin of residual systematics across the sensor, as discussed in Section~\ref{sec:intro}. Accordingly, we adopted a simple low-rank linear model for the systematics $\mathbf{L}=\mathbf{V}\mathbf{C}$ with rank $K=4$. The basis vectors $\mathbf{V}$ were selected at approximately equal spacing as index set (1, 5, 11, 16) from the 16 Kepler cotrending basis vectors (CBV) published for Q10 and channel 30 \citep{stumpe2012kepler}. The coefficients $\mathbf{C}$ were simulated with a random mix of discontinuities and smooth features, broadly comparable to the spatial structure found for the Kepler sensor test data but not constrained to be an exact match, including module gridding (Figure~\ref{fig: pca0}). We chose this spatial structure to be representative but distinct from the Kepler test data. The basis vectors $\mathbf{V}$ and coefficients $\mathbf{C}$ used in the full simulations are shown in Figure~\ref{fig: simulated_systematic}. Median normalization was applied to each simulated systematic term as $\bold{\hat{l}_i}=\bold{l}_i / {\rm med} (\bold{l}_i)-1$, where ${\rm med} (\bold{l}_i)$ is the median of $\bold{l}_i$ for $i \in I$.
We note that a PCA decomposition of the simulated systematics will not be identical. The simulated basis vectors and coefficients are rank $K=4$, however, the simulated systematics are rank $K_s = 5$. An extra basis term is introduced in a small number of light curves due to the median normalization of the simulated systematics. The corresponding basis vector is a constant offset.
We simulated $1500$ light curves in the $30 \times 50$ cell region, chosen to mirror the Kepler sensor region and discretization used in Section \ref{sec: preprocess}.
A total of $300$ astrophysical signals were simulated and are represented as matrix $\mathbf{A}$ as introduced in Section~\ref{sec: exp1}. Equal proportions of sinusoid, flare, and exoplanet transit signal were simulated with random signal parameters drawn as described in Section~\ref{sec: sim_sig}. Each simulated signal was injected into a single randomly-selected light curve in a one-to-one mapping.  The light curves $\mathbf{Y}$ (Section~\ref{sec: sys_model}) were simulated as $\mathbf{Y} = \mathbf{A} + \mathbf{V}\mathbf{C} + \mathbf{N}$, where $\bold{N}$ is iid Gaussian noise  
$[\mathbf{N}]_{n, i} \sim \mathcal{N} (0, \sigma^2)$ where $\sigma^2 = \frac{1}{4 |I|} \sum_{i \in I}  \rm var (\bold{l}_i)$. The simulations were performed over one run as no ensemble was necessary given the sufficient sample size of simulated and injected astrophysical signals.

As in Experiment A (Section~\ref{sec: exp1}), both the PCA method and the spatial systematics algorithm were used to detrend the data $\mathbf{Y}$ using model rank $K=8$, yielding $\mathbf{Y'}_{PCA}$ and $\mathbf{Y'}$ respectively; as before the PCA coefficients were used to initialize the spatial systematics algorithm (Algorithm~\ref{alg: gd}). This experiment deliberately simulates a high level of overfitting as the fitted model rank exceeds the rank of the simulated systematics. 
The performance of the algorithms was evaluated in terms of the accuracy with which the known systematics and astrophysical signals were recovered and the level of residual noise in the detrended light curves, as described in further detail below.

\subsubsection{Experiment C: Comparison with Standard Detrending Methods} \label{sec: compare}

We performed a high-level comparison against several standard detrending methods including the Kepler cotrending basis vector method (CBV, \citealt{stumpe2012kepler, smith2012kepler}), self-flat-fielding (SFF, \citealt{sff}), and pixel-level-decorrelation (PLD, \citealt{deming, Luger_2016, luger18}), all as implemented in the Lightkurve package \citep{lightkurve}. We caution that this is a high-level comparison only and not intended to assess the optimality of any method. For this comparison, the gridded Kepler targets (Figure~\ref{fig: flowchart}) selected for Q10 were used. The spatial systematics method was applied to these data with rank $K = 20$ after pre-processing (Section~\ref{sec: preprocess}; Figure~\ref{fig: flowchart}) and with no injection of astrophysical signals. However for the Lightkurve methods, the data for this target list were downloaded and processed inside that package as is customary. 

In general, for all Lightkurve methods we used the recommended parameters in the Lightkurve tutorial \footnote{\url{https://docs.lightkurve.org/tutorials/index.html\#removing-instrumental-noise}}. Specifically for the CBV and SFF detrending methods SAP pre-processing was performed using the {\it remove$\_$nans} and {\it remove$\_$outliers} functions. The PLD detrending method uses pixel-level data and no additional pre-processing was applied. The Lightkurve module masks bad quality-flagged cadences and removed $\sim 100$  cadences per light curve for these test data. We did not remove these cadences in our processing using the spatial systematics method (Figure \ref{fig: flowchart}). However, in our calculation of performance metrics we use the intersection of non-masked cadences among Lightkurve light curves. This excludes a further $\sim 100$ cadences per light curve. For the Lightkurve CBV detrending we used all single-scale CBVs available $K=16$. For Lightkurve SFF detrending we specified a window of width 401 cadences for pre-processing; this removes long-term variability and flattens each light curve. The Lightkurve SFF detrending was performed using 20 such windows. No other Lightkurve input parameters were specified. The broad relative performance of the spatial systematics algorithm against the comparison Lightkurve methods was evaluated in terms of the estimated level of residual systematics after detrending (Section~\ref{sec: noise_metrics}).

\section{RESULTS}\label{sec: results}
In this section, we describe the results of experiments A, B and C.

\subsection{Performance Metrics} \label{sec: noise_metrics}
The metrics used to numerically evaluate algorithm performance fall into the following two broad categories.

\paragraph{Residual Noise in Detrended Light Curves:}
This section concerns metrics used to assess the level of residual systematic noise in detrended light curves. The combined differential photometric precision (CDPP) over 6 hr is used in the Kepler pipeline and serves as an estimate of the level of white noise remaining after detrending on a transit timescale \citep{cdpp_christ}.

Following \citet{aig2016,aig} we used the procedure described in \citet{Gilliland_2011} to compute an approximate measure of the CDPP in our detrended light curves. A 2D quadratic Savitsky-Golay filter was applied to the detrended light curves to remove low-frequency variability and the data were then averaged over 6-h bins. The approximate CDPP was then computed as the standard deviation of the binned time series and scaled by a value $1.168$. The median CDPP  across the set of detrended light curves is reported. \\

The second metric for residual systematic noise was adopted from the goodness metric defined by \citet{stumpe2012kepler} and is denoted here as $G_{y'}$. This metric is formed as the average absolute cubed pairwise correlation between detrended light curves as $G_{y'}=\frac{2}{N(N-1)}\sum_{i \neq j} |\rho(\mathbf{y}^{'}_i, \mathbf{y}^{'}_j)|^3$. Lower values of $G_{y'}$ suggest a lower residual systematic noise in detrended light curves. The cube down weights correlations which may be spuriously low if all residual astrophysical variability is removed and higher correlations make a more significant contribution to this metric.

\paragraph{Cross-Correlation against Known Signals and Systematics:}

In this section, we consider metrics used to measure the degree to which known astrophysical signals or systematics are recovered in the injection tests or full simulations. 

The injection matrix $\mathbf{A}$ defined above is a null matrix with a subset of columns containing simulated vector light curves that are either sinusoids ($\mathbf{a}_s$), flares ($\mathbf{a}_f$), or exoplanet transit signals ($\mathbf{a}_t$), as defined in Section~\ref{sec: sim_sig}. Each simulated light curve is mapped to a random light curve in a one-to-one mapping. We denote the mean correlation of all simulated astrophysical signals of type $\mathbf{a}_{[m]},\ m \in \{s,t,f\}$ against their associated detrended light curves $\mathbf{y}'_{[m]}$ and averaged over all simulation runs as $\bar{\rho}(\mathbf{a}_{[m]},\bold{y}'_{[m]})$. This is a measure of the degree to which the simulated or injected astrophysical signal has been recovered from the detrended light curve. When averaged across all  signal types this is denoted $\bar{\rho}(\mathbf{a},\bold{y}'_{\mathbf{a}})$. 

We denote the mean absolute correlation of all simulated astrophysical signals of type $\mathbf{a}_{[m]}$ against the set of all detrended light curves where no astrophysical signal was injected $\mathbf{y}'_{\mathbf{a}^c}$ and averaged over all simulation runs, as $\text{\textbar}\bar{\rho}(\bold{a}_{[m]}, \bold{y}'_{\bold{a}^c})\text{\textbar},\ m \in \{s,t,f\}$. The absolute value is used as the simulated signal may be anti-correlated with the estimated systematics. This correlation measures whether a detrending algorithm is corrupting light curves that do {\it not} contain an injected astrophysical signal $(\mathbf{y}'_{\mathbf{a}^c})$. The unprocessed light curves before detrending $(\bold{y}_{\mathbf{a}^c})$ may however be incidentally correlated with an injected astrophysical signal and therefore we normalize by the mean absolute correlation of the non-detrended light curves 
$\frac{\text{\textbar}\bar{\rho}(\bold{a},\bold{y}'_{ \bold{a}^c}) \text{\textbar} }{ |\bar{\rho}(\bold{a}, \bold{y}_{\bold{a}^c})| }$. Values below unity indicate that the detrending algorithm is not introducing spurious correlation with the injected signals beyond that present incidentally in the non-detrended light curves.

In Experiment B, the constituent systematic light curve vectors $\bold{l}_i,\ (i \in I)$ comprising matrix $\bold{L}$ are also known, in addition to the injected astrophysical signals $\bold{a}_{[m]}$, $m \in  \{s,t,f \}$. We denote the estimated systematics vector for light curve $i$ as $\hat{\bold{l}}_i$. The mean correlation of the simulated $\bold{l}_i$ and estimated systematics vectors $\hat{\bold{l}}_i$ averaged over all light curves is denoted $\bar{\rho}(\bold{l}, \hat{\bold{l}})$. If this metric is restricted to light curves where an astrophysical signal $\bold{a}_{[m]}$, $m \in \{s,f, t \}$ was injected, it is denoted $\bar{\rho}(\bold{l}_{\bold{a}}, \hat{\bold{l}}_{\bold{a}})$.

\subsection{Experiment A}

\begin{figure}[htbp!] 
  \centering
  \includegraphics{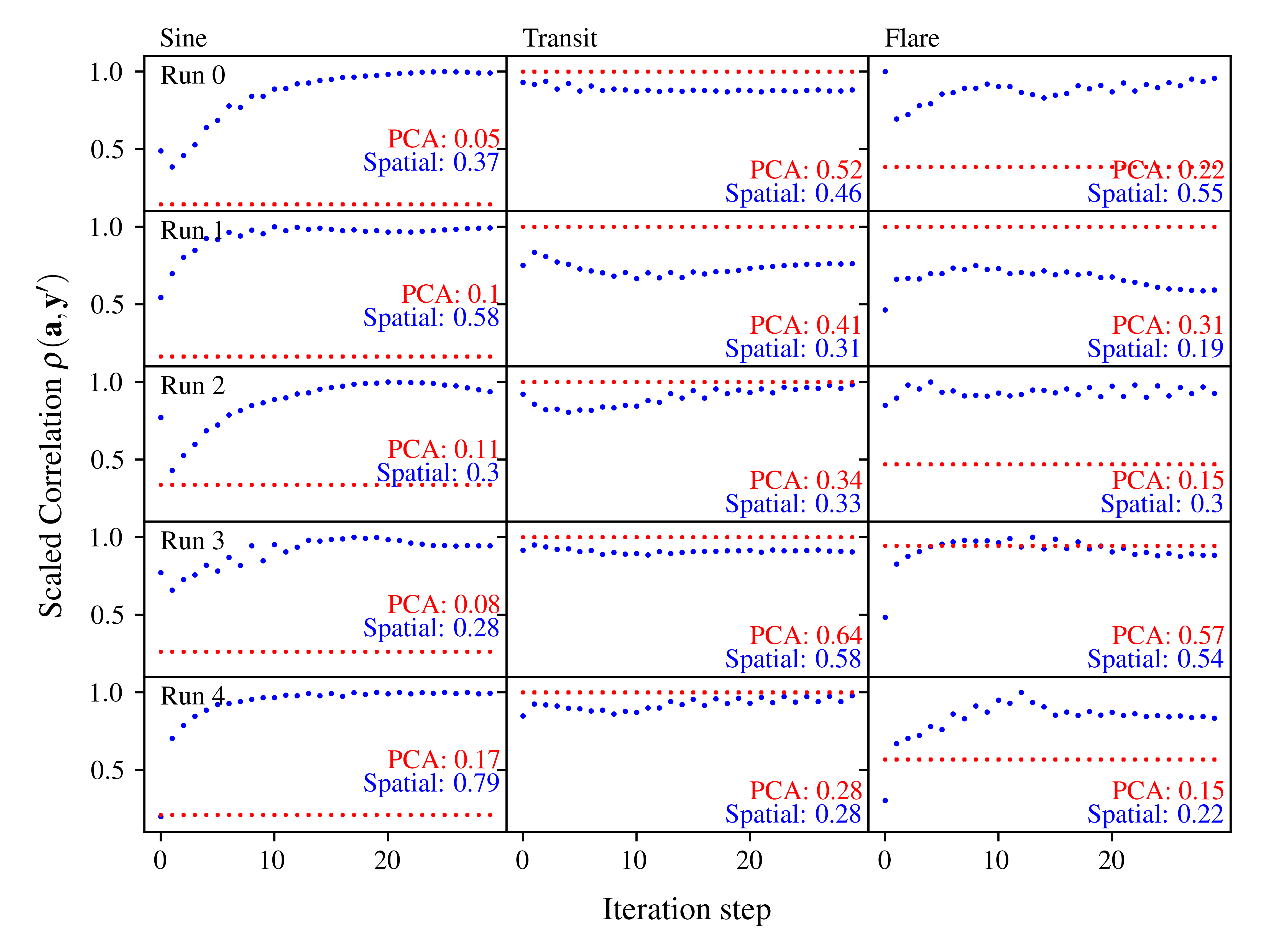}
    \caption{The scaled mean correlation of injected signals and associated detrended light curves $\bar{\rho}(\mathbf{a}_{[m]},\bold{y}'_{[m]})$ for Experiment A for sine, flare, and transit signal types $m \in \{s,t,f\}$ (across columns) for each of five separate simulation runs (across rows). This mean correlation, defined in the main text, is plotted for the spatial systematics algorithm (blue) and the PCA decomposition method (red). The mean correlation is scaled by the maximum value achieved by either method; this maximum is shown in the legend of each subplot. The PCA algorithm is non-iterative and is depicted as a straight line accordingly. The x-axis label is the iteration number of the spatial systematics algorithm. The spatial systematics algorithm is generally convergent and typically has comparable or improved performance relative to PCA but this is not universal for all runs.}  \label{fig: exp1_iter} 
\end{figure}

Experiment A, as described in Section~\ref{sec: exp1}, was performed to numerically evaluate the spatial systematics algorithm in terms of its ability to recover transient and variable signals injected into Kepler test data. A standard PCA decomposition method was used as a comparison. The convergence of the spatial systematics algorithm for five representative runs of this experiment is shown in Figure~\ref{fig: exp1_iter}. This figure plots the mean correlation $\bar{\rho}(\mathbf{a}_{[m]},\bold{y}'_{[m]})$ (over light curve) of the injected astrophysical signal types $m \in \{s,t,f\}$ as a function of spatial systematics iteration number. This mean correlation is a proxy for the degree of recovery of injected astrophysical signals, as described in Section~\ref{sec: noise_metrics}. Figure~\ref{fig: exp1_iter} shows general convergence of the spatial systematics algorithm within $\sim 30$ iterations. The figure shows signal recovery performance comparable to or exceeding the reference PCA method in the majority of cases but not without exception for individual flare and transit runs.

\begin{figure}[htbp!]
  \centering
  \includegraphics{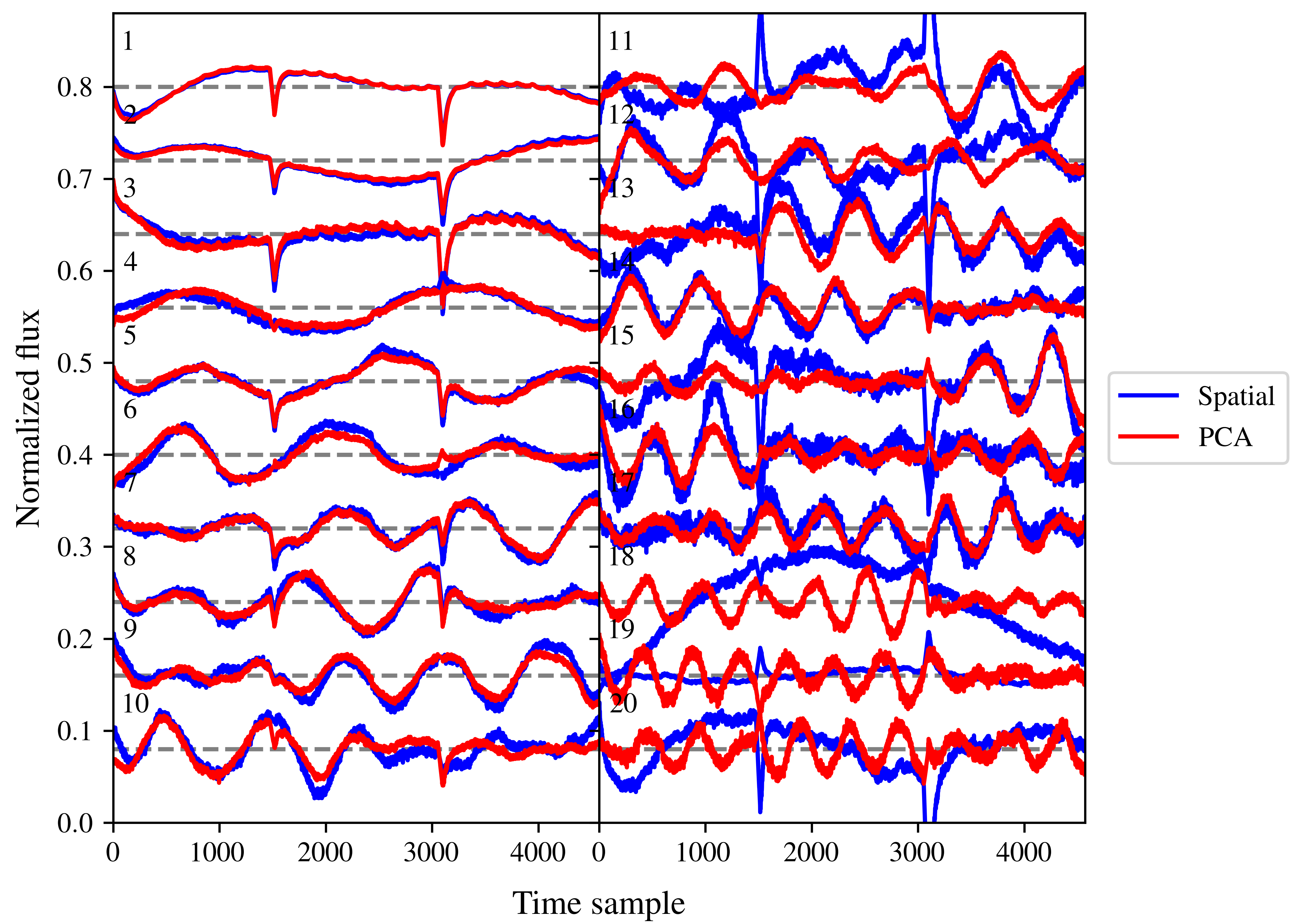}
    \caption{The estimated basis vectors $\bold{v}_k,\ k \in \{1,,20\} (K=20)$ obtained for run \#0, a single representative run of Experiment A, shown here for the spatial systematics algorithm (blue) and the reference PCA method (red). The spatial basis vectors and PCA basis vectors closely overlap for the first two terms.}
      \label{fig: basis_fit} 
\end{figure}

\begin{figure}[htbp!]
  \centering
  \includegraphics{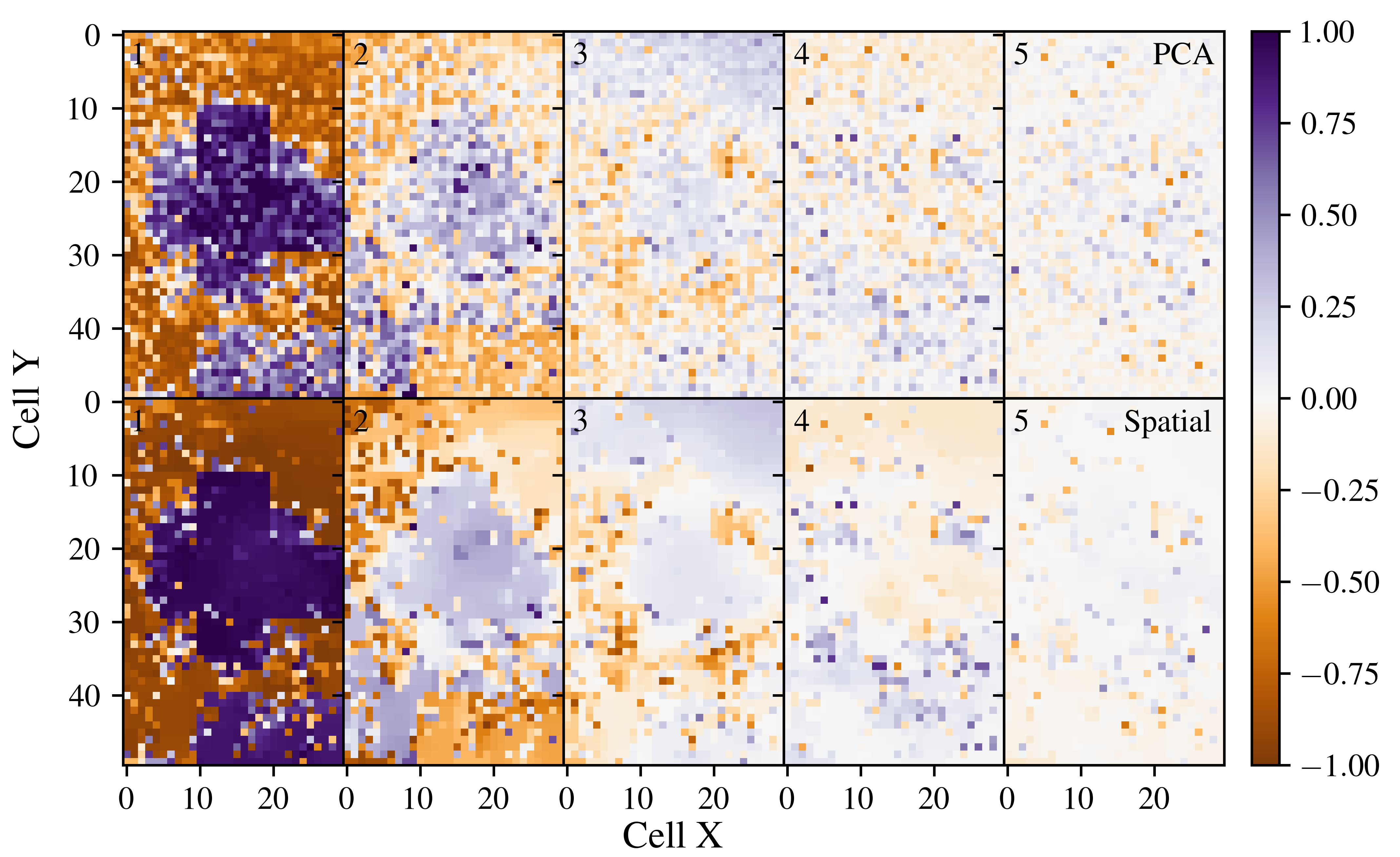}
    \caption{Basis vector coefficients $c^k_i,\ k \in \{1,..,5\} (K=20)$ for light curve $i \to (x,y) \in (X,Y)$ obtained using the spatial systematics (top row) and PCA algorithms (bottom row) for run \#0, a representative run of Experiment A. The coefficient index $k$ is shown in the top left of each sub-figure. The x- and y-axes are in units of gridded spatial cells. 
    \label{fig: coeff_fit}}  
\end{figure}

\begin{figure}[htbp!]
  \centering
  \includegraphics{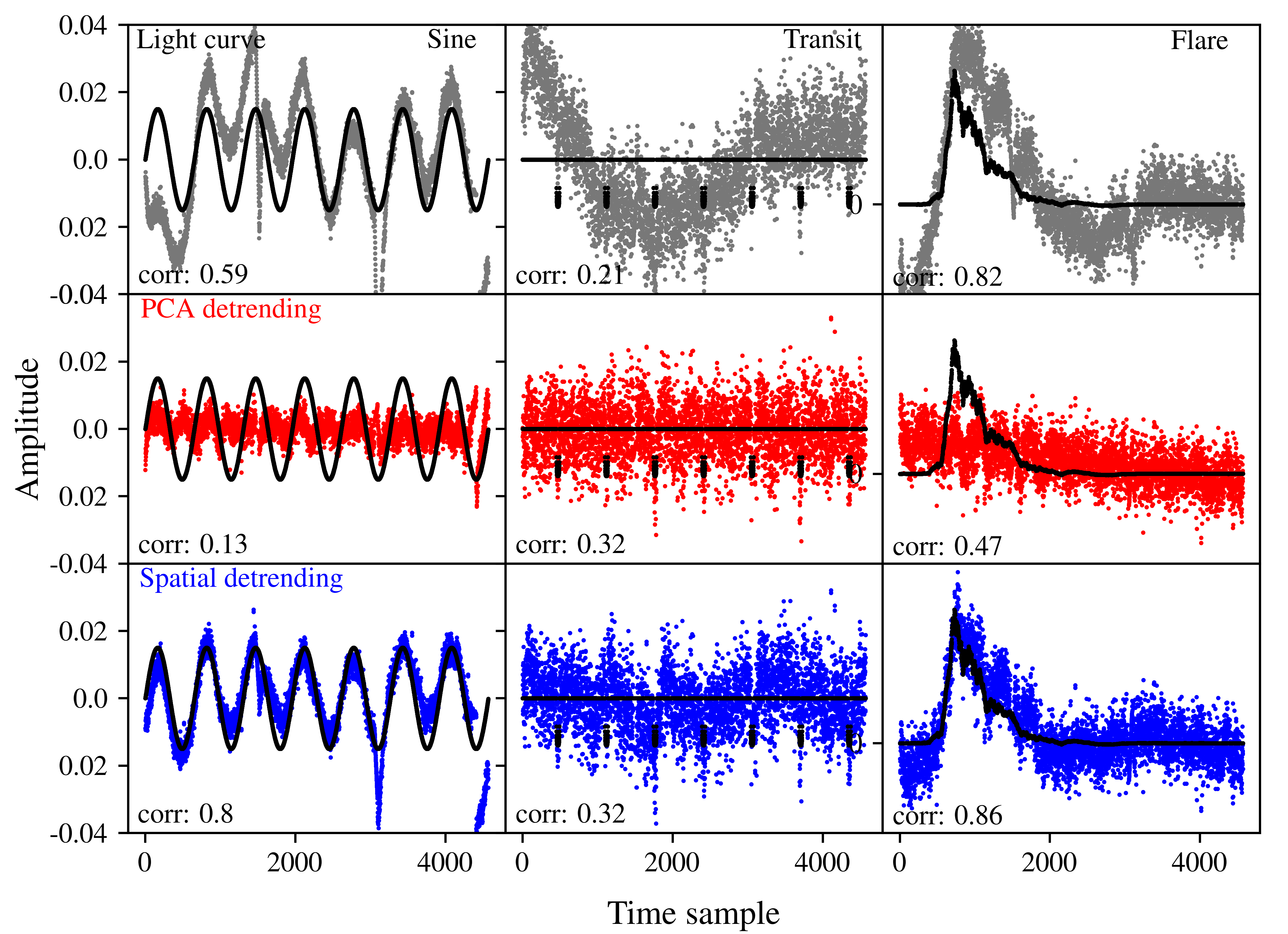}
    \caption{Three sample light curves across injected signal type $\bold{a}_{[m]}$ $m \in \{s, t, f \}$ (sine, transit, and flare) from run \#0, a representative run of Experiment A. The figure shows the pre-processed light curve before detrending $\bold{y}$ (upper row), the detrended light curve using PCA $\bold{y}'_{PCA}$ (middle row), and the detrended light curve using the spatial systematics algorithm $\bold{y}'$ (lower row).Simulated injected signals are overlaid in black. The x-axis is Kepler long-cadence time sample index. This sample shows an instance where PCA detrending removed the injected astrophysical sine and flare signals, while spatial detrending preserved those signals.}

    \label{fig: sim_shock}
\end{figure}

The performance of the spatial systematics algorithm is shown in more detail for a single representative run in Figures~\ref{fig: basis_fit}-\ref{fig: sim_shock}. Figure~\ref{fig: basis_fit} shows the estimated basis vectors $\bold{v}_k$ obtained with the spatial systematics and reference PCA algorithms. The estimated basis vectors $\bold{v}_k$ are generally similar but differences increase at higher-order $k$. Figure~\ref{fig: coeff_fit} shows the fitted coefficients $c^k_i,\ k \in \{1,..,5\}\ (K=20)$ obtained by the spatial systematics and PCA algorithms across the gridded light curves $i \to (x,y) \in (X, Y)$. This figure shows that the total variation constraint in the spatial systematics algorithm enforces smoothness in the coefficients while preserving discontinuities. In Figure \ref{fig: sim_shock} example detrended light curves are shown over injected astrophysical signal type.
In this example, the spatial systematics algorithm has preserved the injected astrophysical variability to a greater degree than the reference PCA method. However, the spatial systematics algorithm has a higher residual scatter in the detrended light curves than PCA, as is visible near cadence 3000 for the light curve injected with a sine signal. 

The summary noise metrics (Section~\ref{sec: noise_metrics}) for Experiment A for the final iteration of the spatial systematics algorithm and averaged over all runs in the ensemble are listed in Table~\ref{tab: overfit}. This table shows that for these simulations the spatial systematics algorithm generally outperforms the reference PCA method with caveats discussed in further detail in Section~\ref{sec: discussion}.

\begin{splitdeluxetable*}{cccccBcccccc}
\tablecaption{Experiment A Results}
\tablehead{
\colhead{} & \multicolumn{1}{c}{Sine recovery}
 & \multicolumn{1}{c}{Sine corruption} 
 & \multicolumn{1}{c}{Transit recovery}  & \multicolumn{1}{c}{Transit corruption}
 & \colhead{}  &\multicolumn{1}{c}{Flare recovery} & \multicolumn{1}{c}{Flare corruption} &   \multicolumn{1}{c}{Signal recovery} & \multicolumn{2}{c}{Residual systematics} \\
 \colhead{} & \colhead{$\bar{\rho}(\bold{a}_{s}, \bold{y}^{'}_{s})$ } &  
$\text{\textbar}\bar{\rho}(\bold{a}_{s},\bold{y}'_{ {\bold{a}^c}}) \text{\textbar} / \text{\textbar}\bar{\rho}(\bold{a}_{s}, \bold{y}_{ \bold{a}^c})\text{\textbar}$
 &  \colhead{$\bar{\rho}(\bold{a}_{t}, \bold{y}'_{t})$}
 &  $\text{\textbar}\bar{\rho}(\bold{a}_{t},\bold{y}'_{ {\bold{a}^c}}) \text{\textbar} / \text{\textbar}\bar{\rho}(\bold{a}_{t}, \bold{y}_{ \bold{a}^c})\text{\textbar}$ & \colhead{}  & \colhead{$\bar{\rho}(\bold{a}_{f}, \bold{y}^{'}_{f})$}  &   $ \text{\textbar}\bar{\rho}(\bold{a}_{f},\bold{y}'_{ {\bold{a}^c}}) \text{\textbar} / \text{\textbar}\bar{\rho}(\bold{a}_{f}, \bold{y}_{\bold{a}^c})\text{\textbar} $  &    \colhead{$\bar{\rho}(\bold{a}, \bold{y}'_{\bold{a}})$}  &  \colhead{$G_{\bold{y}'}$} &  
 \colhead{CDPP$_{6h}$ (ppm) } }
\startdata 
PCA & 0.14 $\pm$ 0.04  & 0.12 $\pm$ 0.02 & 0.45 $\pm$ 0.08 & 0.60 $\pm$  0.11  &
PCA & 0.35 $\pm$ 0.13 & 0.09 $\pm$ 0.03 & 0.31 $\pm$ 0.08 & (6.61 $\pm$ 0.12) $\times$ 10$^{-4}$ & 38.91 $\pm$ 0.03 \\
Spatial & 0.57 $\pm$ 0.11 & 0.85 $\pm$ 0.09 & 0.44 $\pm$ 0.06 & 0.80 $\pm$ 0.12 & 
Spatial & 0.64 $\pm$ 0.15 & 0.54 $\pm$ 0.09 & 0.55 $\pm$ 0.11 & (1.73  $\pm$ 0.04) $\times$ 10$^{-2}$  & 40.33 $\pm$ 0.14  \\
Ratio(PCA/Spatial) &  0.25 & - & 1.01 & - & Ratio(PCA/Spatial) &  0.54 & - &  0.56 &  0.04 &  0.97 \\
\enddata
\tablecomments{Performance metrics for detrended light curves, as defined in Section~\ref{sec: noise_metrics}. As described there, {\it recovery} indicates the degree to which injected signals are preserved correctly in the detrended light curves. {\it Corruption} indicates the degree to which injected signals incorrectly appear in light curves that contain no injected signals. The uncertainties listed are the standard deviation of the value over the ensemble of ten runs. For reference, the correlation of non-detrended, pre-processed and non-injected light curves $\bold{y}_{\bold{a}^c}$ with simulated astrophysical signals $\bold{a}_{[m]}$ have values $\text{\textbar}\bar{\rho}(\bold{a}_{s}, \bold{y}_{ \bold{a}^c})\text{\textbar} = 0.07$, $\text{\textbar}\bar{\rho}(\bold{a}_{t}, \bold{y}_{ \bold{a}^c})\text{\textbar} = 0.04$, $\text{\textbar}\bar{\rho}(\bold{a}_{f}, \bold{y}_{ \bold{a}^c})\text{\textbar} = 0.18$ for sine, transit, and flare signals respectively. For non-detrended pre-processed light curves, the goodness metric is  $G_{\bold{y}}$ is $0.26 \pm 0.01$ and the median CDPP$_{6h}$ is $51.21$ $\pm 0.14$ ppm. \label{tab: overfit}}

\end{splitdeluxetable*} 

\subsection{Experiment B}
Experiment B (Section~\ref{sec: exp2}) was designed to numerically evaluate the spatial systematics algorithm using fully-simulated data generated using only the Kepler test data coordinates. The algorithm was evaluated in terms of its ability to recover the known astrophysical signals and systematics used in simulating the light curves. 

\begin{figure}[htbp!]
  \centering
  \includegraphics{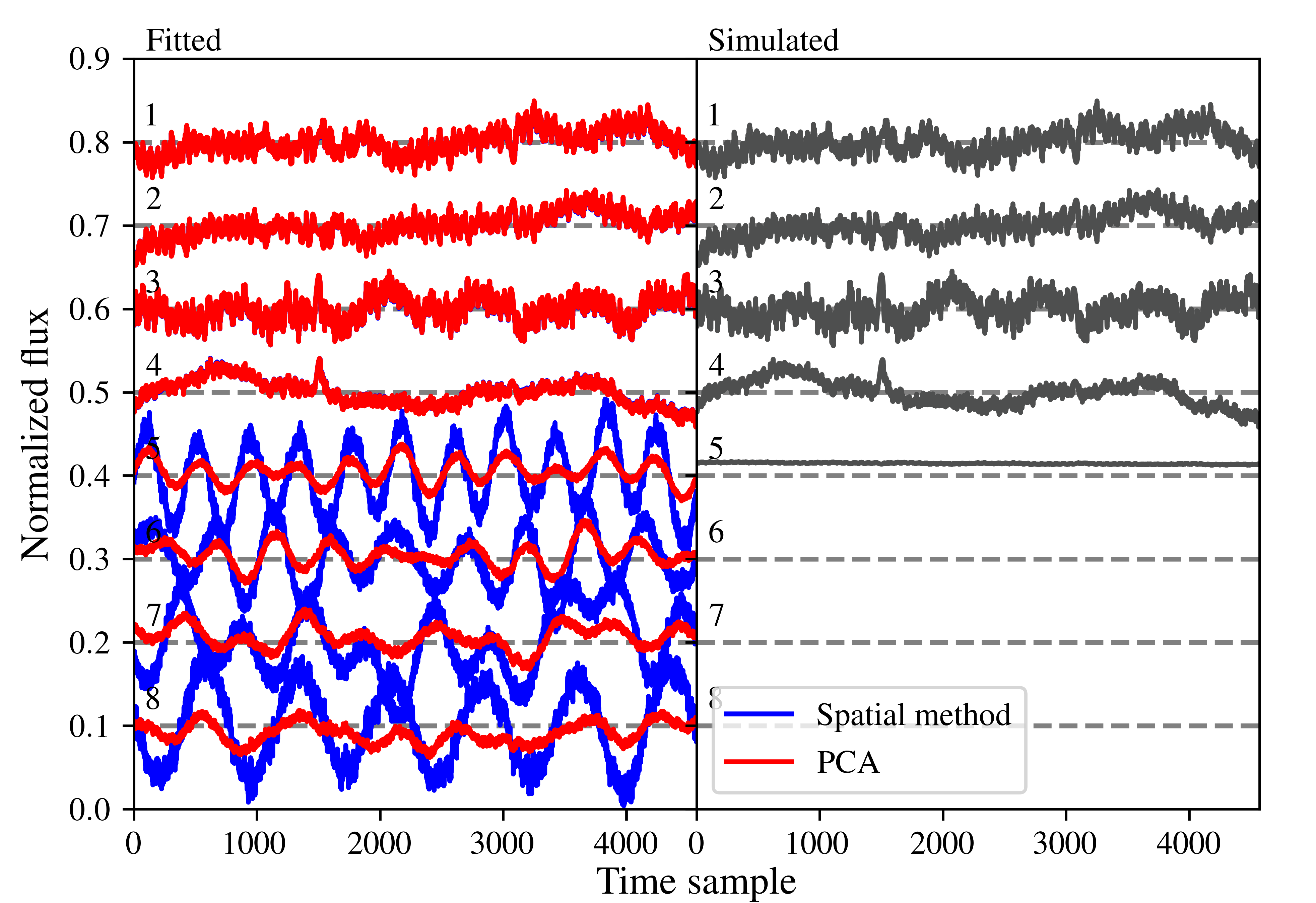}
    \caption{The estimated (left) and known simulated (right) basis vectors $\bold{v}_k,\ k \in \{1,..,8\}$ for Experiment B. The estimated basis vectors (left) are shown for the spatial systematics algorithm (blue) and the reference PCA method (red). The x-axis is in units of Kepler long-cadence time sample.  The estimated spatial and PCA basis vectors (left) for the first four terms overlap one another in these plots.} \label{fig: expb_fitted_basis}
\end{figure}

\begin{figure}[htbp!] 
  \centering
  \includegraphics{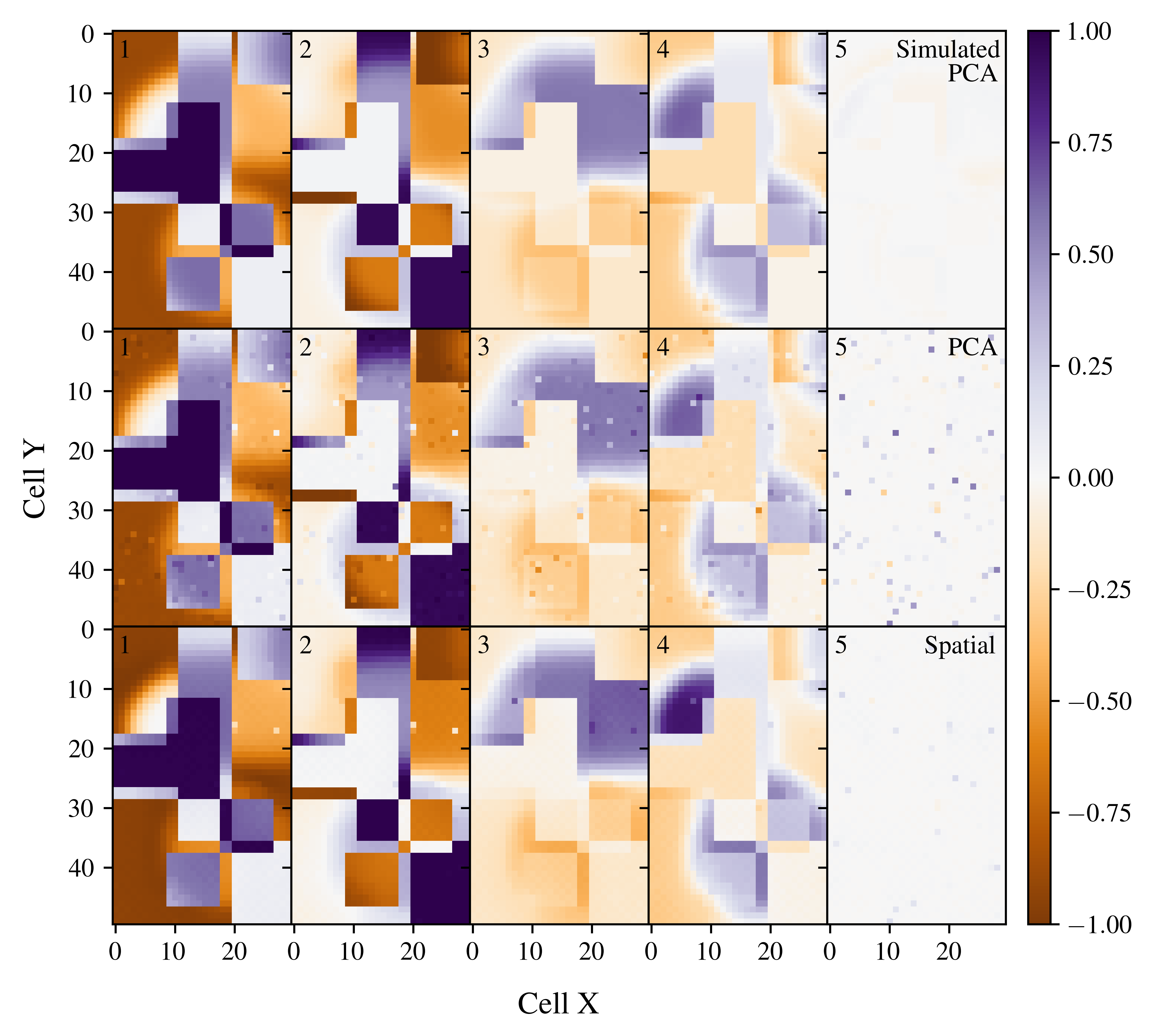}
    \caption{Coefficients $c_i^k,\ k \in \{1,..,5\} (K=8)$ for light curve $i \to (x,y) \in (X,Y)$ for Experiment B. This figure shows the true simulated coefficients (top row), the coefficients obtained using the PCA method (middle row), and the coefficients obtained using the spatial systematics algorithm (bottom row). The coefficient index $k$ is shown in the top left of each sub-figure. The x- and y-axes are in units of gridded spatial cells. Coefficients fitted with PCA show speckling, indicative of the overfitting of injected astrophysical signals.}
    \label{fig: exp2_map}
\end{figure}

The simulated systematics are rank $K_s=5$  but were estimated with rank $K=8$ as described above. The estimated and simulated basis vectors $\bold{v}_k,\ k \in \{1,..,8\}$ for Experiment B are shown in Figure~\ref{fig: expb_fitted_basis}. The first four estimated basis vectors are identical between the spatial systematics and PCA algorithms. Higher-order basis vectors deviate between the two methods and are also poorly constrained. However, fitted coefficients for these basis terms ($K > 4$) are small. For the spatial systematics algorithm these fitted coefficients contribute $\sim 4 \%$ in total magnitude ($\sum_{k = 5}^K \| c_k\| / \sum_{k = 1}^K \| c_k\|$), while for PCA the contribution is $\sim 11 \%$. Figure \ref{fig: exp2_map} shows the true simulated coefficients and those estimated using the PCA and spatial systematics algorithms for the first $5/8$ basis terms. The coefficients estimated using PCA have a more speckled spatial sub-structure than the smoother distribution obtained by the spatial systematics algorithm.

\begin{figure}[htbp!] 
  \centering
  \includegraphics{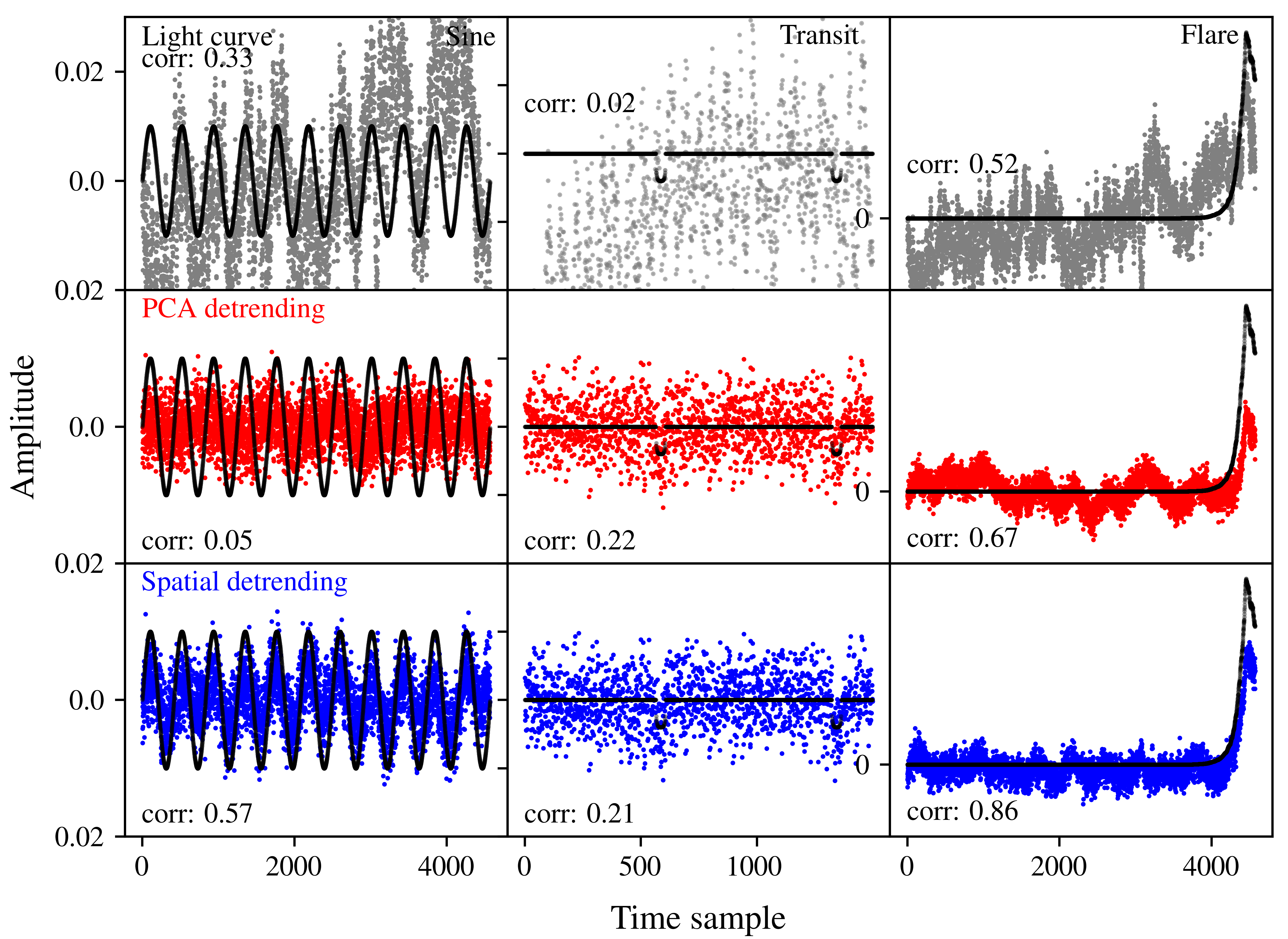}
    \caption{Three sample light curves across injected signal type $\bold{a}_{[m]}$ $m \in \{s, t, f \}$ (sine, transit, and flare) for Experiment B. The figure shows the simulated light curve before detrending $\bold{y}$ (upper row), the detrended light curve using PCA $\bold{y}'_{PCA}$ (middle row), and the detrended light curve using the spatial systematics algorithm $\bold{y}'$ (lower row). Simulated injected signals are overlaid in black. The x-axis is Kepler long-cadence time sample index. This sample shows an instance where PCA detrending moderately removed the injected astrophysical sine and flare signals, while spatial detrending more clearly preserved those signals. \label{fig: sim_detrend}}
\end{figure}

Example individual detrended light curves for sine, transit, and flare simulated signal types are shown in Figure \ref{fig: sim_detrend}. Summary results for Experiment B, in terms of the metrics described in Section~\ref{sec: noise_metrics}, are provided in Table~\ref{tab: exp2}. 

\begin{deluxetable}{ccccccccc} [htbp!]
\tablecaption{Experiment B results}
\tablehead{
\colhead{} & \multicolumn{2}{c}{Systematics recovery} & \multicolumn{4}{l}{Signal recovery} &  \multicolumn{2}{l}{Residual systematics} \\
\colhead{} & \colhead{$\bar{\rho}(\bold{l}_{\bold{a}}, \hat{\bold{l}}_{\bold{a}})$} & \colhead{$\bar{\rho}(\bold{l}, \hat{\bold{l}})$} & \colhead{$\bar{\rho}(\bold{a}_{s}, \bold{y}_{s}')$} & \colhead{$\bar{\rho}(\bold{a}_t, \bold{y}_t')$} & \colhead{$\bar{\rho}(\bold{a}_{f}, \bold{y}_{f}')$} & \colhead{$\bar{\rho}(\bold{a}, \bold{y}_{\bold{a}}^{'})$}  & \colhead{$G_{\bold{y}'}$}  &  \colhead{CDPP$_{6h}$}} 
\startdata
PCA & 0.96 $\pm$ 0.08 & 0.99 $\pm$ 0.03 & 0.52 $\pm$ 0.30 & 0.28 $\pm$ 0.15 & 0.39 $\pm$ 0.17 & 0.39 $\pm$ 0.24 & 1.71 $\times$10$^{-4}$  & 0.39 $\pm$ 0.01 \\
Spatial & 0.98 $\pm$ 0.08 & 0.99 $\pm$ 0.03 & 0.63 $\pm$ 0.22 & 0.28 $\pm$ 0.15 & 0.48 $\pm$ 0.19 & 0.47 $\pm$ 0.24 & 1.55 $\times$10$^{-4}$ & 0.40 $\pm$ 0.01 \\
Ratio $(PCA/Spatial)$ & 0.98 & 0.99 & 0.82 & 0.98 & 0.79 & 0.82 & 1.10 & 0.98
\enddata
\tablecomments{Performance metrics for detrended light curves, as defined in Section~\ref{sec: noise_metrics} and used in Table~\ref{tab: overfit}. As described there, {\it recovery} indicates the degree to which injected signals are preserved correctly in the detrended light curves. The uncertainties listed are the standard deviation of the value over the ensemble of simulated light curves. No uncertainty is listed for $G_{\bold{y}'}$ as it is computed over the ensemble of light curves. The goodness metric value computed over the non-detrended simulated light curves $G_{\bold{y}}$ is $0.23$. The goodness metric value computed between all simulated astrophysical signals $G_{\bold{a}}$ is $2.2 \times 10^{-2}$. } \label{tab: exp2}
\end{deluxetable} 

\subsection{Experiment C} \label{sec: compare_results}
\begin{figure}[htbp!]
  \centering
  \includegraphics[width=.6\linewidth]{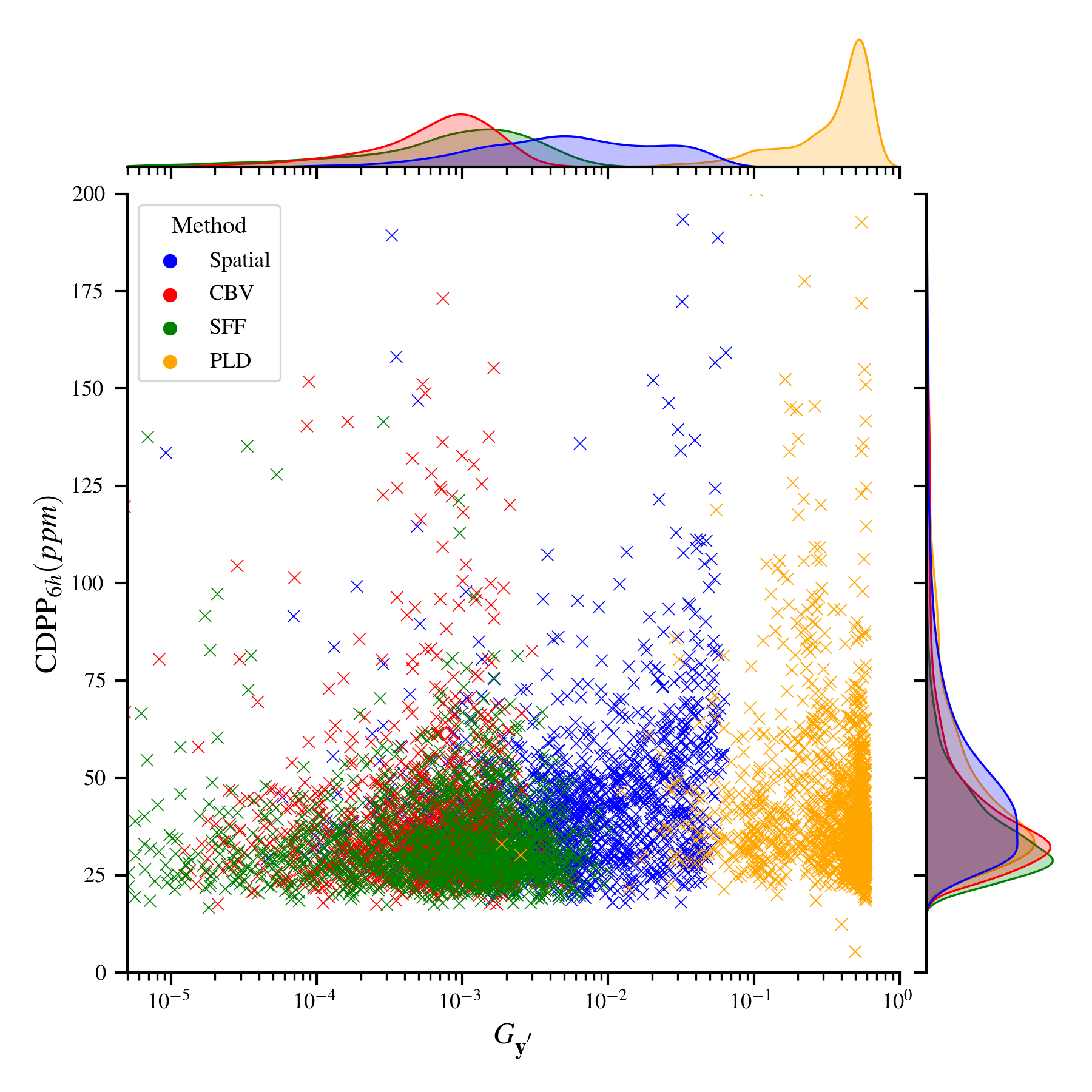}
    \caption{A plot of CDPP$_{6h}$ (ppm) versus goodness metric $G_{\bold{y}'}$ per target light curve shown for the spatial systematics method and comparison standard detrending methods CBV, SFF, and PLD. Empirical distribution functions for each light curve type are plotted over $G_{\bold{y}'}$ (top) and CDPP$_{6h}$ (right). The CDPP$_{6h}$ metric is broadly comparable across method while the $G_{\bold{y}'}$ metric is more strongly dependent on detrending method. }\label{fig: compare_ratio} 
\end{figure}

Experiment C was designed to provide a high-level comparison of the spatial systematics algorithm against the standard CBV, SFF, and PLD methods (Section~\ref{sec: compare}). The performance metrics $G_{\bold{y}'}$ and CDPP$_{6h}$ for the residual noise in the detrended light curves for this experiment are shown in Table~\ref{tab: compare_table}. A composite plot of CDPP$_{6h}$ versus $G_{\bold{y}'}$ across all methods in Experiment C is shown in Figure~\ref{fig: compare_ratio}. These results show that the CDPP$_{6h}$ metric is broadly comparable across all methods while the $G_{\bold{y}'}$ metric has a greater dependence on detrending method. As noted in Section~\ref{sec: compare}, the data used in this comparison have undergone different pre-processing, and furthermore detrending methods have not been optimized in each specified case. The purpose of this comparison is to provide a high-level reference comparison and not to show the general optimality of any single method.

\begin{deluxetable}{cccc}[htbp!]
\tablecaption{Experiment C results}
\tablehead{
\colhead{} & \multicolumn{3}{c}{Residual systematics} \\
\colhead{}  & \colhead{$G_{\bold{y}'}$}  &  \multicolumn{2}{l}{CDPP$_{6h}$ }\\
\colhead{} & \colhead{} & \colhead{$\mu$} & \colhead{$\sigma$}}
\startdata
Spatial & 1.13 $\times$10$^{-2}$ & 38.37 & 60.17\\
CBV & 8.95 $\times$10$^{-4}$ & 33.16 & 97.60 \\
SFF & 1.45 $\times$10$^{-3}$ & 30.78 & 24.16 \\
PLD & 3.92 $\times$10$^{-1}$ & 36.49 & 41.44
\enddata
\tablecomments{Residual systematics metrics for detrended light curves produced in Experiment C. The CDPP$_{6h}$ mean over the ensemble of simulated light curves is denoted $\mu$ with sample standard deviation $\sigma$. No uncertainty is listed for $G_{\bold{y}'}$ as it is computed over the ensemble of light curves. } \label{tab: compare_table}
\end{deluxetable} 

\clearpage

\section{DISCUSSION}\label{sec: discussion}
The numerical evaluation of the spatial systematics algorithm presented here, using injected signals in Kepler data (Experiment A) and full simulations (Experiment B), shows that in these cases the algorithm matches or outperforms the reference PCA method for astrophysical signal recovery and achieves comparable performance for systematics removal. This is indicated by higher values of $\bar{\rho}(\bold{a},\bold{y}'_{\bold{a}})$ and $\bar{\rho}(\bold{l}_{\bold{a}},\hat{\bold{l}}_{\bold{a}})$ respectively in Tables~\ref{tab: overfit} and~\ref{tab: exp2}. 

The improved recovery of injected astrophysical signals by the spatial systematics algorithm is direct evidence that the algorithm is less prone to erroneously absorbing true astrophysical variability in the systematics (overfitting) than the reference PCA method. Astrophysical variability and systematics are not separable a priori and overfitting is thus a foundational concern in the detrending of exoplanet transit light curves \citep{stumpe2012kepler, Smith_2018}. Overfitting may remove true astrophysical variability from detrended light curves and also introduce spurious variability into other detrended light curves in the sample as these corrupted systematics are applied. A number of detrending approaches infer a basis representative of systematic effects from a correlated set of light curves \citet{ Petigura_2012, foreman2015systematic} and Kepler PDC-MAP \citet{smith2012kepler, stumpe2012kepler}. The set of light curves from which a set of basis vectors is constructed can be robustly filtered to exclude outliers and those with clear astrophysical variability as a means to mitigate overfitting, as in PDC-MAP \citep{stumpe2012kepler}. A basis consisting only of systematics (which is difficult to realize) can however still be overfitted to astrophysical features \citep{Smith_2018}, as it is unlikely that a basis is completely orthogonal to all possible astrophysical signals.
Spatial dependence among light curves has been identified by \citet{Petigura_2012} and \citet{moreno}, whereby the latter work explains spatial correlations as time-delayed systematic effects traversing the sensor. The modified total variation prior applies a spatial correlation constraint across the sensor which is well-suited for inference of local effects of this form. Functionally, the spatial systematics model depends on systematic coefficients alone, this model is differentiable, and derived gradients are computationally tractable. The algorithm iteratively refines both the fitted coefficients and the overall systematics model. These algorithm properties help to separate the systematics and astrophysical variability and thereby mitigate overfitting. A position-based prior is used in PDC-MAP to reduce overfitting of cotrending basis vectors to a light curve. The spatial systematics algorithm is novel in that it uses a spatial constraint computed directly from systematic estimates, of total variation form and further, iteratively refines the overall systematics solution for a collection of light curves. We emphasize the comparative performance to standard PCA and while we perform a high-level comparison to specialized detrending methods (Experiment C), we do not claim optimality for this exploratory algorithm development.
The reduced overfitting is most evident in Tables~\ref{tab: overfit} and~\ref{tab: exp2} for sine $\bar{\rho}(\bold{a}_s,\bold{y}'_s)$ and flare $\bar{\rho}(\bold{a}_f,\bold{y}'_f)$ signals and comparable for transit signals $\bar{\rho}(\bold{a}_t,\bold{y}'_t)$. This is also visible in the sample detrended light curves shown in Figures~\ref{fig: sim_shock} and~\ref{fig: sim_detrend}. In summary, the longer-duration astrophysical signals (sine, flare) have reduced overfitting relative to the transit signals, which are of shorter duration. The shorter the duration of an astrophysical signal relative to the timescale of systematics present in Kepler light curves (Figure~\ref{fig: basis_fit}), the less linearly dependent on an inferred systematics basis and therefore the smaller the expected improvement over the reference PCA method. Notably, the spatial systematics algorithm performs well over all astrophysical signal characteristic timescales considered here. The spatial systematics algorithm contains no explicit model for transient or variable astrophysical signals, however, it performs well for the signal types considered here (Section~\ref{sec: sim_sig}). The consistent relative performance between signal types in Experiments A and B further supports the preceding proposed causal link between the signal and characteristic residual systematic timescales, rather than the exact functional form of the astrophysical signal.

\begin{figure}[htbp] 
  \centering
  \includegraphics{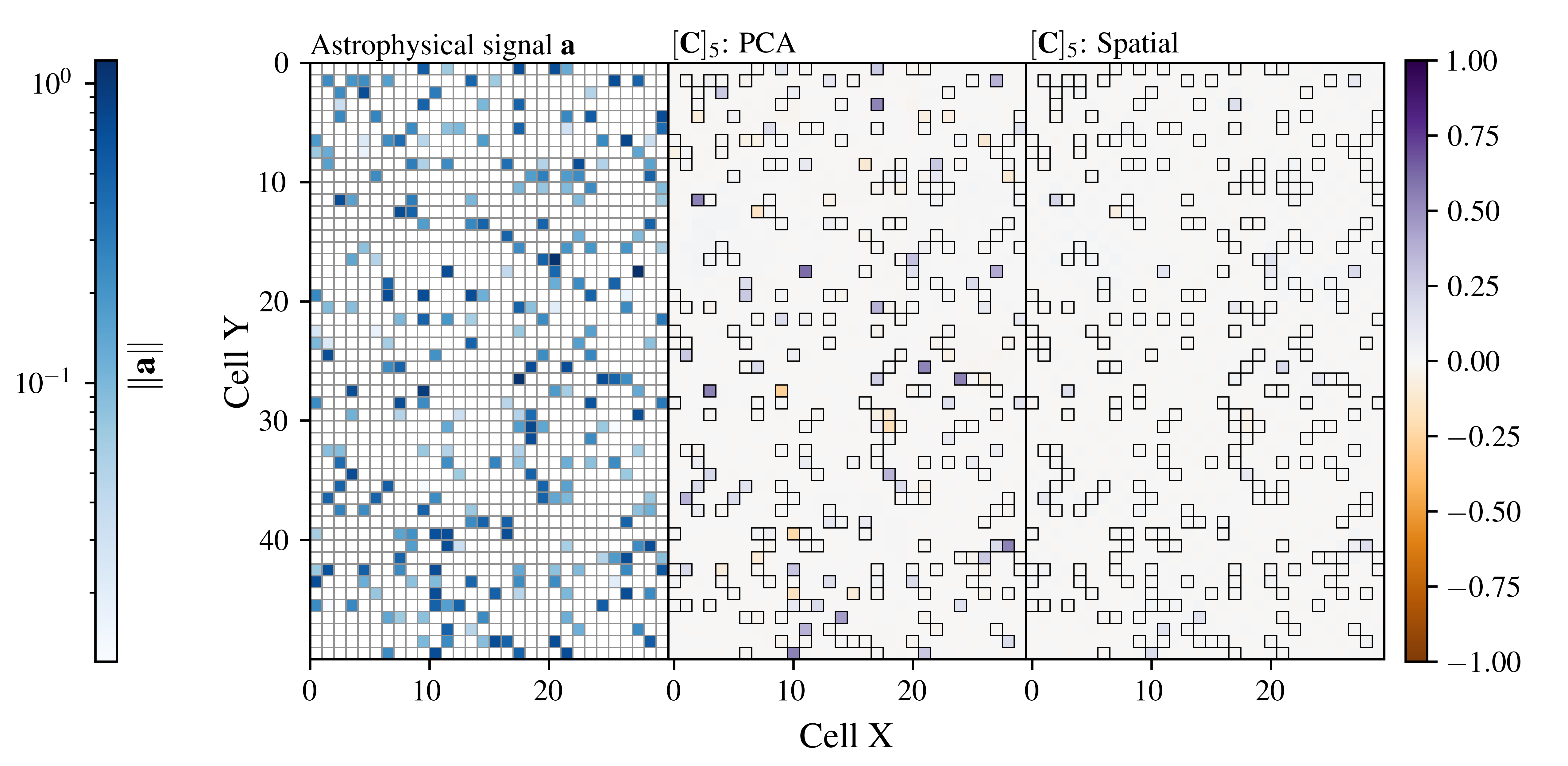}
    \caption{A plot of gridded light curve cells $i \to (x,y) \in (X,Y)$ where a simulated astrophysical signal $\bold{a}_{[m]}$ was injected for Experiment B. The x- and y-axes are in units of gridded spatial cells. The column at left depicts the magnitude of the injected astrophysical signal $\| \bold{a}_{[m]}\|$ as the color intensity of each cell on a log scale (blue color wedge at left). The middle and right columns depict coefficients $c_i^5$ for light curve $i \to (x,y) \in (X,Y)$ at each spatial cell (from Figure~\ref{fig: exp2_map}) for the reference PCA method and the spatial systematics method respectively. For these columns the magnitude of the coefficient is shown as color intensity (color wedge at right). A black outline around a cell indicates that the corresponding lightcurve contained an injected astrophysical signal. The rank 5 coefficient was chosen for clarity of interpretation as this coefficient should be closer to zero. Coefficients fitted with PCA exhibit some overfitting of injected astrophysical signals.}
    \label{fig: expb_inj_cell}
\end{figure}

\begin{figure}[htbp] 
  \centering
  \includegraphics{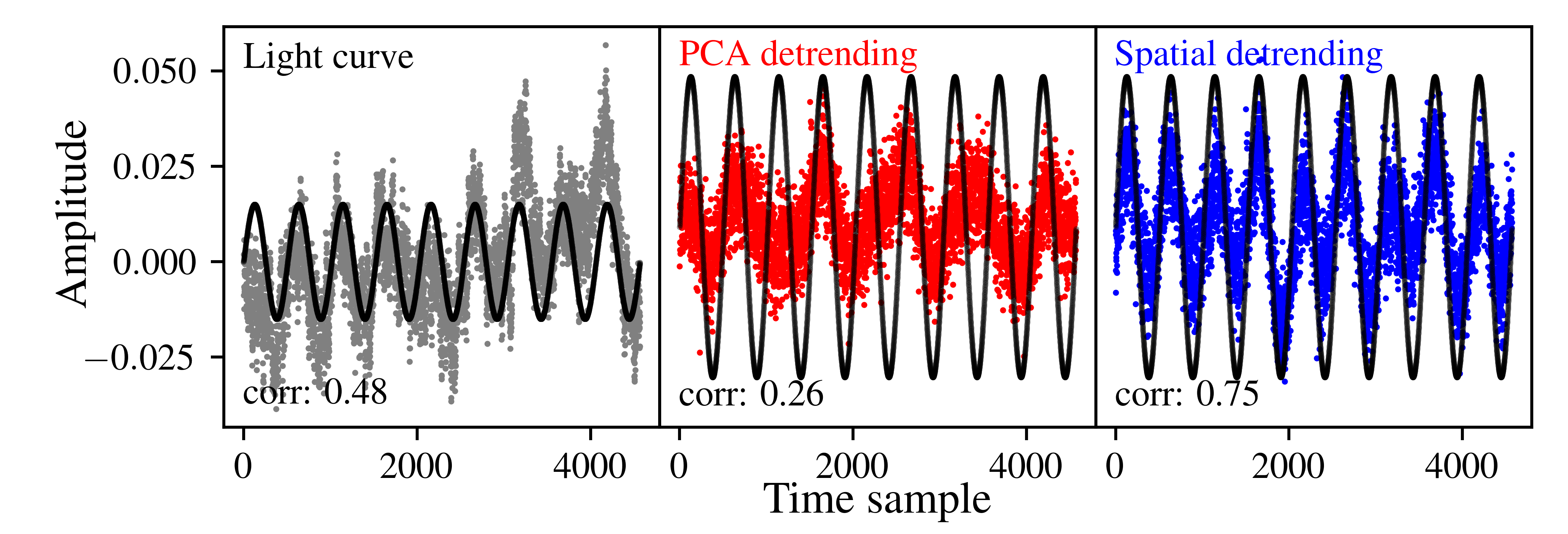}
    \caption{The simulated light curve $\bold{y}$ before detrending (left), the detrended light curve using PCA $\bold{y}'_{PCA}$ (middle), and the detrended light curve using the spatial systematics algorithm $\bold{y}'$ (right), all at the position of the the maximum value of $[\bold{C}_{PCA}]_{5}$ in Figure \ref{fig: expb_inj_cell}. Simulated injected signals are overlaid in black. The x-axis is Kepler long-cadence time sample index. This shows an instance where PCA detrending overfit the injected astrophysical sine signal relative to the spatial systematics algorithm.}
    \label{fig: expb_overfit_cell}
\end{figure}

Further evidence for the improved mitigation of overfitting by the spatial systematics algorithm over the reference PCA method is provided by Experiment B. As a full simulation, this experiment includes no unknown astrophysical variability beyond the simulated astrophysical signals. In Figure~\ref{fig: exp2_map} the true simulated coefficients shown in the top row are spatially smooth but with realistic module discontinuities. In contrast, the coefficients in the second row estimated using the PCA method show speckling in smooth regions. These discrepant coefficient values correspond to cells containing light curves with simulated astrophysical signals that are overfitted by the PCA method. The coefficients estimated by the spatial systematics method shown in the bottom row of this figure generally have reduced speckling and reflect the true coefficients with greater fidelity. Figure \ref{fig: expb_inj_cell} shows the cells where an astrophysical signal was injected into a simulated light curve, alongside fitted coefficients from PCA and the spatial systematics method. This figure supports the argument that PCA has overfit light curves containing astrophysical variability. However, we note that not all individual cells containing an injected astrophysical signals are overfit by PCA. A light curve and the associated detrending results corresponding to an overfit cell in Figure \ref{fig: expb_inj_cell} are shown in Figure \ref{fig: expb_overfit_cell}.
The spatial systematics algorithm varies both the coefficients and basis vectors (Figure \ref{fig: expb_fitted_basis}) in the iterative fit. Empirically, in Experiment B the spatial systematics algorithm estimates systematics $\hat{\bold{l}}$ with a correlation $\bar{\rho}(\bold{l}, \hat{\bold{l}})$, against true simulated systematics $\bold{l}$, that is comparable to the reference PCA method. The correlation $\bar{\rho}(\bold{l}_{\bold{a}}, \hat{\bold{l}}_{\bold{a}})$ against systematics for light curves containing a simulated astrophysical signal, and therefore a higher risk of overfitting, is slightly higher for the spatial systematics algorithm relative to the reference PCA method. This implies the spatial systematics algorithm shows reduced overfitting relative to the reference PCA method. We believe this is due to the spatial total variation constraint appropriately estimating spatially correlated systematics.

\begin{figure}[htbp]
  \centering
  \includegraphics[width=.6\linewidth]{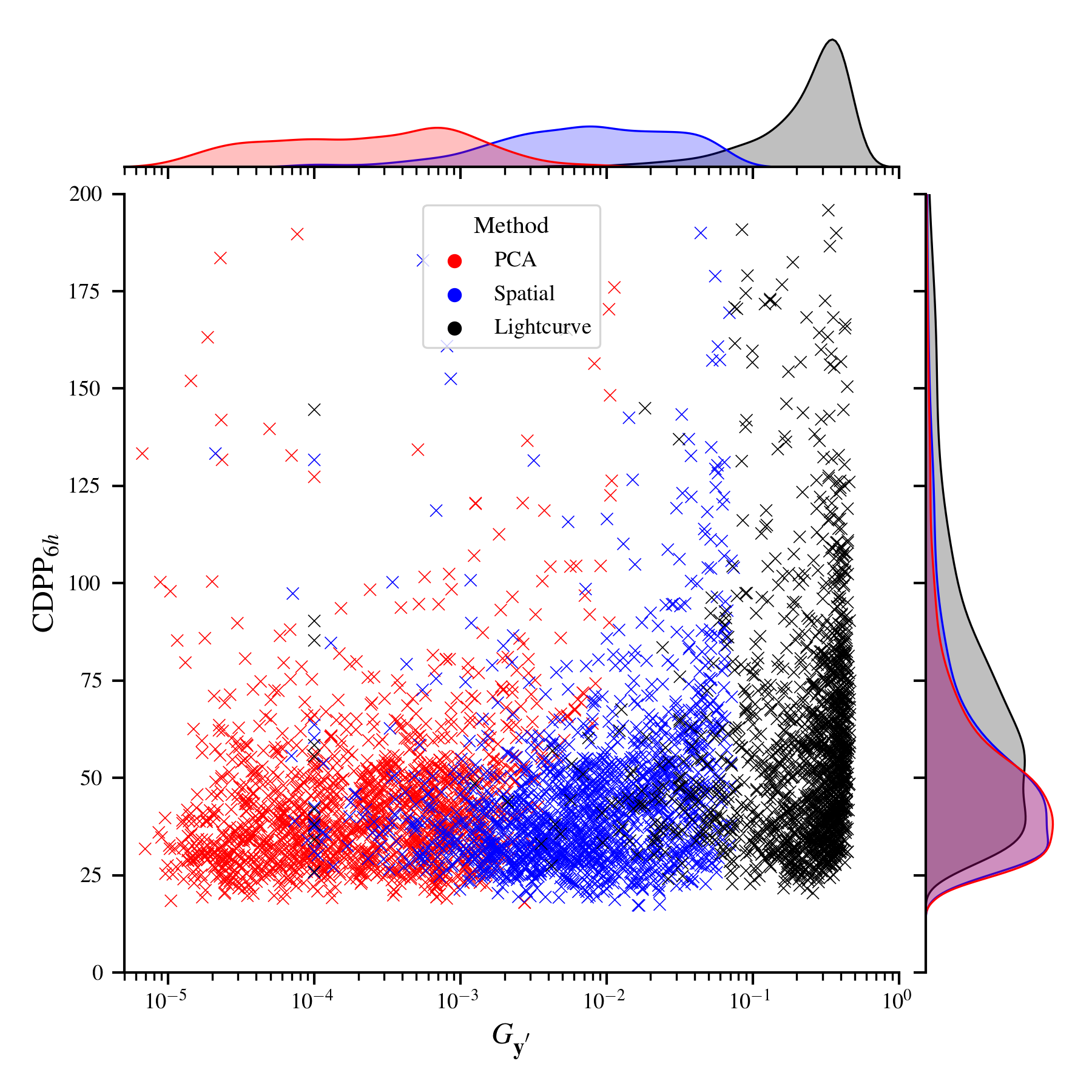}
    \caption{A plot of CDPP$_{6h}$ versus goodness metric $G_{\bold{y}'}$ per target light curve for a single run $\#$0 of Experiment A. Three light curve types are shown, including non-detrended preprocessed light curves (black), light curves detrended using PCA (red), and light curves detrended using the spatial systematics algorithm (blue). Empirical distribution functions for each light curve type are plotted over $G_{\bold{y}'}$ (top) and CDPP$_{6h}$ (right).} \label{fig: goodness_ratio} 
\end{figure}

\begin{figure}[htbp]
  \centering
  \includegraphics[width=\linewidth]{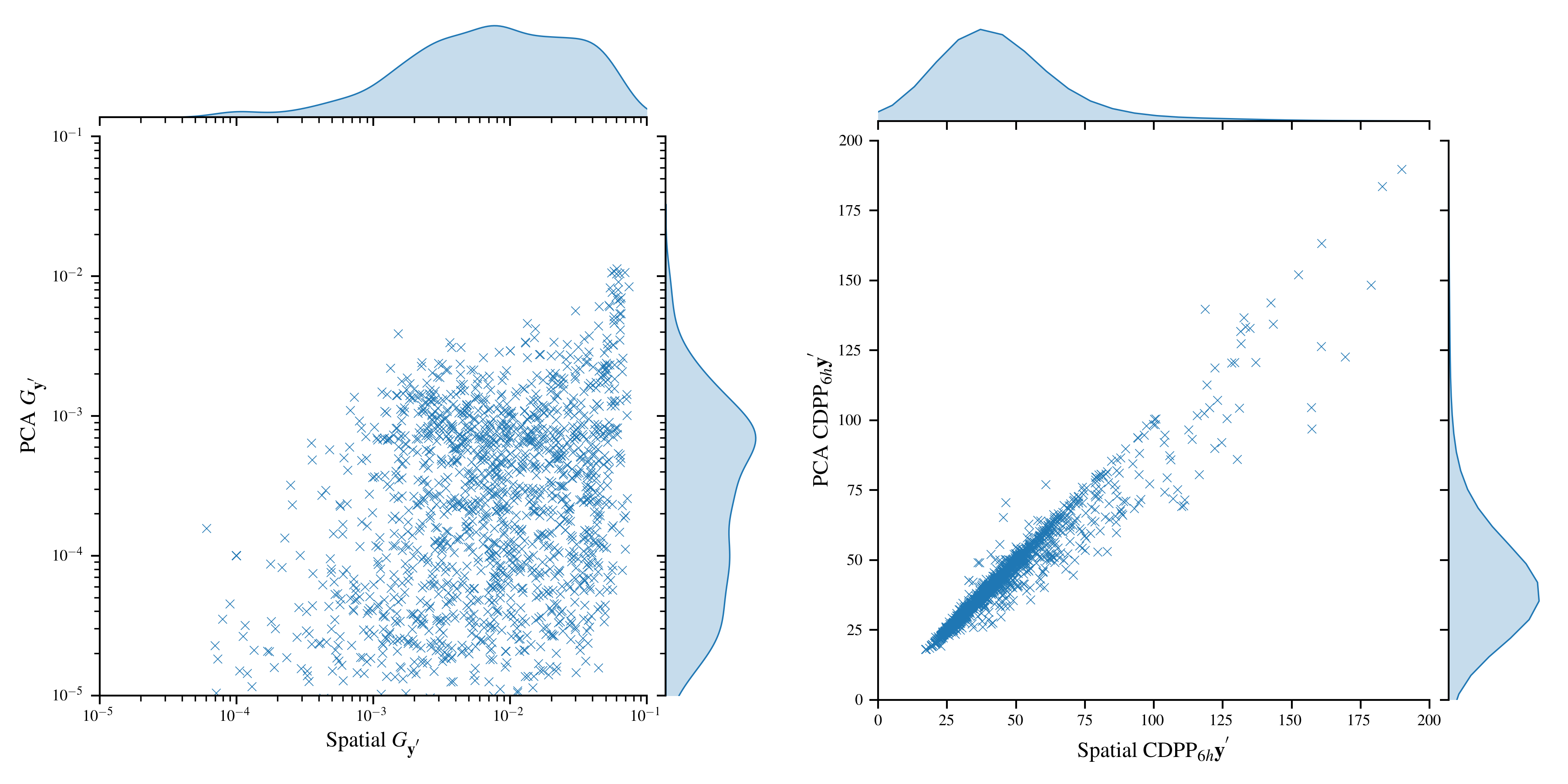}
    \caption{The metrics $G_{\bold{y}'}$ (left) and CDPP$_{6h}$ (right) for the reference PCA method versus the spatial systematics method. Both plots are per target light curve for a single run $\#$0 of Experiment A. Empirical distribution functions for each light curve metric are plotted over the spatial systematics method (horizontal axis) and the PCA method (vertical axis). Although correlated, there is significant scatter in the left plot due to the intrinsic nonlinear and pair-wise nature of the metric $G_{\bold{y}'}$. It is accordingly more sensitive to light curves with poorly-modeled systematics or greater retained astrophysical variability.}\label{fig: expa_gof_cdpp_vs} 

\end{figure}

As a definitional consequence of the reduction in overfitting, the detrended light curves obtained using the spatial systematics algorithm will retain a higher degree of true astrophysical variability relative to those obtained using the reference PCA method in addition to a component for uncorrected systematics, in keeping with all detrending algorithms. We note that PCA is guaranteed to maximally flatten a collection of light curves as the PCA solution is equivalent to a minimal least-squares residual between fitted systematics and light curves (Equation~\ref{eq: pca_lsq}). PCA implicitly assumes a white noise model, and is susceptible to overfitting of astrophysical signals \citep{rob_pca} because astrophysical signals are not appropriately modeled as white noise \citet{pont}. In both experiments, the two algorithms obtained low goodness metric values $G_{\bold{y}'} < 2 \%$ indicating largely successful systematics removal in the detrended light curves. The residual systematics metrics $G_{\bold{y}'}$ and CDPP$_{6h}$ are plotted for run $\#$0 of Experiment A per light curve and over detrending algorithm in Figure~\ref{fig: goodness_ratio} and in Figure~\ref{fig: expa_gof_cdpp_vs}. Figure~\ref{fig: goodness_ratio} shows that the distribution of the goodness metric $G_{\bold{y}'}$ is broadly larger than that for PCA. This metric (Section~\ref{sec: noise_metrics}) is nonlinear and pair-wise, and therefore sensitive to those light curves with poorly-modeled systematics (Figure~\ref{fig: expa_gof_cdpp_vs}). For example, individual light curves may be poorly modeled by the spatial systematics algorithm if their systematics exhibit strong temporal variation insufficiently captured by the fixed spatial model. It is also possible that the spatial systematics algorithm retains a greater degree of astrophysical variability that biases $G_{\bold{y}'}$ through incidental correlation. We lack sufficient evidence to state either claim strongly here. Future generalizations of the constraint terms are described below as potential future work.

In Experiment B, the goodness metric value $G_{\bold{y}'}$ was lower for the spatial method than the reference PCA method. For both experiments (Table~\ref{tab: overfit} and~\ref{tab: exp2}) the two algorithms obtained very similar CDPP$_{6h}$ metrics and a reduction from the non-detrended preprocessed light curves, indicating that both methods were equivalently successful at suppressing transit time-scale systematics.

In Experiment A the measures of residual systematics $G_{y'}$ and CDPP$_{6h}$ are higher for the spatial systematics method compared to the PCA method. 
As noted above, these values for the detrended light curves may be elevated primarily due to improved preservation of true astrophysical variability or alternatively due to unmodeled residual systematics. In Figure \ref{fig: sim_shock} the example detrended light curve from Experiment A obtained by the spatial systematics algorithm in the first column shows an example of large residual systematic scatter not present in the PCA detrended light curve; however the overall variability is dominated in this case by the injected astrophysical signal. We argue, that although the spatial method may retain a greater level of residual systematics, comparably the major component of retained light curve variability is astrophysical. As discussed above, Experiment B demonstrated a more accurate estimation both of systematics $\hat{\bold{l}}$ and of astrophysical signals $\bold{a}$ in light curves containing astrophysical signals. Experiment B, although simulated with a simplified data model, therefore aids the interpretation of metric performance. We note also that the improved preservation of true astrophysical variability in the light curves detrended by the spatial systematics algorithm may reduce the correlations $\bar{\rho}(\bold{a}_{[m]}, \bold{y}_{[m]}'),\ m \in \{s,f,t\}$ in Tables~\ref{tab: overfit} and~\ref{tab: exp2} discussed above and that these may be underestimated as a result.

\begin{figure}[htbp] 
  \centering
  \includegraphics{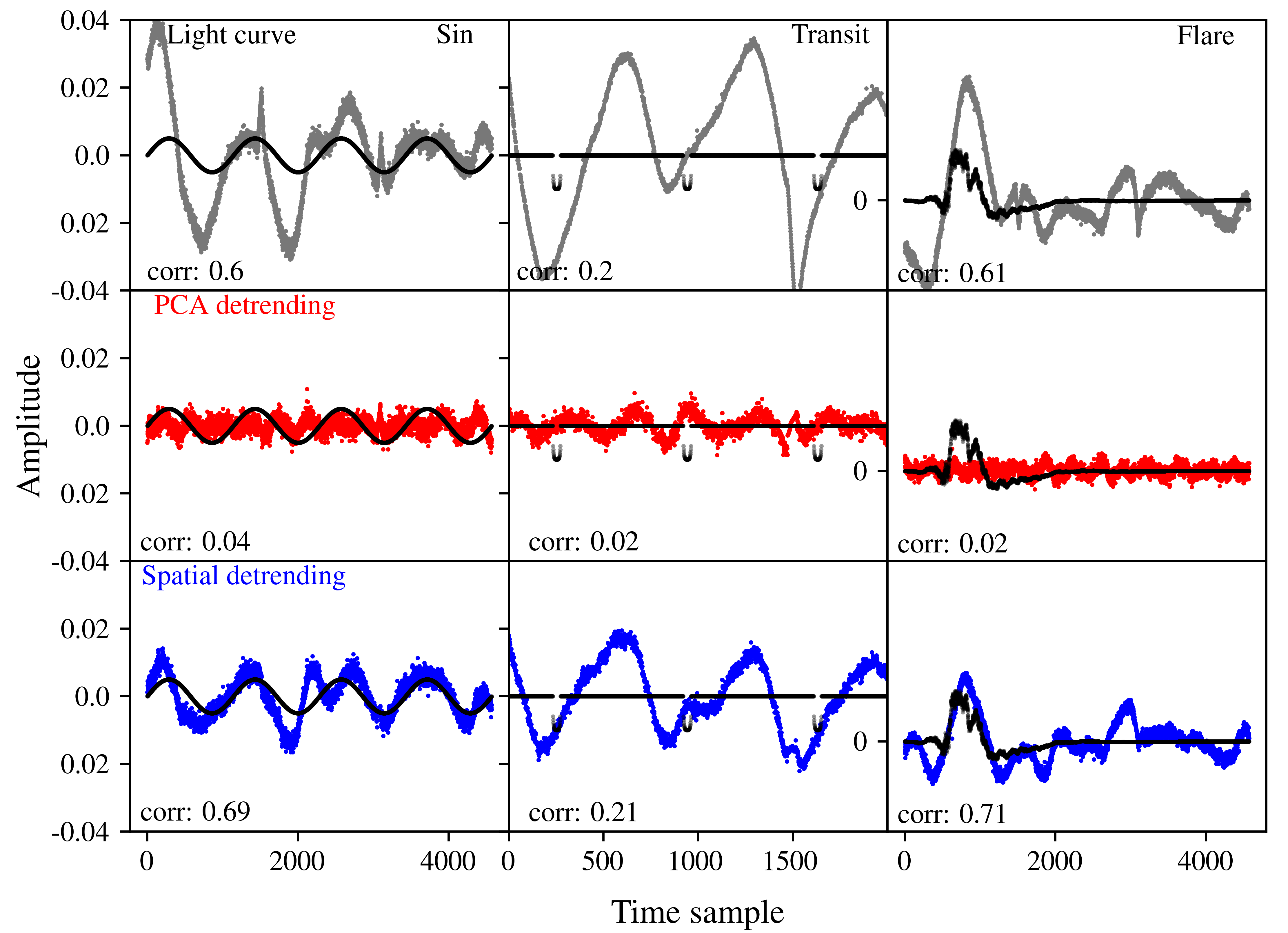}
    \caption{The non-injected light curves $\bold{y}_{{\bold{a}^c}}$ for which the spatial systematics algorithm produces the highest incidental correlation $\max_{\bold{y}_{{\bold{a}^c}}} \rho(\bold{a}_{[m]}, \bold{y}_{{\bold{a}^c}})$, $m \in \{s,t,f\}$ between the non-detrended light curve $\bold{y}_{{\bold{a}^c}}$ and any injected astrophysical signal $\bold{a}_{[m]}$ used in the simulation run. The top row shows the non-detrended light curve  and the astrophysical signal (injected into another light curve) with which the incidental correlation is highest. The detrended light curves $\bold{y}'_{ {\bold{a}^c}}$ are shown as obtained by the reference PCA method (middle row) and the spatial systematics algorithm (bottom row).Simulated injected signals are overlaid in black. In these worst cases it can be seen that the high level of correlation of the spatial detrended lightcurve $\bold{y}'$ with an astrophysical signal $\bold{a}$ is incidental, as the original lightcurve $\bold{y}$ itself is incidentally correlated with $\bold{a}$. This suggests that high values for the metric of corruption $\text{\textbar}\bar{\rho}(\bold{a},\bold{y}'_{ {\bold{a}^c}}) \text{\textbar} / \text{\textbar}\bar{\rho}(\bold{a}, \bold{y}_{ \bold{a}^c})\text{\textbar}$ may result incidentally for the spatial systematics algorithm because it retains a higher level of astrophysical variability.}
    \label{fig: most_corr}
\end{figure}

In Experiment A, both the spatial systematics algorithm and the reference PCA method produce values $\text{\textbar}\bar{\rho}(\bold{a}_{[m]},\bold{y}'_{ {\bold{a}^c}}) \text{\textbar} / \text{\textbar}\bar{\rho}(\bold{a}_{[m]}, \bold{y}_{\bold{a}^c})\text{\textbar} < 1$, for $m \in \{ s, t, f\}$. Therefore, as discussed in Section~\ref{sec: noise_metrics}, this suggests neither algorithm has increased the average incidental correlation between the simulated astrophysical signals and the light curves in which no astrophysical signals were injected (the non-injected subsample). Equivalently, neither method has introduced spurious astrophysical content into this non-injected subsample of light curves. However, the ratio $\text{\textbar}\bar{\rho}(\bold{a}_{[m]},\bold{y}'_{ {\bold{a}^c}}) \text{\textbar} / \text{\textbar}\bar{\rho}(\bold{a}_{[m]}, \bold{y}_{\bold{a}^c})\text{\textbar}$ is higher for the spatial systematics algorithm compared to the PCA method (Tables~\ref{tab: overfit} and~\ref{tab: exp2}) particularly for sine and flare $m \in \{s,f\}$ astrophysical signals, which have longer duration. In Figure \ref{fig: most_corr} we plot the light curves in the non-injected subsample for which the spatial systematics algorithm produces the highest incidental correlation $\max_{\bold{y}_{{\bold{a}^c}}} \text{\textbar} \rho(\bold{a}_{[m]}, \bold{y}_{{\bold{a}^c}})\text{\textbar}$, $m \in \{s,t,f\}$ between the non-injected spatial detrended light curve and any injected astrophysical signal used in the simulation run. This figure shows the worst cases of incidental correlation; in general the incidental correlation average is low. As described above, the spatial systematics algorithm is more successful at preserving astrophysical variability in detrended light curves and will therefore retain rare incidentally correlated signals, particularly if their timescale matches the simulated astrophysical signal. We believe this explains the higher values of this ratio the case of sine and flare signal types.

Section~\ref{sec: compare_results} describes the results of Experiment C, a high-level comparison between the spatial systematics algorithm and the standard detrending methods CBV, SFF, and PLD \citep{lightkurve}. As noted above, the default recommended parameters were used for each method as opposed to a customized optimization. In addition, individual methods apply different default pre-processing steps. Accordingly, this high-level comparison cannot address the optimality of different detrending methods.

In this experiment the CDPP$_{6h}$ residual systematics metric, which is computed per light curve, shows broadly comparable ($\leq 8$ ppm between mean values) detrending performance between the spatial systematics algorithm and the reference detrending methods CBV, SFF, and PLD (Table~\ref{tab: compare_table}; Figure~\ref{fig: compare_ratio}).The spatial systematics method has a slightly increased mean CDPP$_{6h}$ compared to other methods possibly due to instances of residual scatter where the spatial model imperfectly corrects strong temporal systematic variation. The $G_{\bold{y}'}$ residual systematics metric differs in distribution across detrending method (Figure~\ref{fig: compare_ratio}). By definition (Section~\ref{sec: noise_metrics}) and as discussed earlier, this metric is nonlinear and pair-wise; accordingly, light curves with unmodeled systematics are more heavily weighted. The PLD method has the highest mean value of $G_{\bold{y}'}$ (Figure~\ref{fig: compare_ratio}) but this method retains long-term variability, which is fit by a spline term in the data model, and therefore is expected to have an inherently higher $G_{\bold{y}'}$ distribution. In contrast, in pre-processing, SFF light curves are flattened using a Savitsky-Golay filter \citep{lightkurve}, and SFF is therefore broadly comparable to the CBV method in this metric. The spatial systematics algorithm has a higher mean value of $G_{\bold{y}'}$ and broader distribution than the CBV and SFF methods (Figure~\ref{fig: compare_ratio}). As discussed earlier for Experiments A and B, this may be due to the spatial systematics method retaining a higher level of astrophysical variability or due to the nonlinear contribution to $G_{\bold{y}'}$ of light curves that do not fit the spatial systematics data model assumptions. Generalizations to make the constraint terms more complete are discussed below as potential future work.

Our current work has several limitations that would benefit from exploration in future work. As described in Section~\ref{sec: init_val}, we chose tunable algorithm parameters, including model rank, convergence criteria, optimization step size, weighting matrix, and initial coefficient values, amongst other parameters, based on feasibility and empirical evaluation alone. It would be beneficial to sample this algorithm parameter space over a broader range, for completeness. In conjunction, our numerical evaluations relied on an associated set of injected astrophysical signal types (Section~\ref{sec: sim_sig}), which could be expanded. Generalizing the sensor spatial discretization to non-uniform layouts is left for future work. It has been noted that light curves at small spatial separations may suffer from blending of astrophysical signals \citep{Kov_cs_2005, Hattori_2022} requiring future consideration when using the spatial systematics algorithm on densely-populated fields. Although some light curves in our data sample have small spatial separation, this is not a major concern in the current work, as the cell separation on the spatial grid used significantly exceeds the PSF width. The sparse population of the field is likely due to the narrow (and bright) magnitude range selected. There is scope to generalize the constraint term to incorporate non-neighbouring pixel spatial correlations or weightings of the spatial prior to exclude targets within a pixel separation where astrophysical blending could occur. In addition, the constraint terms may be generalized to include the known magnitude correlation \citep{stumpe2012kepler, smith2012kepler} and other parametric dependencies \citep{moreno}. As noted earlier, the Kepler test data were selected in a narrow magnitude range to isolate the effect of magnitude dependence in the current work.

\section{CONCLUSIONS}\label{sec: conclusions}

In this paper we present an exploratory algorithm for detrending light curves obtained in wide-field exoplanet transit surveys. This spatial systematics algorithm fits a low-rank linear systematics model to a collection of light curves while also including a total variation spatial constraint across the sensor at a foundational level. The resulting objective function is reduced using variable elimination \citep{Golub2007TheDO} which also stabilizes the solution \citep{Golub_2003, Shearer_2013}. An approximate closed-form gradient was developed for minimization by gradient descent; this particular formulation is a modification of total variation optimization approaches \citep{vogel_book}. The spatial systematics algorithm was numerically evaluated relative to a reference PCA method using both injection tests with Kepler data (Experiment A) and full simulations including simulated systematics, astrophysical signals, and statistical noise within the same Kepler coordinate framework (Experiment B).

The principal conclusions of the paper are:
\begin{enumerate}
    \item The spatial systematics algorithm showed reduced overfitting of instrinsic astrophysical variability relative to the PCA method in the two numerical evaluation studies. The reduced overfitting was demonstrated by comparable or significantly higher correlation between the known astrophysical signals and the detrended light curves in both experiments. In Experiment B the known simulated systematics were comparably or more accurately estimated relative to the PCA method, particularly for light curves containing injected signals. We argue that the reduced overfitting likely arises from the physically-realistic total variation constraint. In addition, the algorithm simulataneously varies both the basis vectors and their coefficient weights. Both factors likely contribute to the improved separation of systematics and astrophysical signals.

    \item A marked reduction in overfitting relative to PCA was found for slowly-varying astrophysical signals while comparable performance was achieved for shorter exoplanet transit signals. We argue that the longer-duration astrophysical signals overlap more significantly with the timescale of residual systematics in the Kepler data and therefore allow the algorithm to achieve a clearer separation of systematics for these signals. 
    \item In terms of estimates of residual systematics in the detrended light curves, both methods achieved comparable CDPP$_{6h}$ values in Experiment A suggesting largely equivalent success in suppressing transit timescale systematic noise. However, the goodness metric $G_{\bold{y}'}$ was comparatively larger in this experiment for the spatial systematics algorithm relative to PCA indicative of residual systematics. 
    \item Neither method increased the incidental correlation between the simulated astrophysical signals and non-injected subsample of light curves in the numerical evaluations. Equivalently, neither method added spurious astrophysical variability into the detrended light curves.

\end{enumerate}

\begin{acknowledgments}

This paper includes data collected by the Kepler mission. Funding for the Kepler mission is provided by the NASA Science Mission Directorate. STScI is operated by the Association of Universities for Research in Astronomy, Inc., under NASA contract NAS 5–26555.
All of the data presented in this paper were obtained from the Mikulski Archive for Space Telescopes (MAST) at the Space Telescope Science Institute. The specific observations analyzed can be accessed via \dataset[DOI 10.17909/T9388B]{https://archive.stsci.edu/missions/kepler/lightcurves/tarfiles/DOI_LINKS/Q6_LC/}, \dataset[DOI 10.17909/T9K88P]{https://archive.stsci.edu/missions/kepler/lightcurves/tarfiles/DOI_LINKS/Q10_LC/}, and \dataset[DOI 10.17909/T92C7P]{https://archive.stsci.edu/missions/kepler/lightcurves/tarfiles/DOI_LINKS/Q14_LC/}. We thank the anonymous referee of the paper for their suggested revisions; these significantly improved the paper.

\software{This work used the following external packages: Astropy \citep{astropy:2013, astropy:2018, astropy:2022}, Scikit-learn \citep{scikit-learn}, Matplotlib \citep{Hunter:2007}, Seaborn \citep{Waskom2021}, NumPy \citep{harris2020array}, SciPy \citep{2020SciPy}, and Lightkurve \citep{lightkurve}. The algorithm described in this paper is available as a public Python package on the Github repository\footnote{\url{github.com/xiaziyna/spatial-detrend}} or via PyPi under package name {\it spatial-detrend}\ \footnote{\url{https://pypi.org/project/spatial-detrend}}. }
\end{acknowledgments}

\appendix
\section{List of symbols} \label{ap: symbols}

A list of important symbols and mathematical notation used in the paper is provided in Table~\ref{tab:symbols}.
\begin{table}[htbp!]
\caption{List of symbols}
\begin{center}
\begin{tabular}{r l p{10cm} }
\toprule
$[\bold{M}]_{i, j}$ or $M_{i,j}$ & Element at $i$-th row and $j$-th column of matrix $\bold{M}$ \\
$\mathds{1}_N$ & Identity matrix of size $N \times N$ \\
$\bold{M}^\dagger$ & Pseudo-inverse matrix operator\\
$\bold{M}^T$ & Matrix transpose operator\\
$\lVert \bold{x} \rVert_p$ & $L_p$ norm of vector $\bold{x}$ \\
$\lVert \bold{M} \rVert_{p,q}$ & Combined $L_{p,q}$ norm for matrix $\bold{M}$ \\
$\lVert \bold{M} \rVert_F$ & Frobenius norm of matrix $\bold{M}$ \\
$i \in I$ & Target light curve index set\\
$i \to (x, y) \in X, Y;\ X, Y \in \mathbb{Z}^+$ & Gridded target position on sensor \\
$\bold{y}_i$ & Vector light curve for target index $i$ \\
$\bold{n}_i$ & Vector statistical noise term for light curve target index $i$\\
$\bold{l}_i$ & Vector systematics term for light curve target index $i$\\
$\bold{Y}$ & Matrix of light curves; for vectors $\mathbf{y}$  \\
$\bold{Y}'$ & Matrix of detrended lightcurves; for vectors $\mathbf{y}'$ \\
$\bold{N}$ & Matrix of statistical noise \\
$\bold{a}_{[m]} \; : \; m \in \{s, t, f\}$ & Simulated astrophysical signal of type: $s$ sine, $t$ transit, $f$ flare \\
$\bold{A}$ & Matrix of simulated astrophysical signals \\
$K$ & Rank of systematic noise model \\
$\bold{v}_k$ & Basis vector $k \in K$ for systematic noise model \\
$\bold{L} = \bold{V} \bold{C}$ & Matrix of low-rank systematic noise ($\bold{V}$ basis vectors, $\bold{C}$ coefficient matrix) \\
$\bold{\bar{C}}$ & Column-normalized coefficient matrix \\
$c_i^k$ & Coefficient weighting of $\bold{v}_k$ for light curve $i \in I$ \\
$\bold{c}_i = [c_i^1, \dots, c_i^K]^T$ & Coefficient vector for light curve $i$ \\
$\bold{W}$ & Weight matrix\\
$\bold{D}$ & Difference operator ($\bold{D}_{\bold{W}}$ difference operator with weights $\bold{W}$) \\
$\bold{\alpha}^t$ & Step size at iteration $t$\\
$f(\mathbf{V},\mathbf{C})$ & Least-square residual between $\mathbf{Y}$ and $\mathbf{VC}$\\
$h(\mathbf{C})$ & For fixed $\mathbf{C}$, minimizing value $\mathbf{V}$ of $f(\mathbf{V},\mathbf{C})$ \\
$g(\mathbf{C})$ & Total variation penalty constraint\\
$w(.)$ & Variable-reduced objective function\\
$\nabla f'(.)$ & Gradient of $f(h(\mathbf{C}),\mathbf{C})$\\
$\nabla g(.)$ & Gradient of $f(\mathbf{C})$\\
$\mathbf{P}_{R(\mathbf{C})^T}$ & Projection onto the range space of $\mathbf{C}^T$\\
$\rho(\bold{y}_i, \bold{y}_j)$ & Correlation between vectors $\bold{y}_i$ and $\bold{y}_j$ \\
$\bar{\rho}(\bold{a}, \bold{y}'_{\bold{a}})$ & Averaged correlation of astrophysical signals $\bold{a}$ and detrended lightcurves with an injected signal $\bold{y}'_{\bold{a}}$\\
$ \{ \bold{y}_{\bold{a}^c}$ \} & Set of lightcurves excluding those with a simulated astrophysical signal \\
$G_{y'}$ & Residual detrended systematics: goodness metric  \\
CDPP$_{6h}$ & Residual detrended systematics: 6 hr combined differential photometric precision\\
\bottomrule
\end{tabular}
\end{center}
\label{tab:symbols}
\end{table}

\section{Principal Component Analysis}\label{ap: PCA}
Singular Value Decomposition (SVD) and Principal Component Analysis (PCA) are summarized here, as foundations of the cotrending method. A complete reference for SVD and PCA is provided by \citet[chap.~14.5]{Hastie2009} and \citet{golub2013matrix}. Every matrix $\bold{Y} \in \mathbb{R}^{n \times m}$ admits a SVD: $\bold{Y} = \bold{V}\bold{\Sigma}\bold{U^T}$ with unitary matrices $\bold{V} \in \mathbb{R}^{n \times n} \; : \; \bold{V}^T \bold{V} = \mathds{1}_n$ and  $\bold{U} \in \mathbb{R}^{m \times m} \; : \; \bold{U}^T \bold{U} = \mathds{1}_m$ and a diagonal matrix $\bold{\Sigma} \in \mathbb{R}^{n \times m}$ of ordered non-negative values. The columns of $\bold{V}$ and $\bold{U}$ are the left and right singular vectors respectively, the values $\bold{\Sigma} = \diag(\sigma_1, ... \sigma_{\min(m, n)})$ are the singular values, and $\mathds{1}$ denotes an identity matrix. The number of non-zero singular values determines the rank $r$ of $\bold{Y}$. The SVD may be equivalently defined using only non-zero singular values, in which case $\bold{V} \in \mathbb{R}^{n \times r} \; : \; \bold{V}^T \bold{V} = \mathds{1}_r$, $\bold{U} \in \mathbb{R}^{m \times r} \; : \; \bold{U}^T \bold{U} = \mathds{1}_r$ and $\bold{\Sigma} = \diag(\sigma_1, ... \sigma_r)$. This is termed compact SVD and is the form used in this work. There are a number of efficient algorithms to compute a SVD \citep{golub2013matrix} making it extremely practical for analysis. \\

 Consider the data model $\bold{Y}=\bold{L}+\bold{N}$, defined in Section~\ref{sec:Model}, where $\bold{Y} \in  \mathbb{R}^{N \times I}$, $\bold{L} \in \mathbb{R}^{N \times I}$, and $\bold{N} \in \mathbb{R}^{N \times I}$ represent the light curves, systematic noise, and statistical noise respectively. The Eckart–Young–Mirsky theorem \citep{eckart} states that the optimal rank-$K$ matrix approximation of $\bold{Y}$ which minimizes the least-squares residual $||\bold{Y}-\bold{L}||_F^2$ (subsctipt $F$ denoting the  Frobenius norm \citep{golub2013matrix}) is obtained by retaining only the top $K$ singular values and associated singular vectors, of the SVD of $\bold{Y}$; this is known as rank thresholding:

\begin{align} \label{eq: SVD}
    \argmin_{\bold{L}: \rank(\bold{L}) \leq K} ||\bold{Y}-\bold{L}||_F^2 = \bold{V}_K \bold{\Sigma}_K \bold{U}_K^T
\end{align}
with $\bold{\Sigma} = \diag(\sigma_1, ..., \sigma_K)$, $\bold{V}_K$ and $\bold{U}_K$ denote the associated leading $K$ left and right singular vectors respectively. PCA is also the solution to the objective function minimization in Equation \ref{eq: SVD} with  form $\bold{L}=\bold{V}_K \bold{C}_K$, $\bold{V}_K \in  \mathbb{R}^{n \times K},\ \bold{C}_K \in \mathbb{R}^{K \times m}$, and $\bold{V}_K^T\bold{V}_K = \mathds{1}_K$. The PCA solution can be obtained from the compact SVD solution, using $\bold{V}_K$ and setting $\bold{C}_K = \bold{\Sigma}_K \bold{U}_K^T$. Any $\bold{V}_K' = \bold{V}_K \bold{\bar{U}}^T$ and $\bold{C}_K' = \bold{\bar{U}} \bold{C}_K$ is also an equivalent low-rank solution $\bold{L} = \bold{V}_K'\bold{C}_K' = \bold{V}_K\bold{C}_K$, where $\bold{\bar{U}} \in \mathbb{R}^{K \times K}$ is any unitary matrix such that $\bold{\bar{U}}^T \bold{\bar{U}} = \bold{\bar{U}} \bold{\bar{U}}^T  =  \bold{\bar{U}} \bold{\bar{U}}^{-1} = \mathds{1}_K$. 

\section{Total Variation Measure} \label{ap: tv}

An overview of total variation is provided in the monograph by \cite{vogel_book} and in \citet[Chapter~3.6]{karl2005regularization}. In the continuous limit, a Lebesque-integrable \citep{edwards1994} function $f(x,y) \in L^1$ with vector gradient $\nabla f = \left(\frac{\partial f}{\partial x}, \frac{\partial f}{\partial y}\right)$ has total variation:
\begin{align}
    TV(f) = \int \int \|\nabla f \|_2\ dx dy 
\end{align}
where $\lVert . \rVert_2$ denotes an $L_2$ vector norm. Functions of bounded total variation have  $TV(f) \leq \infty$. The total variation measure $TV(f)$ can be used as a regularization or penalty constraint on $f$ to enforce spatial uniformity while preserving limited function discontinuities \citep{vogel_book}.\\

A discretized approximation of the total variation measure $TV_d(f)$ for $f$ sampled at points $X \times Y$ is given by:

\begin{align} \label{eq: sum_tv}
    TV_{d}(f) = \sum_{i \in X} \sum_{j \in Y} ||(D_x f)_{i,j} , (D_y f) _{i,j}||_2
\end{align} 
where $(D_x f)_{i,j} = f_{i,j} - f_{i-1,j}$ and $(D_y f) _{i,j} = f_{i,j} - f_{i,j-1}$ are unweighted function differences. If matrix $\bold{D}\bold{f}$ is constructed with $\mvec(D_x f)^T$ and $\mvec(D_y f)^T$ as the first and second rows respectively (where $\mvec(.)$ denotes vectorization \citep{golub2013matrix}), then: 
\begin{align}
    TV_{d}(f) = ||\bold{D}\bold{f}||_{2,1}^1
\end{align}

This may be generalized to: $||\bold{D}\bold{f}||_{2, p}^p : p \in [1,2]$, where $p$ controls the degree of spatial uniformity. 
\subsection{Total Variation Measure for $p=2$:} \label{ap: corr_same}
We show here that minimizing the generalized total variation measure $TV_d(f)=||\bold{D}\bold{f}||_{2, p}^p$ for the case $p=2$ is equivalent to maximizing the correlation $\rho_{ij}$ between neighboring normalized light curve coefficients $\bar{\bold{c}}_i$ and $\bar{\bold{c}}_j$ $(i \in I, j \in I, i,j \in n(I))$, where $n(I)$ is the set of neighboring light curve tuples. The maximum neighboring coefficient correlation $\rho_n$ takes the form:

\begin{align}
    \rho_n=\argmax_{\bold{V}, \bold{C}\; : \; \rank(\bold{V}\bold{C}) \leq K} \sum_{(i,j) \in n(I)} \bar{\bold{c}}_{i}^T \bar{\bold{c}}_{j} 
\end{align}
Since $|| \bar{\bold{c}}_j ||_2 = || \bar{\bold{c}}_i ||_2 = 1$ and $|| \bar{\bold{c}}_j - \bar{\bold{c}}_i ||_2^2 = || \bar{\bold{c}}_j ||_2^2 + || \bar{\bold{c}}_i ||_2^2 - 2 \bar{\bold{c}}_j^T \bar{\bold{c}}_i = 2 -  2 \bar{\bold{c}}_j^T \bar{\bold{c}}_i$, an equivalent formulation is given by:
\begin{align} \label{eq: sumterms}
    \rho_n = \argmin_{\bold{V}, \bold{C}\; : \; \rank(\bold{V}\bold{C}) \leq K} \sum_{(i,j) \in n(I)}|| \bar{\bold{c}}_i - \bar{\bold{c}}_{j} ||_2^2 
\end{align}
The summation can be replaced by a linear difference operator $\bold{D}$ (here unweighted for simplicity) acting on columns of $\bold{\bar{C}}$. This difference operator was introduced above in the definition of the generalized total variation measure. Therefore:
\begin{align} \label{eq: l2pen}
    \rho_n = \argmin_{\bold{V}, \bold{C}\; : \; \rank(\bold{V}\bold{C}) \leq K}|| \bold{D} \bold{\bar{C}} ||_F^2 = \argmin_{\bold{V}, \bold{C}\; : \; \rank(\bold{V}\bold{C}) \leq K} TV_d(\bold{\bar{C}})
\end{align}

We note that even if the minimum is not achieved either objective should produce an equivalent result.

\section{Probabilistic view of the objective} \label{ap: prob}
Although the spatial systematics algorithm was not derived within a Bayesian framework, a probabilistic interpretation of the objective function can be shown. An overview of Bayesian estimation and low rank models is provided in the monograph by \citet{murphy2020machine}. We adopt the nomenclature of Section~\ref{sec: sys_model} here. In Bayesian MAP estimation the posterior probability is maximized: 
\begin{align}
     \argmax_{\bold{C}} p(\bold{C} | \bold{Y}) \propto \argmax_{\bold{C}} p(\bold{Y}|\bold{C}) p(\bold{C}) \\
\end{align}
The signal model in Equation \ref{eq: decompose} is $\bold{Y} = \bold{V}\bold{C} + \bold{N}$, where $\bold{N}$ is white Gaussian noise, and 
therefore $p(\bold{Y}|\bold{C}, \bold{V}) \sim \mathcal{N} (\bold{V}\bold{C}, \mathds{1})$. The conditional basis vectors are $\bold{V}_\bold{|C}  = \argmin_{\bold{V}} p(\bold{Y}, \bold{V}|\bold{C})$ and $p(\bold{Y}| \bold{C}) \sim \mathcal{N} (\bold{V}_{|\bold{C}}\bold{C}, \mathds{1})$. Therefore $\ln(p(\bold{Y}|\bold{C})) = \|\bold{Y} - \bold{V}_{|\bold{C}}\bold{C} \|_F^2$.\\
The total variation prior implies that $\ln (p(\bold{C})) \propto \| \bold{D}_W\bold{\bar{C}}\|_{2, p}^p$, whereby $p(\bold{C}) \propto \exp{\|\bold{D}_{\bold{W}} \bold{\bar{C}} \|_{2,p}^p}$. For $p=2$ this is equivalent to a Gaussian prior on the product $\bold{D}_{\bold{W}} \bold{\bar{C}}$. 

\section{Relation between Coefficient and Systematics Correlations} 
\label{ap: corr_distortion}

We consider here the relation between the correlation  $\bold{\bar{c}}_i^T \bold{\bar{c}}_j$ of the normalized coefficients $\bold{\bar{c}}_i \in \mathbb{R}^K$ and the correlation $\bold{\bar{l}}_i^T\bold{\bar{l}}_j$ of the normalized systematics $\bar{\bold{l}}_i = \frac{\bold{V}\bold{c}_i}{|| \bold{V}\bold{c}_i ||_2^2}\; : \; \bold{\bar{l}}_i \in \mathbb{R}^N$.

The maximum deviation of $\bar{\bold{l}}_i^T \bar{\bold{l}}_j$ from $\bold{\bar{c}}_i^T \bold{\bar{c}}_j$ can be bounded and understood in terms of $\bold{V} \in \mathbb{R}^{N \times K}$. In the simplest case, if $\bold{V}$ is orthonormal then $\|\bold{V}\bold{c}\| = \|\bold{c}\|$ \citep{Gentle2010}, and:
\begin{align}
    \|\bold{\bar{l}}_i - \bold{\bar{l}}_j \|_2^2 =
    \left\lVert \frac{\bold{V}\bold{c}_i}{|| \bold{V}\bold{c}_i ||_2^2} - \frac{\bold{V}\bold{c}_j}{|| \bold{V}\bold{c}_j ||_2^2} \right\rVert_2^2 = 
    \left\lVert\bold{V} \left(\frac{\bold{c}_i}{\|\bold{c}_i\|} - \frac{\bold{c}_j}{\|\bold{c}_j\|}\right) \right\rVert_2^2 = 
    \| \bold{\bar{c}}_i - \bold{\bar{c}}_j \|_2^2\\
    \implies \bold{\bar{l}}_i^T \bold{\bar{l}}_j = \bold{\bar{c}}_i^T \bold{\bar{c}}_j
\end{align}
However $\bold{V}$ is not restricted to be orthonormal, as discussed in Section~\ref{sec: sys_model}. In the general case, the worst case deviation between $\bar{\bold{l}}_i^T \bar{\bold{l}}_j$ and $\bar{\bold{c}}_i^T \bar{\bold{c}}_j$ can be understood in terms of the singular values of $\bold{V}$; we follow the approach of \citet{whuber}. Denoting the rank $K$ singular value decomposition as $\bold{V} = \bold{A}\bold{\Sigma}\bold{B}^T$ (Appendix~\ref{ap: PCA}), $\bold{\Sigma}$ is comprised of singular values of descending magnitude $\sigma_1 \geq \sigma_2... \geq \sigma_K \geq 0$ and where, without loss of generality, we adopt $\sigma_1 = 1$. Matrix $\bold{A}$ contains the left singular vectors $\{ \bold{a}_k : k \in K \}$ and matrix $\bold{B}$ the right singular vectors $\{ \bold{b}_k : k \in K \}$. The systematics correlation $ \bar{\bold{l}}_i^T \bar{\bold{l}}_j$ can be expanded as:
\begin{align} \label{eq: sys_corr}
        \bar{\bold{l}}_i^T \bar{\bold{l}}_j= \frac{\bold{c}_i^T \bold{V}^T \bold{V} \bold{c}_j}{\|\bold{V}\bold{c}_i \| \|\bold{V}\bold{c}_j \|}  = \frac{\bold{c}_i^T \bold{B} \bold{\Sigma} \bold{A}^T\bold{A} \bold{\Sigma} \bold{B}^T \bold{c}_j}{\|\bold{A}\bold{\Sigma} \bold{B}^T \bold{c}_i \| \|\bold{A}\bold{\Sigma} \bold{B}^T\bold{c}_j \|} \\
         = \frac{\bold{c}_i^T \bold{B} \bold{\Sigma}^2 \bold{B}^T \bold{c}_j}{\|\bold{\Sigma} \bold{B}^T \bold{c}_i \| \|\bold{\Sigma} \bold{B}^T\bold{c}_j \|} 
\end{align}
since $\bold{A}$ is orthonormal $\bold{A}^T\bold{A} = \mathds{1}_K$ and therefore $\|\bold{A}\bold{\Sigma} \bold{B}^T\bold{c} \| =  \|\bold{\Sigma} \bold{B}^T\bold{c} \|$. 
By the cosine similarity \citep{phillips2021}, $\bold{\bar{c}}_i^T \bold{\bar{c}}_j = \cos(\phi)$ where  $\phi$ is the angle between the normalized coefficient vectors in $K$-dimensional space. The correlation $\bold{\bar{l}}_i^T \bold{\bar{l}}_j$ is similarly determined by the angle $\psi$ between unit length vectors $\frac{\bold{\Sigma}\bold{B}^T\bold{c}_i}{ \| \bold{\Sigma}\bold{B}^T\bold{c}_i \| } $ and $\frac{\bold{\Sigma}\bold{B}^T\bold{c}_j}{ \| \bold{\Sigma}\bold{B}^T\bold{c}_j \| } $. The deviation of $\bar{\bold{l}}_i^T \bar{\bold{l}}_j$ from $\bold{\bar{c}}_i^T \bold{\bar{c}}_j$ is defined as $|\phi-\psi|$.\\

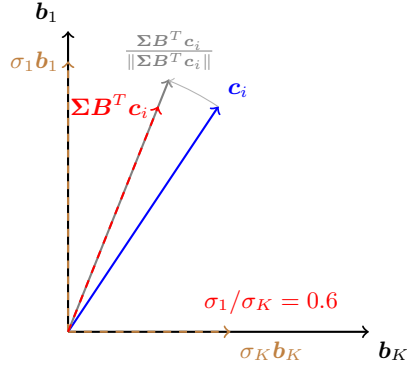
\begin{figure}[!htb]
\centering
\begin{tikzpicture}
\dimendef\prevdepth=0

\draw[thick,->] (0,0) -- (4,0) node[anchor=north west] {$\boldsymbol{b}_K$};
\draw[thick,->] (0,0) -- (0,4) node[anchor=south east] {$\boldsymbol{b}_1$};

\draw[thick,->,blue] (0,0) -- (2,3) node[anchor=south west] {$\boldsymbol{c}_i$};

\draw[thick,->,dashed,brown] (0,0) -- (3.6*.6,0) node[anchor=north west] {$\sigma_K\boldsymbol{b}_K$};
\draw[thick,->,dashed,brown] (0,0) -- (0,3.6) node[anchor=east] {$\sigma_1\boldsymbol{b}_1$};

\coordinate (origin) at (0,0);
\coordinate (a) at (2,3);
\coordinate (b) at (1.2*1.12,3*1.12);

\draw[thick,->,gray] (0,0) -- (1.2*1.12,3*1.12) node[anchor=south] {$\frac{\boldsymbol{\Sigma}\boldsymbol{B}^T\boldsymbol{c}_i}{\|\boldsymbol{\Sigma}\boldsymbol{B}^T\boldsymbol{c}_i\|}$};

\draw[thick,->,red,dashed] (0,0) -- (1.2,3) node[anchor=east] {$\boldsymbol{\Sigma}\boldsymbol{B}^T\boldsymbol{c}_i$};

\pgfgetlastxy{\baseunitx}{\baseunity}

\pic [draw, ->, color=gray!60, angle radius = {3 * \baseunitx}, angle eccentricity=1.5] {angle = a--origin--b};

\node[text=red] at (2.7,.4) {$\sigma_1 / \sigma_K = 0.6$};
\end{tikzpicture}

\caption{This illustration shows an example coefficient vector $\bold{c}_i = 3\bold{b}_1 + 2\bold{b}_K$ (blue), linearly transformed to vector  $\bold{\Sigma}\bold{B}^T\bold{c}_i = 3\bold{b}_1 + \sigma_K 2\bold{b}_K$ (red), plotted on the 2D space spanned by $\bold{b}_1$ and $\bold{b}_K$ (brown). The normalized transformed vector is also shown (gray). A value $\sigma_K = 0.6$ was adopted for this example (recall $\sigma_1=1$). } \label{fig: rot}

\end{figure}

The transform $\bold{\Sigma}\bold{B}^T \bold{c}_i$ projects $\bold{c}_i$ along each $\bold{b}_k$ and scales by $\sigma_k$:  $\bold{\Sigma}\bold{B}^T\bold{c}_i = [\sigma_1 \bold{b}_1^T \bold{c}_i, \dots, \sigma_K \bold{b}_K^T \bold{c}_i]^T $. Any $\bold{c}_i \in \mathbb{R}^K$ must lie in the span of orthonormal vectors $\{ \bold{b}_k : k \in K \}$ as the $\bold{b}_k$ are a basis for $\mathbb{R}^K$. For simplicity, assume that $\bold{c}_i$ and $\bold{c}_j$ lie in the 2D span of two of the basis vectors $\bold{b}_m$ and $\bold{b}_n$ in the form $\bold{c}_i=\alpha_i \bold{b}_m + \beta_i \bold{b}_n$ and $\bold{c}_j=\alpha_j \bold{b}_m + \beta_j \bold{b}_n$. Then  $\bold{\Sigma}\bold{B}^T\bold{c}_i = \sigma_m \alpha_i \bold{b}_m + \sigma_n \beta_i \bold{b}_n$ and $\bold{\Sigma}\bold{B}^T\bold{c}_j = \sigma_m \alpha_j \bold{b}_m + \sigma_n \beta_j \bold{b}_n$. Then $\phi=\arctan{\beta_j/\alpha_j}-\arctan{\beta_i/\alpha_i}$ and $\psi=\arctan{\frac{\sigma_n \beta_j}{\sigma_m \alpha_j}}-\arctan{\frac{\sigma_n \beta_i}{ \sigma_m \alpha_i}}$. Therefore the maximum differential rotation $|\phi-\psi|$ occurs when the ratio between singular values $\frac{\sigma_m}{\sigma_n}$ is greatest, i.e. when $m=1$ and $n=K$. In this case, the transformed vector component along $\bold{b}_1$ is unchanged (as $\sigma_1=1$) while the transformed component along orthogonal vector $\bold{b}_K$ is scaled by the smallest relative value $\sigma_K$; this maximizes rotation. The rotation of a single coefficient vector $\bold{c}_i$ under the linear transformation $\bold{\Sigma}\bold{B}^T$ is illustrated in Figure~\ref{fig: rot}.

\begin{figure}[!htb]
\centering
\begin{tikzpicture}
\dimendef\prevdepth=0

    \draw[thick,<->] (-5/3,0) -- (5/3,0) node[anchor=north west] {$\boldsymbol{b}_K$};
    \draw[thick,<->] (0,-5/3) -- (0,5/3) node[anchor=south east] {$\boldsymbol{b}_1$};

    \draw[thick,->,blue] (0,0) -- (4/3,2/3) node[anchor=west] {$\boldsymbol{c}_i$};
    \draw[thick,->,blue] (0,0) -- (2/3,4/3) node[anchor=east] {$\boldsymbol{c}_j$};
    \draw[thick,-,gray] (0,0) -- (1.2,1.2) node[anchor=south west] {Bisector};

    \coordinate (origin) at (0,0);
    \coordinate (b1) at (0,5/3);
    \coordinate (bisector) at (1.2,1.2);
    \pic [draw, <-, "$\theta$", angle eccentricity=1.5] {angle = bisector--origin--b1};
\end{tikzpicture}
\caption{A bisector at angle $\theta$ between coefficient vectors $\bold{c}_i$ and $\bold{c}_j$ in the 2D space spanned by basis vectors $\bold{b}_1$ and $\bold{b}_K$.} \label{fig: bisect}
\end{figure}
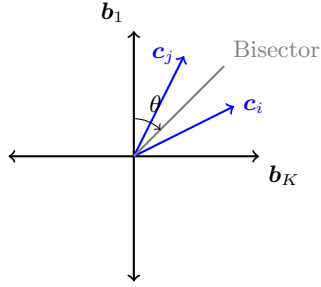

The maximum differential rotation $|\phi-\psi|$ depends on the relative position of the coefficient vectors $\bold{c}_i$ and $\bold{c}_j$ before linear transformation. In this 2D case we denote their bisector angle as $\theta$, as depicted in Figure~\ref{fig: bisect}. By geometric inspection and by considering the algebra above for the 2D case, the maximal differential rotation $|\phi-\psi|$ occurs when $\theta = 0$ and the coefficient vectors are bisected by $\bold{b}_1$. The minimal differential rotation occurs when $\theta = \frac{\pi}{2}$ and the coefficient vectors are bisected by $\bold{b}_K$. For these values of $\theta$ and following the trigonometric approach by \citet{whuber}, the upper and lower limits for the correlation $\bold{\bar{l}}_i^T\ \bold{\bar{l}}_j$ of the normalized systematics vectors can be expressed the form: 

\begin{align} \label{eq: bound_dev}
    \frac{\sigma_K^2 - \tan^2(\phi)}{\tan^2(\phi) + \sigma_K^2} \leq \bar{\bold{l}}_i^T \bar{\bold{l}}_j\leq \frac{1-\sigma_K^2 \tan^2(\phi)}{1 + \sigma_K^2 \tan^2 (\phi)}
\end{align}

\begin{figure}[ht!]
  \centering
  \includegraphics{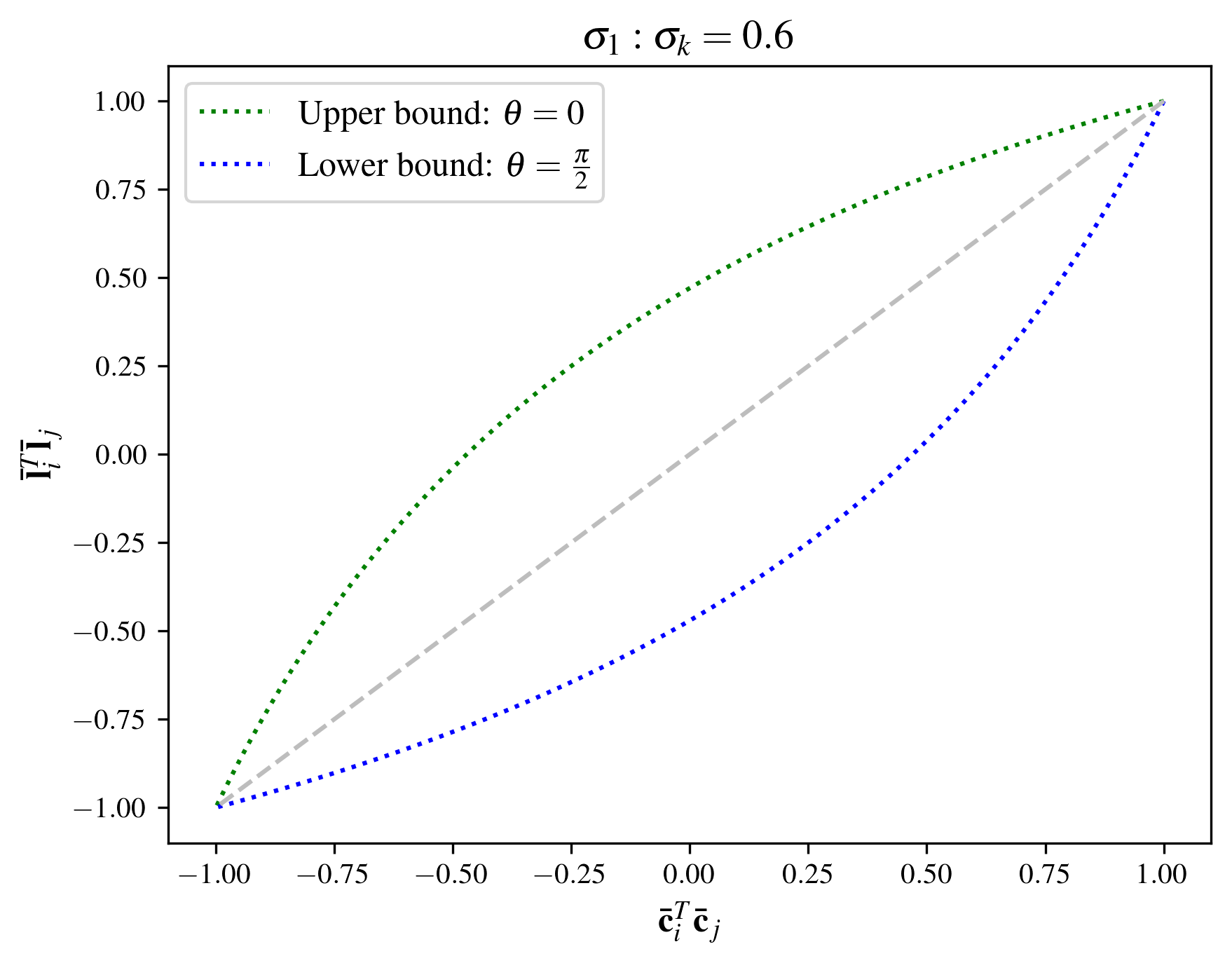}\caption{Deviation of the systematics correlation $\bold{\bar{l}}_i^T \bold{\bar{l}}_j$, upper bound (green) and lower bound (blue), from the coefficient correlation $\bold{\bar{c}}_i^T\bold{\bar{c}}_j$ (gray) for $\sigma_1: \sigma_K = 0.6$.\label{fig:corr_sys_dev}}
\end{figure}

As the correlation $\bold{\bar{c}}_i^T\bold{\bar{c}}_j \to 1$ implies $\phi \to 0$, the lower and upper bounds smoothly converge to unity and $\bar{\bold{l}}_i^T \bar{\bold{l}}_j \to 1$ in this limit. This implies that the deviation between $\bar{\bold{l}}_i^T \bar{\bold{l}}_j$ and $\bar{\bold{c}}_i^T \bar{\bold{c}}_j$ is lower when the absolute value of the coefficient correlation is high. Furthermore the worst case deviation depends on the ratio of the smallest and largest singular values of $\bold{V}$; when $\sigma_K = \sigma_1$, $\bold{V}$ is orthonormal and there is no deviation, also shown above. In Figure \ref{fig:corr_sys_dev} the deviation bounds are shown as a function of the coefficient correlation for a hypothetical value $\sigma_K = 0.6\ (\sigma_1=1)$.

\section{Spatial systematics optimization method}
In this section we derive gradient descent steps for minimizing the objective in Equation \ref{eq: obj1}, repeated here for convenience: 
\begin{align}
    \argmin_{\bold{V}, \bold{C}\; : \; rank(\bold{V}\bold{C}) \leq K} ||\bold{Y} - \bold{V}\bold{C} ||_F^2  + ||\bold{D}  \bold{\bar{C}} ||_{2, p}^p 
\end{align}
As above, the least-squares penalty is denoted as $f(\bold{V}, \bold{C}) = ||\bold{Y} - \bold{V}\bold{C} ||_F^2 $ and the total variation spatial constraint as $g(\bold{C}) = ||\bold{D}  \bold{\bar{C}} ||_{2, p}^p$. For clarity of presentation we use an unweighted difference operator $\bold{D}$ instead of $\bold{D}_{\bold{W}}$ (Equation~\ref{eq: obj1}).

\subsection{Variable Projection} \label{ap: vp}
In this section we use variable projection to eliminate the dependence of the least-squares penalty on $\bold{V}$. An overview of variable projection applied to separable non-linear least-squares problems is provided by \citet{Golub_2003}. For completeness we note that alternating iterative minimization over $\bold{V}$ and $\bold{C}$ could also be used, however this would not necessarily be superior to variable projection used here. For simplicity, we work with a transposed version of the least-squares penalty: 
\begin{align}
    f(\bold{V}, \bold{C}) = \|\bold{Y}-\bold{V}\bold{C} \|_F^2 = \|\bold{Y}^T - \bold{C}^T \bold{V}^T \|_F^2
\end{align}
We denote as $h(\bold{C}) = \argmin_{\bold{V}} \|\bold{Y}^T - \bold{C}^T \bold{V}^T \|_F^2$ the value $\bold{V}$ that minimizes this penalty for a particular $\bold{C}$. Here, and afterwards $h(\bold{\cdot})^T$ denotes the transpose of $h(\cdot)$. The minimum $\bold{V}^T$ is the minimum-norm least-squares solution, which is the pseudoinverse of $\bold{C}^T$ applied to $\bold{Y}^T$ \citep{golub2013matrix}. Denoting the pseudoinverse of $\bold{C}^T$ as $\bold{C}^{\dagger} = (\bold{C}\bold{C}^T)^{-1}\bold{C}$ we obtain $h(\bold{C})^T = \bold{C}^\dagger \bold{Y}^T$. 
Rewriting the least-squares term using the conditional solution:
\begin{align}
    f(h(\bold{C}), \bold{C}) = ||\bold{Y}^T - \bold{C}^T h(\bold{C})^T ||_F^2 \\
    = ||\bold{Y}^T - \bold{C}^T\bold{C}^\dagger\bold{Y}^T ||_F^2\\
    = || (\mathds{1}_{I} -  \bold{C}^T\bold{C}^\dagger)\bold{Y}^T ||_F^2
\end{align}
As $\bold{C}^T\bold{C}^\dagger = \mathbf{P}_{R(\bold{C}^T)}$ is the projection onto the range space of $\bold{C}^T$ \citep{Gentle2010}, then $(\mathds{1}_{I} - \mathbf{P}_{R(\bold{C}^T)}) =  \mathbf{P}_{R(\bold{C}^T)}^{\perp}$ is the projection onto the orthogonal complement to the range space of $\bold{C}^T$. The variable-reduced least-squares penalty is denoted as $f'(\bold{C})$:
\begin{align}
   f'(\bold{C}) = f(h(\bold{C}), \bold{C}) = || \mathbf{P}_{R(\bold{C}^T)}^{\perp} \bold{Y}^T||_F^2
\end{align}

\subsection{Gradient Descent} \label{ap: grad}
We use gradient descent to minimize $w(\bold{C}) = ||(\mathds{1}_{I} - \bold{C}^T\bold{C}^\dagger)\bold{Y}^T||_F^2  + ||\bold{D}  \bold{\bar{C}}^T ||_{2,p}^p=f'(\bold{C})+ g(\bold{C})$ with respect to $\bold{C}^T \in \mathbb{R}^{I \times K}$.
The gradient $\nabla w(\bold{C}^T) \in  \mathbb{R}^{I \times K}$ is the matrix of partial derivatives. We denote the gradient of the variable projection least-squares term as $\nabla f'({\bold{C}^T})$ and the gradient of the total-variation constraint as $\nabla g({\bold{C}^T})$. The total gradient is:
\begin{align}
    \nabla w({\bold{C}^T}) = \nabla f'({\bold{C}^T}) + \nabla g({\bold{C}^T})
\end{align}
where $\nabla w(\bold{C})^T = \nabla w(\bold{C}^T)$.

\subsubsection{Variable Projection Least-Squares Gradient}
For clarity in computing $\nabla f' (\bold{C}^T)$, we expand the variable-projection least-squares penalty as $|| \mathbf{P}_{R(\bold{C}^T)}^{\perp} \bold{Y}^T||^2_F = \sum_{N} ||\mathbf{P}_{R(\bold{C}^T)}^{\perp} [\bold{Y}]_n ||_F^2 $ where $[\bold{Y}]_n$ is the $n^{th}$ row of $\bold{Y} \in  \mathbb{R}^{N \times I}$. 
Noting the Frobenius matrix norm may be expressed as a Frobenius matrix inner product, as $\|\bold{A}\|_F^2 = \ip{\bold{A}}{\bold{A}}$, where the Frobenius matrix inner product is defined as $\ip{\bold{A}}{\bold{B}} = tr(\bold{A}^T \bold{B})$ and $tr$ denotes a trace \citep{matrixcook}. By application of the chain rule:
\begin{align}
d ||\mathbf{P}_{R(\bold{C}^T)}^{\perp} [\bold{Y}]_n ||_F^2 = \ip{2 \mathbf{P}_{R(\bold{C}^T)}^{\perp} [\bold{Y}]_n }{d \mathbf{P}_{R(\bold{C}^T)}^\perp [\bold{Y}]_n }\\
= \ip{2 \mathbf{P}_{R(\bold{C}^T)}^{\perp} [\bold{Y}]_n [\bold{Y}]_n^T }{d \mathbf{P}_{R(\bold{C}^T)}^\perp}  \label{eq: eq_1_var}
\end{align}
The derivative of the projection matrix is shown due to \citet{harville2008matrix} (Theorem 15.11.1):
\begin{align}
d \mathbf{P}_{R(\bold{C}^T)}^\perp = - \mathbf{P}_{R(\bold{C}^T)}^{\perp} (d \bold{C}^T ) \bold{C}^\dagger - (\bold{C}^\dagger)^T (d \bold{C}) \mathbf{P}_{R(\bold{C}^T)}^{\perp}
\end{align} 
Substituting this result for $d \mathbf{P}_{R(\bold{C}^T)}^\perp $ into \ref{eq: eq_1_var}:
\begin{align}
d ||\mathbf{P}_{R(\bold{C}^T)}^{\perp} [\bold{Y}]_n ||_F^2 
= - 2 \ip{\mathbf{P}_{R(\bold{C}^T)}^{\perp} [\bold{Y}]_n [\bold{Y}]_n^T}{  \mathbf{P}_{R(\bold{C}^T)}^{\perp} (d \bold{C}^T) \bold{C}^\dagger + (\bold{C}^\dagger)^T (d \bold{C}) \mathbf{P}_{R(\bold{C}^T)}^{\perp}} \\
=  - 2 \ip{\mathbf{P}_{R(\bold{C}^T)}^{\perp \; T} \mathbf{P}_{R(\bold{C}^T)}^{\perp} [\bold{Y}]_n [\bold{Y}]_n^T (\bold{C}^\dagger)^T}{ d \bold{C}^T} - 2 \ip{ \bold{C}^\dagger \mathbf{P}_{R(\bold{C}^T)}^{\perp} [\bold{Y}]_n [\bold{Y}]_n^T}{(d \bold{C}) \mathbf{P}_{R(\bold{C}^T)}^{\perp}}  \label{eq: var_proj_grad}
\end{align}
Noting that an orthogonal projection matrix $\mathbf{P}$ is symmetric $\mathbf{P}^T = \mathbf{P}$ and furthermore, idempotent $\mathbf{P} \mathbf{P} = \mathbf{P}$ \citep{golub2013matrix}, the first term in Equation $\ref{eq: var_proj_grad}$ may be further simplified. 
Consider $\bold{C}^\dagger \mathbf{P}_{R(\bold{C}^T)}^{\perp} = \left( \mathbf{P}_{R(\bold{C}^T)}^{\perp} (\bold{C}^\dagger )^T \right)^T$. It can be seen that the range space of $(\bold{C}^\dagger )^T$ is the range space of $\bold{C}^T$, therefore $\mathbf{P}_{R(\bold{C}^T)}^{\perp} (\bold{C}^\dagger )^T = 0$ and $\bold{C}^\dagger \mathbf{P}_{R(\bold{C}^T)}^{\perp} = 0$, so that the second term in Equation $\ref{eq: var_proj_grad}$ is dropped: 
\begin{align} 
d ||\mathbf{P}_{R(\bold{C}^T)}^{\perp} [\bold{Y}]_n ||_F^2 
=  - 2 \ip{ \mathbf{P}_{R(\bold{C}^T)}^{\perp} [\bold{Y}]_n [\bold{Y}]_n^T (\bold{C}^\dagger)^T}{ d \bold{C}^T}
\end{align}
Re-arranging and summing together the derivatives as $\sum_{N} \frac{d}{d \bold{C}^T} ||\mathbf{P}_{R(\bold{C}^T)}^{\perp} [\bold{Y}]_n ||_F^2 $, the gradient $ \nabla f' (\bold{C}^T) \in \mathbb{R}^{I \times K}$ is:
\begin{align}
    \nabla f'(\bold{C}^T) = -2 \mathbf{P}_{R(\bold{C}^T)}^{\perp} \left(\sum_n[\bold{Y}]_n[\bold{Y}]_n^T \right) (\bold{C}^{\dagger})^T
 \end{align}
This can be compactly expressed as:
\begin{align}
    \nabla f'(\bold{C}^T) = - 2 \mathbf{P}_{R(\bold{C}^T)}^{\perp} \bold{Y}^T \bold{Y} (\bold{C}^{\dagger})^T 
 \end{align}
\subsubsection{Total-Variation Gradient}
The derivative of $||\bold{D}  \bold{\bar{C}}^T ||_{2,p}^p$ can be computed by application of the chain rule: 
\begin{align} \label{eq: chain}
    \nabla g({\bold{C}^T}) = \frac{d}{d \bold{C}^T} ||\bold{D} \bold{\bar{C}}^T ||_{2,p}^p =  \frac{d}{d \bold{\bar{C}}^T} ||\bold{D} \bold{\bar{C}}^T ||_{2,p}^p \cdot \frac{d \bold{\bar{C}}^T}{d \bold{C}^T}
\end{align}  

The matrix $\bold{D} \bold{\bar{C}}^T \in \mathbb{R}^{2 \times K\cdot X\cdot Y}$ has columns for each coefficient $k \in K$ and each pixel $j \to (x,y) \in X \times Y$, formed as $[[\bold{D}_x \bold{\bar{C}}^T]_{j,k}, [\bold{D}_y \bold{\bar{C}}^T]_{j,k}] = [\bar{c}_{x,y}^k - \bar{c}^k_{x+1,y},  \bar{c}_{x,y}^k - \bar{c}_{x,y+1}^k]$, where $\bold{D}_x \in \mathbb{R}^{ (X-1)Y\times XY}$ and $\bold{D}_y \in \mathbb{R}^{ X(Y-1)\times XY}$ are difference matrices in $x$ and $y$ respectively.  Note, for the edge pixels in top row and right-most column, difference values are only included along X or Y respectively. \\

The total variation penalty computes the $L_p^p$ norm over the $L_2$ norm of each column of $\bold{D}\bold{\bar{C}}^T$, given by $ \|\bold{D}\bold{\bar{C}}^T\|_{2,p}^p = \sum_{n} ||[\bold{D}\bold{\bar{C}}^T]_{\cdot, n}||_2^p = \sum_k \sum_j \| [[\bold{D}_x \bold{\bar{C}}^T]_{j,k}, [\bold{D}_y \bold{\bar{C}}^T]_{j,k}] \|_2^{p}$. Here each index $n$ maps to some $(j, k)$, but the ordering does not affect the computation.\\

The $L_p$ norm is non-differentiable at zero for $p \in [1,2)$ and a differentiable approximation of the $L_p$ norm is used instead in the form of the Huber functional \citep{vogel_book}. 
The Huber functional avoids zero values by adding a very small positive value $\delta$ to every element. 

We apply the Huber functional to arrive at the modified spatial constraint:
\begin{align}
    ||\bold{D} \bold{\bar{C}}^T ||_{2,p}^p \approx \sum_{n} (||[\bold{D}\bold{\bar{C}}^T]_{\cdot, n}||_2^2 + \delta)^\frac{p}{2}
\end{align}
where
\begin{align}
 ||[\bold{D}\bold{\bar{C}}^T]_{\cdot, n}||_2^2 = [\bold{D}_x \bold{\bar{C}}^T]_{n \to j,k}^2 + [\bold{D}_y \bold{\bar{C}}^T]_{n \to j,k}^2
\end{align}
The derivative is decomposable over $k$ and for simplicity we find the derivative with respect to each column of $[\bold{\bar{C}}^T]_{\cdot, k}$. Using
$\frac{d [\bold{D}_x \bold{\bar{C}}^T]^2_{j,k}}{d [\bold{\bar{C}^T}]_{\cdot, k}} = 2 [\bold{D}_x \bold{\bar{C}}^T]_{j,k}[\bold{D}_x]_j^T $, and by applying the chain rule we derive: 

\begin{align}
    \frac{d  ||\bold{D} \bold{\bar{C}}^T ||_{2,p}^p}{d [ \bold{\bar{C}^T}]_{\cdot, k}}= p
    \sum_j \left([\bold{D}_x \bold{\bar{C}}^T]_{j,k}^2 + [\bold{D}_y \bold{\bar{C}}^T]_{j,k}^2 + \delta \right)^{\frac{p}{2}-1}  \left( [\bold{D}_x \bold{\bar{C}}^T]_{j,k}[\bold{D}_x]_j^T + [\bold{D}_y \bold{\bar{C}}^T]_{j,k}[\bold{D}_y]_j^T  \right) 
\end{align}

We denote $\bold{L}_k = \diag \left( p([\bold{D}_x \bold{\bar{C}}^T]_{j,k}^2 + [\bold{D}_y \bold{\bar{C}}^T]_{j,k}^2  + \delta )^{\frac{p}{2}-1}\ \forall\ j\right)$. The sum over $j$ allows the expression to be written compactly as: 

\begin{align}
    \frac{d  ||\bold{D} \bold{\bar{C}}^T ||_{2,p}^p}{d [ \bold{\bar{C}^T}]_{\cdot, k}} = (\bold{D}_x^T \bold{L}_k \bold{D}_x + \bold{D}_y^T \bold{L}_k \bold{D}_y ) [\bold{\bar{C}}^T]_{,k}
\end{align}

The final term in Equation~\ref{eq: chain} is the derivative due to the row-wise normalization of $\bold{\bar{C}}^T$. 
\begin{align}
    \frac{d [\bold{\bar{C}}^T]_i}{d [\bold{C}^T]_j} =
    \begin{cases}
     (\mathds{1}_K - [\bar{\bold{C}}^T]_i [\bar{\bold{C}}^T]_i^T) \frac{1}{||[{\bold{C}}^T]_i ||} \quad i=j \\
    0 \quad i \neq j
    \end{cases}
\end{align}
Each row $i$ of the final derivative is the product of $ \left[ \frac{d  ||\bold{D} \bold{\bar{C}}^T ||_p}{d [ \bold{\bar{C}^T}]} \right]_{i}$ and $\frac{d [\bold{\bar{C}}^T]_i}{d [\bold{C}^T]_i}$ such that 
\begin{align}
    [\nabla g(\bold{C}^T)]_{i} = [\bold{B}_1 [\bold{\bar{C}}^T]_{\cdot,1},\dots\bold{B}_K [\bold{\bar{C}}^T]_{\cdot,K} ]_{i} (\mathds{1}_K - [\bar{\bold{C}}^T]_i [\bar{\bold{C}}^T]_i^T) \frac{1}{||[{\bold{C}}^T]_i ||}
\end{align}
where $\bold{B}_k =\bold{D}_x^T \bold{L}_k \bold{D}_x + \bold{D}_y^T \bold{L}_k \bold{D}_y$. \\

For the case $p=2$ an analytic solution can be computed without the use of the Huber functional approximation. 

\begin{align} 
    [\nabla g(\bold{C}^T)]_i = [(\bold{D}_x^T \bold{D}_x + \bold{D}_y^T \bold{D}_y)\bar{\bold{C}}^T]_i (\mathds{1}_K - [\bar{\bold{C}}^T]_i [\bar{\bold{C}}^T]_i^T) \frac{1}{||[{\bold{C}}^T]_i ||}
\end{align}
It can be seen that the effect of the $L_p$ norm for $p \in [1,2)$ is contained in the diagonal weightings $\bold{L}_k$. 

\bibliography{sample}{}
\bibliographystyle{aasjournal}

\end{document}